\documentclass{aa}  
\usepackage{graphicx}
\usepackage{txfonts}
\usepackage{lipsum}
\usepackage{subcaption}
\usepackage{lscape}
\usepackage{placeins}
\usepackage{hyperref}
\hypersetup{
    colorlinks=true,
    linkcolor=blue,
    filecolor=blue,      
    urlcolor=blue,
    citecolor=blue
}

\begin{document}

   \title{The mass distribution of stars stripped in binaries: The effect of metallicity}

   \author{B. Hovis-Afflerbach\inst{1,2,3,4}
        \and Y. G\"{o}tberg\inst{5,4}
        \and A. Schootemeijer\inst{6}
        \and J. Klencki\inst{7,8}
        \and A. L. Strom\inst{1,2}
        \and B. A. Ludwig\inst{9}
        \and M. R. Drout\inst{9}
        }

   \institute{Department of Physics and Astronomy, Northwestern University, 2145 Sheridan Rd, Evanston, IL 60208, USA\\
            \email{beryl@u.northwestern.edu}
            \and Center for Interdisciplinary Exploration and Research in Astrophysics (CIERA), Northwestern University, 1800 Sherman Ave, Evanston, IL 60201, USA
            \and Division of Physics, Mathematics, and Astronomy, California Institute of Technology, Pasadena, CA 91125, USA
            \and The Observatories of the Carnegie Institution for Science, 813 Santa Barbara Street, Pasadena, CA 91101, USA
            \and Institute of Science and Technology Austria (ISTA), Am Campus 1, 3400 Klosterneuburg, Austria\\
            \email{ylva.gotberg@ista.ac.at}
            \and Argelander-Institut f\"{u}r Astronomie, Universit\"{a}t Bonn, Auf dem H\"{u}gel 71, 53121 Bonn, Germany
            \and Max Planck Institute for Astrophysics, Karl-Schwarzschild-Strasse 1, 85748 Garching, Germany
            \and European Southern Observatory, Karl-Schwarzschild-Strasse 2, 85738 Garching bei M\"{u}nchen, Germany
            \and David A. Dunlap Department of Astronomy and Astrophysics, University of Toronto, 50 St. George Street, Toronto, ON, M5S3H4, Canada}

   \date{Received 28 November 2024 /
Accepted 14 March 2025 }

  \abstract
  {
  Stars stripped of their hydrogen-rich envelopes through binary interaction are thought to be responsible for both hydrogen-poor supernovae and the hard ionizing radiation observed in low-$Z$ galaxies. A population of these stars was recently observed for the first time, but their prevalence remains unknown. In preparation for such measurements, we estimate the mass distribution of hot, stripped stars using a population synthesis code that interpolates over detailed single and binary stellar evolution tracks. We predict that for a constant star formation rate of $1 \,M_\odot$/yr and regardless of metallicity, a scalable model population contains $\sim$30,000 stripped stars with mass $M_{\rm strip}>1M_\odot$ and $\sim$4,000 stripped stars that are sufficiently massive to explode ($M_{\rm strip}>2.6M_\odot$). Below $M_{\rm strip}=5M_\odot$, the distribution is metallicity-independent and can be described by a power law with the exponent $\alpha \sim -2$. At higher masses and lower metallicity ($Z \lesssim 0.002$), the mass distribution exhibits a drop. This originates from the prediction, frequently seen in evolutionary models, that massive low-metallicity stars do not expand substantially until central helium burning or later and therefore cannot form long-lived stripped stars. With weaker line-driven winds at low metallicity, this suggests that neither binary interaction nor wind mass loss can efficiently strip massive stars at low metallicity. As a result, a ``helium-star desert'' emerges around $M_{\rm strip} =15\, M_\odot$ at $Z=0.002$, covering an increasingly large mass range with decreasing metallicity. We note that these high-mass stars are those that potentially boost a galaxy's He$^+$-ionizing radiation and that participate in the formation of merging black holes. This ``helium-star desert'' therefore merits further study.}
  

   \maketitle

\nolinenumbers
\section{Introduction} \label{sec:intro}

Metallicity has a significant impact on stellar evolution and structure, with lower metallicity typically leading to hotter stars with harder ionizing spectra \citep{2002MNRAS.337.1309S, 2016MNRAS.456..485S}.
Stars in high-redshift galaxies formed at a time when most of the metals in existence today had not yet been created, and accordingly, high-redshift galaxies have been shown to generally have lower stellar metallicity than galaxies today \citep[e.g.,][]{2001ApJ...554..981P, 2002A&A...391...21P, 2003ApJ...588...65S, 2016ApJ...826..159S, 2018ApJ...868..117S, 2019MNRAS.487.2038C, 2020MNRAS.499.1652T}. 
This has been confirmed with new data from the James Webb Space Telescope (JWST), which has further shown that galaxies at redshift $z \gtrsim 2$ have very different chemical abundances from those nearby to our Milky Way \citep[e.g.,][]{2024ApJ...964L..12R, 2024MNRAS.529.3301T}.

Observations of high-redshift galaxies have also revealed that they have hard ionizing spectra, with more photons emitted at higher energies \citep[e.g.,][]{2020A&A...636A..47S, 2023A&A...677A..88B}.
A hard ionizing spectrum is often inferred from nebular He II lines at 1640 \AA \ or 4686~\AA, because only high energy photons with $> 54.4$ eV are able to ionize helium.
This has also been observed in nearby low-metallicity, star-forming galaxies \citep{2019MNRAS.488.3492S, 2019ApJ...878L...3B, 2022ApJS..261...31B}.
The source of this ionizing radiation remains unclear, but many potential sources have been proposed, including X-ray binaries, shocks, massive stars, and binary products \citep{2016MNRAS.458..624C, 2019A&A...622L..10S, 2020MNRAS.494..941S}.

Wolf-Rayet (WR, an observationally motivated classification) stars are perhaps the most efficient stellar source of ionizing radiation.
However, in the classical prescription for WR star formation, they are predominantly formed through strong stellar winds that are thought to be driven by metal-rich stars \citep[][see however \citealt{2020A&A...634A..79S}]{2007ARA&A..45..177C, 2014ARA&A..52..487S}.
The confirmed existence of WR stars in local metal-poor star-forming galaxies \citep{2003ARA&A..41...15M} is therefore puzzling, and it has been suggested that these stars are formed instead by envelope-stripping in binaries \citep[][see however also \citealt{2024A&A...689A.157S}]{1967AcA....17..355P,1998NewA....3..443V}.

Indeed, envelope-stripping in binary stars has been proposed to substantially boost a stellar population's ionizing photon emission, but with a delay of $\sim 10 - 100$ Myr after a starburst \citep{2016MNRAS.459.3614M, 2019A&A...629A.134G}.
This is because the stripping of the hydrogen-rich envelope reveals the hot and compact helium core, but only after the star has experienced the main sequence   \citep{2017A&A...608A..11G,2018A&A...615A..78G}.
Because the relatively long-lasting central helium burning phase still remains, such stripped stars should be sufficiently long-lived and exist in sufficient numbers to have an observable impact on the population's ionizing spectrum \citep{2020A&A...634A.134G}.

Although the existence of binary-stripped massive stars was proposed over half a century ago \citep{1967ZA.....65..251K}, the first sample of hot and intermediate-mass binary-stripped stars was observationally confirmed only recently \citep[][see also \citealt{2008A&A...485..245G} and \citealt{2023Sci...381..761S}]{2023Sci...382.1287D,2023ApJ...959..125G}.
This discovery highlights the importance of better understanding these objects, both for their impact on full stellar populations in distant galaxies and for massive star evolution in general.

While several WR stars have been found orbiting companion stars \citep{2016A&A...591A..22S,2017MNRAS.464.2066S,2020MNRAS.492.4430S}, \cite{2022A&A...662A..56K} showed that massive stars ($\sim20-50 M_\odot$) at low metallicity ($Z\lesssim 0.005$) may be less likely to create hot, binary-stripped stars. Instead of detaching and contracting to small sizes and high surface temperatures, such stars are predicted to either remain in the mass-transfer phase throughout the rest of their lives \citep[e.g.,][]{2024MNRAS.529.1104B} or only experience partial envelope-stripping, resulting in relatively cool stripped stars (with effective temperatures $T_{\rm eff} \lesssim 30,000$ K). While these cooler surface temperatures could give them a similar appearance to helium giants \citep[e.g.,][]{2020A&A...637A...6L}, they are expected to be substantially longer lived since they undergo central helium burning at these cooler temperatures. In addition, it is possible that their surface gravities also differ substantially. Recent observations support the suggestion that partial stripping could occur \citep[][]{2023MNRAS.525.5121V, 2023A&A...674L..12R}.

Current stellar evolution models predict that massive stars at low metallicity do not become fully stripped because a large intermediate convection zone develops above the helium core \citep{2019A&A...625A.132S, 2020A&A...638A..55K}.
The intermediate convection zone efficiently mixes the chemical elements which, combined with the low opacity of stars at low  metallicity, allows the envelope to remain small until advanced stages of core-helium burning \citep[see][for detailed descriptions of the process]{2022MNRAS.512.4116F}.
When such a star enters into mass transfer, its helium core is already at (or near) thermal equilibrium, which allows for earlier detachment from mass transfer and partial envelope stripping.
This late expansion effect is seen in many different stellar evolution models, including MESA \citep[Modules for Experiments in Stellar Astrophysics, e.g.,][]{2019A&A...625A.132S}, PARSEC \citep[PAdova and TRieste Stellar Evolution Code, e.g.,][]{2014MNRAS.445.4287T}, GENEC \citep[The Geneva Code, e.g.,][]{2013A&A...558A.103G, 2019A&A...627A..24G}, and STARS \citep[e.g.,][]{1998MNRAS.298..525P}, and it becomes increasingly pronounced as metallicity decreases \citep{2020A&A...638A..55K}.
Despite agreement across different models, the late expansion effect is sensitive to uncertain internal mixing processes \citep{2019A&A...625A.132S}. As a result, this evolutionary phase is an uncertain prediction of stellar models, with potentially large impacts. However, if this late expansion influences the number of massive binary-stripped stars that can form, then observations of the intrinsic rate of such systems could help constrain the internal mixing and radial expansion of massive stars.

The ability (or not) to produce massive binary-stripped stars at low metallicity will impact more than just our understanding of the ionizing radiation from stellar populations. For example, the first evolutionary step in the well-established isolated binary evolution pathway that leads to merging binary black holes and neutron stars as observed by the LIGO-Virgo-Kagra collaboration \citep[Laser Interferometer Gravitational-Wave Observatory,][]{2023PhRvX..13d1039A}, is successful envelope-stripping through mass transfer in a massive, metal-poor binary star resulting in the creation of a hot, massive helium star \citep{2017NatCo...814906S, 2018MNRAS.481.1908K}. Most rate predictions for merging binary black holes are made with rapid binary population synthesis codes \citep{2022LRR....25....1M}, which limits the accuracy with which late-stage mass transfer can be modeled. Furthermore, many of the population models are based on the same single star evolutionary model grid \citep{1998MNRAS.298..525P, 2000MNRAS.315..543H}, which necessitates extrapolations for stars more massive than $50 \, M_\odot$, that is, the progenitors of most merging black holes observed in gravitational wave merger events.
While a detailed modeling study for each separate evolutionary step would be justified, we focus here on the effect of late expansion on the distribution of binary-stripped stars. 

In this paper, we modeled the number and mass distribution of binary-stripped stars as function of metallicity using a binary population synthesis code based on detailed single and binary evolutionary models.
In Sect. \ref{sec:models}, we describe our binary population synthesis models and our single and binary stellar evolution models that we use in the population models.
In Sect. \ref{sec:results}, we present our theoretical prediction for the number of stripped stars expected as a function of mass and metallicity---the present-day mass function of stripped stars. 
Section \ref{sec:nearby_pops} contains estimates for the stripped star populations in the Milky Way, Large Magellanic Cloud (LMC), and Small Magellanic Cloud (SMC), while in Sect. \ref{sec:discussion}, we  discuss the implications of our results for high-redshift galaxies and compact object mergers, and we relate our results to other products of massive stars and binaries.
Finally, we summarize our conclusions in Sect. \ref{sec:conclusions}.

\section{Models} \label{sec:models}

This study is based on binary population synthesis results that rely on binary and single stellar evolution models. In Sect. \ref{subsec:mesa}, we describe these binary and single stellar evolution models. In Sect. \ref{subsec:popsynth}, we describe the population synthesis code, which simulates a population of stars with properties drawn from observed distributions, then interpolates from the stellar evolution models to determine the properties of the resulting stripped stars.

\subsection{Stellar evolution models} \label{subsec:mesa}

We used the stellar evolution code MESA (version 8118, \citealt{2011ApJS..192....3P, 2013ApJS..208....4P, 2015ApJS..220...15P, 2018ApJS..234...34P, 2019ApJS..243...10P, 2023ApJS..265...15J}) to compute two sets of models, one for single stars and one for binaries. These models were first presented in \cite{2018A&A...615A..78G}. For an in-depth discussion of the MESA models, we refer to that manuscript. Here we summarize the most relevant parameters for the current study.

\subsubsection{Evolutionary model inputs}\label{subsubsec:inputs}

\paragraph{Convective overshoot and semiconvection}

Both convective overshoot and semiconvection influence whether massive metal-poor stars swell up prior to helium core burning \citep{2019A&A...625A.132S}. However, \cite[][see their Fig. A.1]{2019A&A...625A.132S} show that for their models with metallicity $Z = 0.002$ (approximately 1/7th solar) and without shell overshoot, very inefficient semiconvection or very extensive convective overshoot was required to prevent inflation to the red supergiant stage prior to helium core burning \citep[cf.][]{2001A&A...371..152M}.

Here, we adopted a common set-up, which instead results in expansion to the red supergiant stage after helium core burning \citep[see for example][but also \citealt{2013A&A...558A.103G, 2019A&A...627A..24G}]{2023A&A...672A.198S}. We assumed the Ledoux criterion for semiconvection, setting the efficiency parameter $\alpha_{\rm semiconvection} = 1$ \citep{1985A&A...145..179L, 1991A&A...252..669L}. We implemented a step-like convective overshoot, which extends 0.335 pressure scale heights above all convective regions, following the calibrations of \cite{2011A&A...530A.115B} for the extent of the massive star main sequence.
To model this in MESA, we used the mixing efficiency from 0.05 pressure scale heights inside the convective region (\texttt{f = 0.385, f\_0 = 0.05}).

\paragraph{Stellar winds}

We adopted standard wind prescriptions during the main-sequence phase (the ``Dutch scheme'' in MESA). In addition, especially relevant for our models is the wind mass loss during the stripped star phase. For this phase, we extrapolated from the empirical mass loss prescription for WR stars from \cite{2000A&A...360..227N}, again according to the ``Dutch scheme'' in MESA. \cite{2019MNRAS.486.4451G} studied the effect of adopting the more realistic mass loss rate from \cite{2017A&A...607L...8V} and found that the stripped star could retain more of its hydrogen envelope. When comparing with the results from \citet{2019MNRAS.486.4451G}, we found that the stripped star mass could be up to $\sim 0.3-0.8 M_\odot$ higher than our mass estimates, since we used the stronger mass loss scheme from \citet{2000A&A...360..227N}. Additionally, at stripped star masses $M_{\rm strip} \gtrsim 10 M_\odot$, which are contained in our single star model grid but which are more massive than those produced by our binary evolution models, higher mass loss rates than expected from \cite{2017A&A...607L...8V} could be suitable \citep[see][]{2020MNRAS.499..873S}. We discuss uncertainties that these assumptions could introduce in Sect. \ref{subsubsec:uncertainties}.

\paragraph{Additional energy transport}

For massive stars with initial masses greater than 20 M$_\odot$, we used the MLT++ mechanism (available in the MESA version 8118 through the control \texttt{okay\_to\_reduce\_gradT\_excess = .true.}), an artificial energy transport mechanism used to promote convergence in models with sharp density gradients.
MLT++ likely causes our models to under-predict expansion for stars with initial masses $> 40 M_\odot$ \citep{2020A&A...638A..55K}. However, we expect that the stars with initial masses between 20 and 40 $M_\odot$ are unaffected. 
In addition, as described below, the initial masses of the stars in our binary grids do not exceed $18.17 M_\odot$. As a result, this artificial energy transport only occurs in the single star grids.

\subsubsection{Model grid setup}\label{sec:model_grid_setup}

\paragraph{Single star grids}

There are four grids of single star models, each computed at a different metallicity: $Z = Z_\odot = 0.014$, $Z=0.006$, $Z=0.002$, and $Z=0.0002$. These are designed to resemble stars in the Milky Way, the LMC, the SMC, and the extremely metal-poor star-forming galaxy IZw18, respectively \citep[see, e.g.,][]{2015A&A...581A..15S}.
Their compositions are scaled relative to that of the Sun \citep{1998SSRv...85..161G}.

Each grid is made up of 40 models, with initial masses logarithmically spaced between $M_{\rm init} =$ 2 and 100 M$_\odot$. They were computed at least until central helium depletion, thus covering $\sim 99\%$ of the stellar lifetime. 

\begin{figure}
    \centering
    \includegraphics[width=\columnwidth]{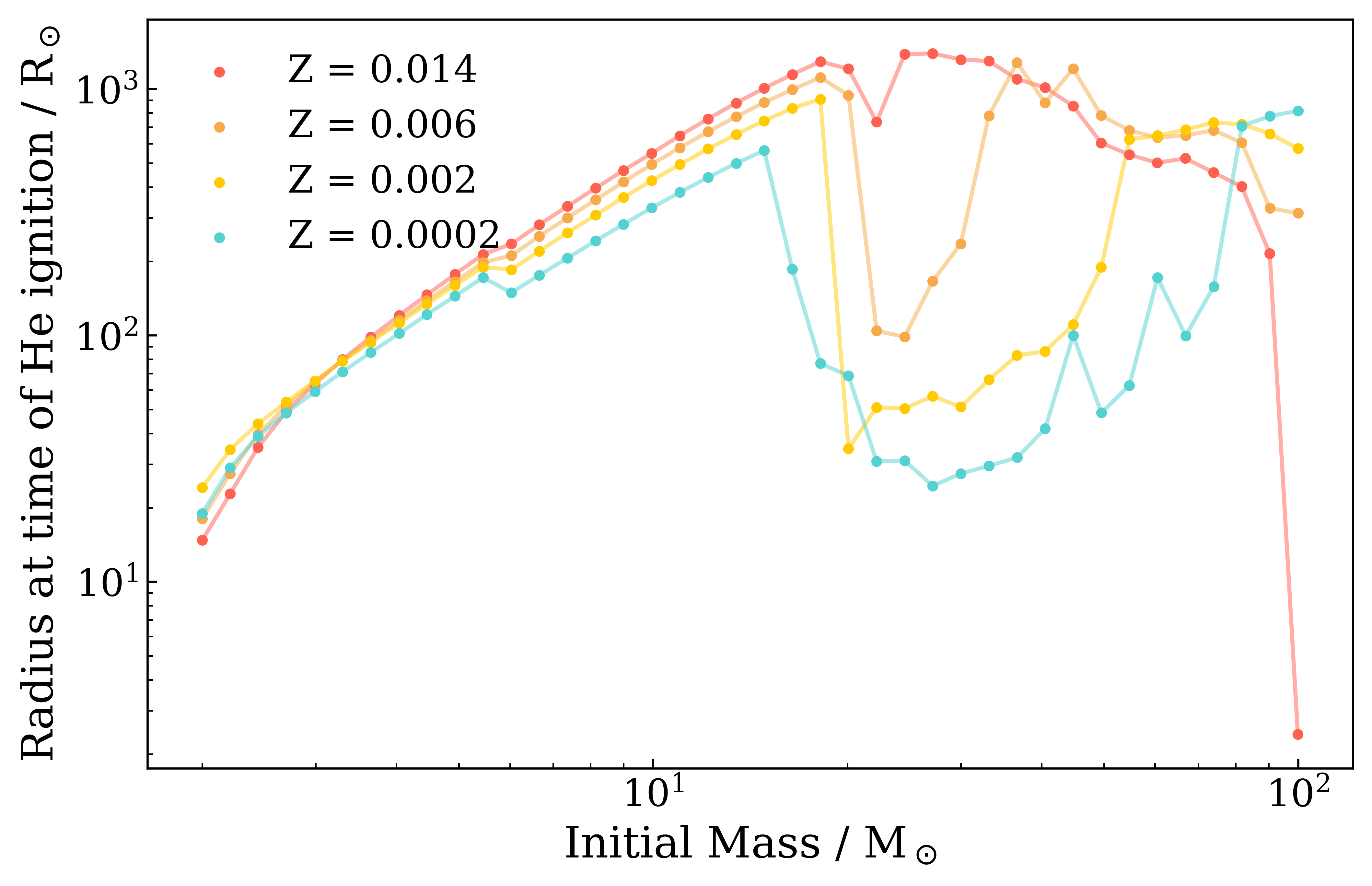}
    \caption{Radius of each of our single stellar evolution models at the time when they begin burning helium in their cores. At $Z=0.014$ (red), higher-mass stars expand to radii $\gtrsim 10^3 R_\odot$ prior to core helium burning. At lower metallicities, however, higher-mass stars expand late and have radii almost an order of magnitude smaller at this time. They will therefore struggle to create long-lived stripped stars.}
    \label{fig:R_at_He}
\end{figure}

To demonstrate late expansion at low metallicity using these single star models, we display in Fig.~\ref{fig:R_at_He} the radius of single stellar evolution models at the time when helium ignites in the core (see Sect. \ref{sec:models}, here we define the helium ignition as when the central helium mass fraction drops below 0.97). While the higher-mass ($\sim 15-40 M_\odot$), high-metallicity ($Z=0.014$) models expand to radii $\gtrsim 10^3 R_\odot$ prior to helium ignition, lower metallicity ($Z\lesssim 0.002$) stars of the same masses do not expand until later, remaining small with radii $\lesssim 10^2 R_\odot$ at this stage. This is similar to the results presented in \citet{2022A&A...662A..56K}.

At high enough masses, even low metallicity stars are no longer quite so affected by late expansion, and they are able to reach larger radii prior to the onset of core helium burning. Additionally, at these high masses, the higher metallicity stars have very strong winds, causing their evolutionary track in the Hertzsprung Russell diagram to turn around before they reach the Hayashi track, on the way to becoming WR stars. This is why the highest mass stars in Fig. \ref{fig:R_at_He} have similar radii regardless of metallicity.

We note that the number of stripped stars estimated in this work is dependent on the radius evolution of the underlying evolutionary model. The radius evolution for high mass stars ($M>40 M_\odot$) can be affected by luminous blue variable-like eruptions and the Humphreys-Davidson limit \citep[e.g.,][]{2021MNRAS.503.1884G, 2024ApJ...974..270C}, which we did not account for in this work.

\paragraph{Binary star grids}

We used the MESA binary module to compute grids of 23 binary models at each of the four metallicities that were used for the single stars.
For each grid, the initial mass of the most massive star in each binary system ($M_{\rm 1, init}$) is logarithmically spaced between 2 and 18.17 M$_\odot$. For numerical reasons, we chose not to compute binary models at higher masses \citep[see, however,][]{2022A&A...662A..56K}.

For our binary star grids, we modeled the formation of stripped stars through mass transfer initiated shortly after the donor star finishes the main sequence, early during the Hertzsprung gap. We therefore chose initial periods that range between 3 and 31.7 days and an initial mass ratio of $q_{\rm init} \equiv M_{\rm 2, init} / M_{\rm 1, init} = 0.8$, where $M_{\rm 2, init}$ is the initial mass of the less massive star in the system.
This resulted in stripped stars with masses ($M_{\rm strip}$) between 0.35 and 8 M$_\odot$, depending on the initial mass and metallicity. Although the MESA binary models are limited to one formation channel of stripped stars, we used them to approximate the result of envelope-stripping through other evolutionary channels as well (see Sect. \ref{subsec:popsynth}).

In this study, we used these models to represent all stripped stars that originate from stars with masses between 2 and 18.17 M$_\odot$. For solar metallicity, we do not expect a significant difference in the predicted stripped star mass depending on initial orbital period or initial mass ratio \citep[see e.g.,][]{2024A&A...683A..37Y}. However, at low metallicity, these models could result in a slight under-prediction of the stripped star mass \citep{2017ApJ...840...10Y}.

\subsection{Binary stellar population synthesis} \label{subsec:popsynth}

We modeled binary stellar populations using a Monte Carlo code that was developed specifically for studies of stripped stars and was first presented in \citet{2019A&A...629A.134G}. This code uses single and binary stellar evolution models as input, which allows for flexibility in adopting suitable evolutionary models \citep[see also][]{2020MNRAS.497.4549A, 2023MNRAS.525..933A}. This method also provides more accuracy in the mass and properties of the stripped stars, compared to the estimate from pure helium star models which are often used in binary population synthesis \citep{2000MNRAS.315..543H, 2002MNRAS.329..897H}. 

The set-up we used here is similar to what was described in \citet{2020ApJ...904...56G}.
We briefly summarize our assumptions below.

\paragraph{Initial distributions}
We first drew initial masses ($M_{\rm init}$) between 0.1 and 100 M$_{\odot}$, following the initial mass function of \cite{2001MNRAS.322..231K}, in which $\frac{dN}{dM} \propto M^{-\alpha}$, with $\alpha = 1.3$ for $0.08 < M_{\rm init} < 0.5 M_\odot$ and $\alpha = 2.3$ for $M_{\rm init} > 0.5 M_\odot$.
We determined whether to assign a binary companion to each of these stars following the mass-dependent binary fraction of \citet[][see their Fig. 42]{2017ApJS..230...15M}.
For the binary stars, the initial mass of the less massive star in the system, $M_{\rm 2, init}$, was then sampled from a uniform mass ratio distribution, where $q \equiv M_{\rm 2, init}/M_{\rm 1, init}$ ranges from 0.1 to 1, in agreement with observed mass ratios \citep[][however, see also \citealt{2017ApJS..230...15M}]{2012Sci...337..444S, 2012ApJ...751....4K}.
The initial orbital period ($P_{\rm init}$) was sampled from a distribution which is uniform in $\log_{10} P_{\rm init}$ for binaries with $M_{\rm 1, init} < 15 M_\odot$ \citep{1924PTarO..25f...1O} and favors short initial periods when $M_{\rm 1, init} > 15 M_\odot$ \citep[$\frac{dN}{d \log_{10} P} \propto (\log_{10} P)^{-0.55}$,][]{2012Sci...337..444S}. The minimum initial orbital period sampled is the shortest period for which neither star overflows its Roche lobe at the zero-age main sequence, and the maximum period is $10^{3.7}$ days \citep{2017ApJS..230...15M}.

\paragraph{Onset of binary interaction}
We determined when binary interaction is initiated by measuring when the stellar radius, obtained by interpolating MESA models over the initial mass, exceeds the Roche radius \citep{1983ApJ...268..368E}.
For simplicity, here we focused on stripped stars produced from the initially most massive star in the system. These should be the most common progenitors for stripped stars, since binaries can be easily disrupted by the creation of the first compact object \citep[e.g.,][]{2019A&A...624A..66R}. We note that stripped stars can be formed from the initially less massive star in a binary system, but they are expected to constitute a smaller fraction \citep[$\sim 7\%$,][]{2017ApJ...842..125Z}, and since their evolutionary pathways require more detailed scrutiny, we chose to omit their implementation from the population model for this work. 
Furthermore, we only considered hot stripped stars produced when the Roche lobe is filled prior to core helium burning of the donor star, which we define as when the helium mass fraction in the center first decreases below 0.98. This is a reasonable assumption because envelope-stripping initiated during central helium burning is likely to result in nuclear-timescale mass transfer or cool, partially stripped, puffed-up stars \citep[][and see Sect. \ref{subsec:partial_strip} for further discussion]{2022A&A...662A..56K}, and envelope-stripping initiated after central helium depletion will result in very short-lived objects, as only about $1\%$ of the stellar lifetime remains.

\begin{figure}
\centering
\includegraphics[width=\columnwidth]{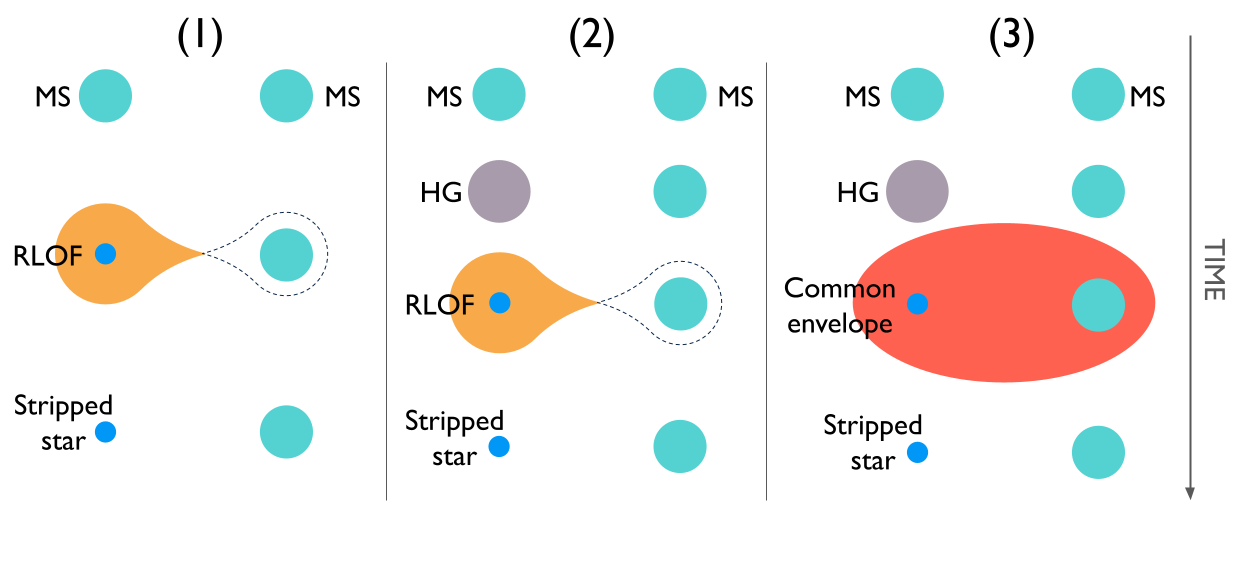}
\caption{Three channels for stripped star formation accounted for in the stellar population models in this work. (1) mass transfer initiated during the main-sequence evolution of the donor star (Case A), (2) mass transfer initiated during the Hertzsprung gap evolution of the donor star (Case B), and (3) the successful ejection of a common envelope initiated during the Hertzsprung gap evolution of the donor star (Case B CEE).}
\label{fig:cartoon}
\end{figure}

\paragraph{Outcome of binary interaction}
For simplicity, we inferred the outcome of binary interaction based on limits for the mass ratio and initial period (e.g., POSYDON, POpulation SYnthesis with Detailed binary-evolution simulatiONs, \citealt{2023ApJS..264...45F}; and BPASS, Binary Population and Spectral Synthesis, \citealt{2017PASA...34...58E}, see also \citealt{2019A&A...631A...5Z}).
We identified three pathways to the creation of a stripped star, which we depict in Fig.~\ref{fig:cartoon}.
To ensure long-lived stripped stars are created, we only considered donor stars that fill their Roche lobes during either main-sequence evolution (central hydrogen fusion), or during the following rapid expansion phase known as the Hertzsprung gap (hydrogen shell burning after central hydrogen depletion).
The three pathways we considered are: (1) stable mass transfer initiated during the main-sequence evolution of the donor star (also known as Case A), (2) stable mass transfer initiated during the Hertzsprung gap evolution of the donor star (Case B), and (3) the successful ejection of a common envelope initiated during the Hertzsprung gap evolution of the donor star (Case B CEE).
The Hertzsprung gap typically includes a radial expansion by at least an order of magnitude (see Fig.~\ref{fig:R_at_He}), making it one of the most common evolutionary stages for binary interaction.

We adopted the following approximate approach to determine what systems result in the creation of a stripped star. 
If the initial mass ratio is $>0.65$ and $>0.4$ for mass transfer initiated during the main sequence and Hertzsprung gap evolution of the donor star, respectively, we assumed that a stripped star can be formed (\citealt{2007A&A...467.1181D, 2002MNRAS.329..897H}, see also \citealt{2019A&A...631A...5Z}).
We assumed that common envelopes develop when the donor fills the Roche lobe during the Hertzsprung gap and $q_{\rm init} < 0.4$ \citep{2002MNRAS.329..897H}. Common envelopes that develop earlier in the donor star's evolution (i.e., on the main sequence) were assumed to lead to coalescence. Common envelopes that develop later in the donor star's evolution were neglected, meaning that our predicted contribution from the common envelope channel could be conservative. To emulate the result of the common envelope evolution, we adopted the $\alpha$-formalism \citep{1984ApJ...277..355W}, setting $\alpha = 1$ \citep[e.g.,][]{2002MNRAS.329..897H} and $\lambda = 0.5$ \citep{2000A&A...360.1043D, 2001A&A...369..170T}.
We note that our treatment for common envelope ejection is simplified, since the binding energy of a stellar envelope evolves during expansion \citep[see e.g.,][]{2021A&A...645A..54K}. This means that our predicted contribution from common envelope ejections is somewhat uncertain (see Sect. \ref{subsubsec:uncertainties}).

The fraction of stripped stars that result from each of these formation channels is discussed further in Appendix~\ref{appendix:channels}. We note that for initial masses $\lesssim 20 M_\odot$, mass transfer initiated on the Hertzsprung gap dominates the production of stripped stars, while at higher masses, mass transfer initiated during main sequence evolution contributes significantly (see Fig.~\ref{fig:channels}). Common envelope evolution provides a minor contribution, regardless of mass.

\paragraph{Initial to stripped star mass relation}

\begin{figure}
    \centering
    \includegraphics[width=\columnwidth]{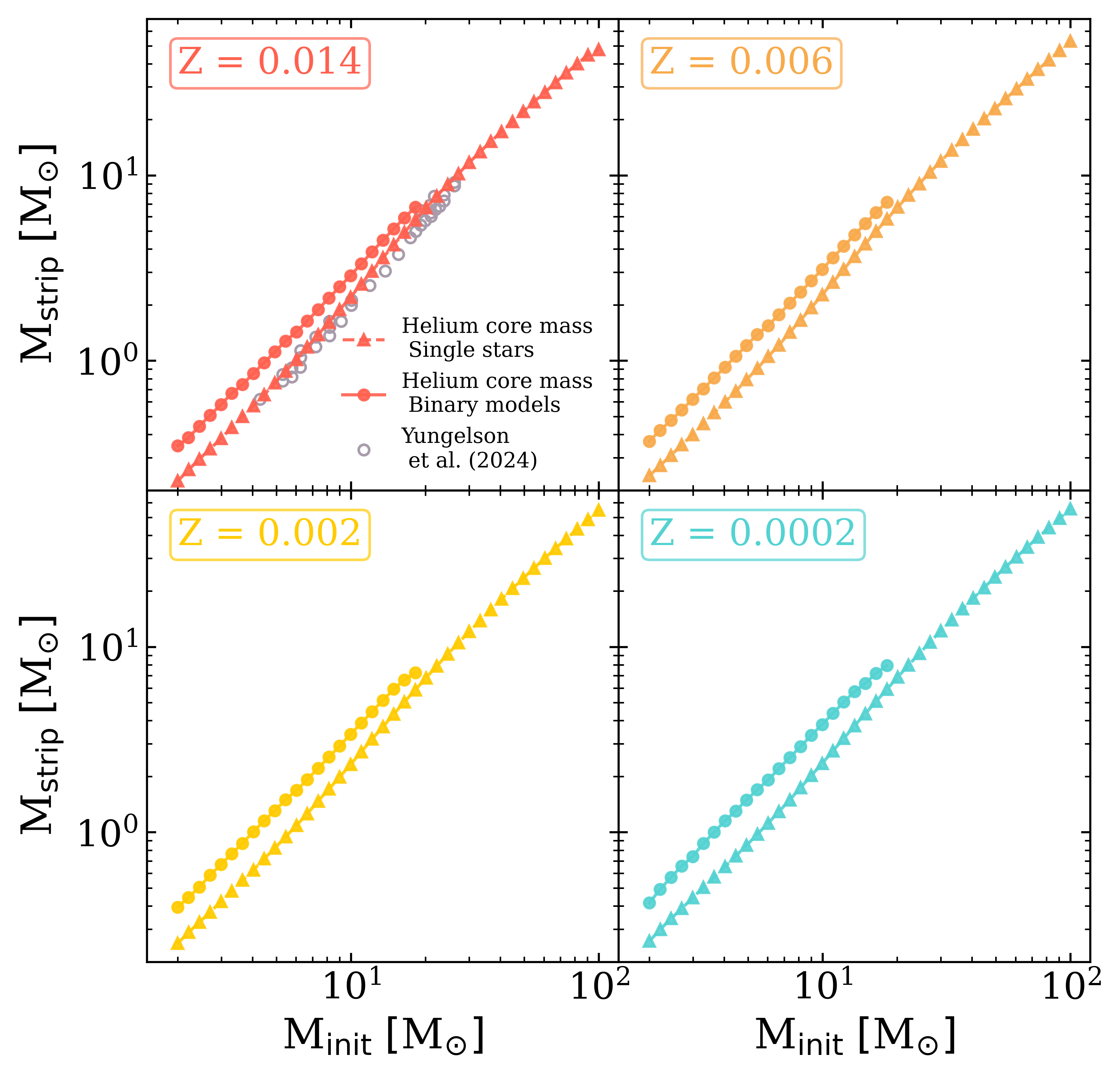}
    \caption{Relation between the initial mass (horizontal axis) and the eventual stripped star mass (vertical axis). The circular markers indicate the stripped star mass measured halfway through central helium burning from binary evolutionary modeling (see Sect.~\ref{subsec:mesa}), while the triangular markers indicate the helium core mass at terminal age main sequence of single stars as defined in MESA. We note that our binary evolution grid does not extend above $M_{\rm init} > 20 M_\odot$. For these massive stars, we instead represent stripped star masses with the helium core masses. The panels show models for different metallicity as labeled at the top of each panel. Gray circles show a subset of the stripped star models from \citet{2024A&A...683A..37Y} that originate from mass transfer initiated during the Hertzsprung gap of the donor star.}
    \label{fig:Minit_Mstrip_relation}
\end{figure}

We produced a mapping from initial to stripped star mass from our MESA models and display it in Fig. \ref{fig:Minit_Mstrip_relation}.
The different panels show the initial to stripped star mass relation for the different metallicities, as labeled in each top left corner. As shown in the figure, the mapping is somewhat metallicity-dependent. The reason for this metallicity dependence is the lower opacity and weaker winds at low metallicity, which leads to both higher convective core masses during the main sequence and a larger amount of hydrogen left after mass transfer \citep{2017ApJ...840...10Y, 2017A&A...608A..11G}. 

For stars with initial masses less than 20 M$_\odot$, we used our binary-stripped star model grids to determine the mass of the resulting stripped stars (colored, filled circles in Fig.~\ref{fig:Minit_Mstrip_relation} and Sect. \ref{sec:model_grid_setup}). In this case, we adopted the mass of the stripped star halfway through the central helium burning, which we defined as when the central helium mass fraction decreases below 0.5. For stars with higher initial masses, we instead used the helium core masses of the single star models at the terminal-age main-sequence (defined as the first time the central hydrogen mass fraction decreases below 0.01), which are marked with colored triangles in Fig.~\ref{fig:Minit_Mstrip_relation} (see Sect. \ref{sec:model_grid_setup}). This estimate gives rise to a maximum stripped star mass of $\sim 50 M_{\odot}$. We chose to use the terminal-age main-sequence core mass because the helium core grows in single stars, while this growth is not prominent for stripped stars. The standard MESA definition for the helium core mass was adopted here, that is, the location in the stellar interior where the hydrogen mass fraction decreases below 0.1. We note that models of binary-stripped stars typically have more leftover hydrogen on their surfaces, but also that wind mass loss could affect the masses especially at the high-mass end.

In this paper, we did not account for formation of stripped stars through strong stellar winds, that is, the canonical WR star formation. This suggests that we may under-predict the number of massive helium stars. We expect that wind-stripping would primarily contribute in the high-metallicity regime, where winds are known to be strong \citep{2007ARA&A..45..177C}.   

While in our population synthesis approach we allowed for the creation of stripped stars through Case A, Case B, and Case B CEE evolution, our binary MESA models simulated only Case B mass transfer, and the mass derived from the single star MESA models is most appropriate for stripped stars created through Case B and Case B CEE evolution. Our use of the same scaling from $M_{\rm init}$ to $M_{\rm strip}$ for Case A evolution is therefore an approximation and introduces some uncertainty.
When envelope-stripping is initiated during the main sequence of the donor star (Case A, pathway 1 in Fig. \ref{fig:cartoon}), the stripped star that is produced should have a slightly lower mass than what we adopted, as the core is not yet fully formed when mass transfer begins \citep[e.g.,][]{2010ApJ...725..940Y}. 
Accounting for this effect would require an expansion of the binary star model grids, which could be a topic of a future study. Here we attempt to quantify the effect on our results (see Appendix \ref{appendix:channels}). Approximately 20\% of the stripped stars we model go through Case A evolution, which is dominant at the high-mass end.
Better accounting for main-sequence mass transfer systems would therefore slightly decrease the number of high-mass stripped stars \citep[see][]{2023A&A...672A.198S}. In future studies, the new mapping from \cite{2024A&A...690A.282S} could be implemented to better represent the resulting stripped star masses.

In Fig. \ref{fig:Minit_Mstrip_relation}, we use gray circles to plot the initial-to-stripped-star mass relation from \cite{2024A&A...683A..37Y}, who modeled stripped stars at solar metallicity. Their predicted stripped star masses are systematically lower compared to our models. For example, our 10 $M_\odot$ progenitor results in a $\sim 3 M_\odot$ stripped star (similar to \citealt{2017ApJ...840...10Y}), while \cite{2024A&A...683A..37Y} predict it should result in a $\sim 2 M_\odot$ stripped star. 
The initial mass to core mass mapping is relatively unknown and can be affected by convective overshoot, as found by \cite{2024ApJ...964..170J}, who show that convective overshoot can alter the helium core mass by up to 50\%.
This large uncertainty, further discussed in Sect. \ref{subsubsec:uncertainties}, emphasizes the importance of finding observable constraints for core mass ratios, which we aim to do with this study.

\paragraph{Interpolation of stripped star properties}
To obtain stellar properties for the stars in the modeled population, we interpolated properties obtained from the underlying grid of evolutionary models over the logarithm of the stellar mass.
We interpolated the properties for main-sequence stars, stripped stars, and post-main sequence stars separately.
Because the absolute age of a star is sensitive to its initial mass, for each of these evolutionary phases, we measured the duration of the phase and normalized the star's evolution by the total duration to reach a smooth interpolation.
In this study, we focused on the stripped star mass, the time that envelope-stripping is initiated, and the duration of the stripped star phase. These properties are easy to interpolate over the binary model grids (when $M_{\rm init} < 20 M_\odot$) and can be interpolated from the single star grids by adopting the helium core mass and timescales derived from the main sequence and post-main sequence evolution of the single stars.

\paragraph{Model setup}
We ran the code at all four metallicities, adopting a constant star formation rate of $1 M_\odot / {\rm yr}$. This makes it possible to compare our results with an observed population undergoing constant star formation by multiplying by its star formation rate.
In our results, we present the number of stripped stars per $1 M_\odot / {\rm yr}$ rather than the fraction of stripped stars. This choice is based on the way our population synthesis code works, as the star formation rate provides a more useful and accurate scaling factor for our results than the total number of stars. We evaluated the stellar population after 1 Gyr.
Because 1 Gyr is approximately the lifetime of a $2 M_\odot$ star, it is more than sufficient to ensure that the population of stars that is the focus of this work---those with initial masses $\gtrsim 4 M_\odot$ ($M_{\rm strip} \sim 1 M_\odot$)---will be in equilibrium.

\paragraph{Model output}
We chose to explore the mass distribution of stripped stars with stripped star mass $> 1 M_\odot$ (corresponding to $M_{\rm 1, init} \gtrsim 3.5-4.5 M_{\odot}$, depending on metallicity). As described above, these systems are expected to be in equilibrium. In addition, our binary population synthesis set-up does not correctly account for the evolution of low-mass binaries ($M_{\rm 1, init} < 2 M_\odot$) and we therefore significantly underpredict the number of low-mass subdwarfs (see however \citealt{2002MNRAS.336..449H, 2003MNRAS.341..669H}).
To capture stochastic variations, we ran the code 100 times for each metallicity. We ensured that 100 runs is sufficient by running a second set of 100 runs at each metallicity, and we found that the mean number of stripped stars in each mass bin presented in Sect. \ref{subsec:mdist} is within $0.5 \sigma$ for $M_{\rm strip} > 20 M_\odot$, within $0.2 \sigma$ for $M_{\rm strip} = 7- 20 M_\odot$, and $<0.1 \sigma$ for lower masses.
\footnote{Results from the model runs are available at \url{https://doi.org/10.5281/zenodo.14811517}.}.

\section{Stripped star population predictions} \label{sec:results}

\subsection{The stripped star mass distribution}\label{subsec:mdist}

\begin{figure}
    \centering
    \includegraphics[width=0.79\columnwidth]{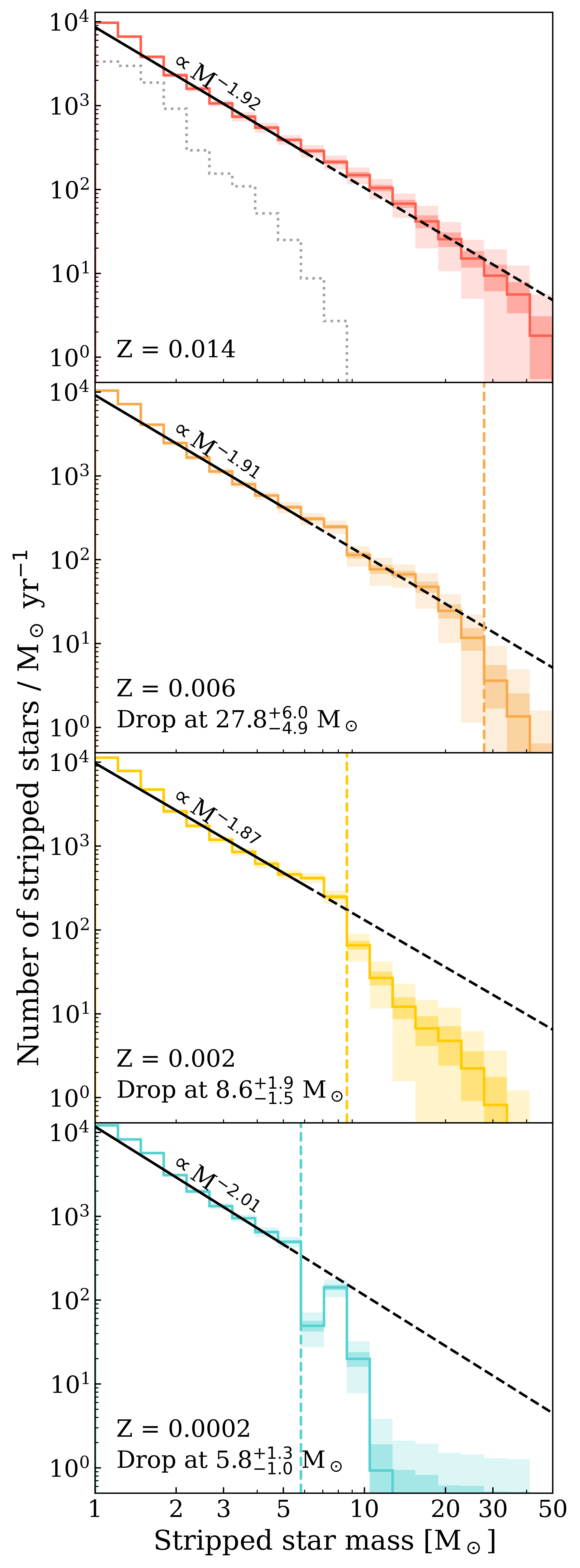}
    \caption{Present-day mass function for hot stars stripped in binaries, adopting a constant star formation of 1 $M_\odot$/yr. The solid, colored lines indicate the mean number of stripped stars expected within each mass bin from 100 model runs, and the shaded regions show 1 and 3 $\sigma$ above and below each line. The dashed black line was fit to the lower-mass end of each distribution, and the range of masses used for the fit is shown by the solid portion of the black line. The ``drop'' in each mass distribution is determined as the point where the upper $3\sigma$ line drops below the dashed line of best fit, and its uncertainty is taken to be the width of one mass bin in each direction. 
    The gray dotted line in the solar metallicity plot shows the mass distribution from \cite{2024A&A...683A..37Y}.}
    \label{fig:mdist}
\end{figure}

In Fig. \ref{fig:mdist}, we plot the mass distributions from our population synthesis models for the four metallicities considered, showing how many stripped stars are expected as a function of stripped star mass.
The distributions are plotted in red, orange, yellow, and blue, in order of decreasing metallicity.
The mass functions displayed in Fig. \ref{fig:mdist} effectively convolve the initial mass function with mass-dependent effects such as the stellar lifetime and binary fraction, resulting in a different slope than the initial mass function \citep[see also][]{2014ApJ...780..117S}. We emphasize that these distributions encompass predictions for hot stripped stars (which initiated mass transfer prior to core helium burning). They would exclude partially stripped stars that may result from later interactions (see Sect. \ref{subsec:partial_strip}).

For the mass distributions, we use 20 mass bins, logarithmically spaced between 1 and 50 M$_\odot$, corresponding to a bin width of $\Delta \log (M_{\rm strip} / M_\odot) = 0.085$.
Because the predicted number of stripped stars per mass bin is normally distributed for all the runs, we use the mean to mark the distribution and the standard deviation to shade possible variations.
The solid line in each distribution denotes the mean number of stars in each mass bin over the 100 model runs.
The accompanying colored regions denote 1 and 3 standard deviations above and below the mean.

\subsubsection{A power law slope at the low-mass end}\label{subsubsec:powerlaw}

For all metallicities, the lower-mass end of the distribution is approximately a power law.
For each mean distribution, we fit a line with the same functional form as the initial mass function, $\frac{dN}{dM} \propto M^{-\alpha}$, to the lower-mass ($M_{\rm strip}$ $<7$ M$_\odot$) end of the distribution. This corresponds to the first 10 out of 20 mass bins.
However, for the $Z=0.0002$ distribution, the slope becomes inconsistent and the number of stripped stars drops off before $M_{\rm strip} = 7 M_\odot$, so for this metallicity, we fit the line to the part of the distribution with $M_{\rm strip} < 5 M_\odot$.
For each metallicity, the best fit is shown as a black dashed line in Fig. \ref{fig:mdist}, and the solid portion of the line corresponds to the mass range used for the fit.

Across the four metallicities, we find a negligible (no more than $1.2 \sigma$) difference between the slopes ($\alpha = 1.92 \pm 0.06, 1.91 \pm 0.07, 1.87 \pm 0.10$, and $2.01 \pm 0.06$, in order of decreasing metallicity) and the absolute numbers of stripped stars with mass $M_{\rm strip} \lesssim 5 M_\odot$, suggesting that the shape of the intermediate mass regime of the stripped star mass distribution is only marginally affected by metallicity.

A number of mass-dependent factors affect how the slope of the mass function differs from that of the initial mass function.
For example, the shorter lifetimes of massive stars cause the slope of the present-day mass function to become more steep.
By contrast, the higher proportion of massive stars that are in close binaries results in a shallower slope.
Further, the more massive a star, the larger fraction of its mass will be in the stripped star it produces (i.e.\ $M_{\rm strip} / M_{\rm 1,init}$ increases with increasing $M_{\rm 1,init}$). This also causes the slope of the present-day mass function to become shallower.

The best fit slopes in Fig. \ref{fig:mdist} are all shallower than that of the input initial mass function, where we assumed $\alpha = 2.3$. This suggests that the fraction of stars that become stripped and the larger relative core masses dominate over the longer lifetimes of the lower mass stars. 

In the solar-metallicity panel, we also display the population predictions from \citet{2024A&A...683A..37Y}, which reaches up to $\sim 8 M_\odot$, using a gray dotted line. Notably, their predicted slope is steeper, and they predict significantly fewer stripped stars overall. This is likely due to our different treatments of convective overshoot, which leads to their predicted lower mass stripped stars (see Fig. \ref{fig:Minit_Mstrip_relation}), as well as different initial period distributions and our inclusion of stripped stars formed through the common envelope channel. However, when accounting for these, we find that our results for stripped stars with $1 M_\odot < M < 7 M_\odot$ are in reasonable agreement.

\subsubsection{A drop at the high-mass end of the low-metallicity mass distributions} \label{subsubsec:drop}

In all three sub-solar populations shown in Fig. \ref{fig:mdist}, the distribution departs from the line fitted to the low-mass end, and there is a drop in the mass distribution at a specific threshold in mass, above which very few stripped stars are created through binary interactions. This drop occurs because of the late expansion of massive stars at lower metallicity, causing the Roche lobe in many binary systems not to be filled until during or after helium-core burning has started, resulting in cool or short-lived stripped stars \citep[see Fig. \ref{fig:R_at_He} and Sect. \ref{sec:intro} along with ][]{2022A&A...662A..56K}.

We quantify the location of the drop as the highest mass bin before which the upper $3 \sigma$ line drops below the dashed, best-fit line in Fig. \ref{fig:mdist}.
We determine the uncertainty on this measurement to be plus or minus the width of one mass bin in each direction, which can be between $1 - 6 M_\odot$, depending on the location of the drop.
The drop occurs at $27.8 ^{+ 6.0}_{- 4.9}$ M$_\odot$ for $Z = 0.006$, at $8.6 ^{+ 1.9}_{- 1.5}$ M$_\odot$ for $Z = 0.002$, and at $5.8^{+ 1.3}_{- 1.0}$ M$_\odot$ for $Z = 0.0002$. As expected, we do not identify a drop for the solar metallicity population. 

We note that the lowest metallicity distribution ($Z=0.0002$) exhibits a non-monotonic drop, with a peak at $7.1 M_\odot$ following the drop at $5.8 M_\odot$. This is a numerical effect occurring due to the somewhat larger difference between the stripped star masses predicted from the binary evolution models and the terminal-age main sequence core masses (see Fig. \ref{fig:Minit_Mstrip_relation}).

\subsubsection{Uncertainties influencing the shape of the mass distribution}\label{subsubsec:uncertainties}

Several uncertain physical mechanisms influence the shape of the predicted mass distribution. Here we discuss the primary sources of uncertainty.

In this study, we adopted a relatively high wind mass loss rate for the intermediate and lower mass stripped stars (see Sect. \ref{subsubsec:inputs}), which could result in an underestimate of the number of lower mass stripped stars. On the other hand, for more massive stripped stars, we used an approximate approach and assumed the stripped star mass to be that of the terminal age main sequence core (see Sect. \ref{subsec:popsynth} and Fig. \ref{fig:Minit_Mstrip_relation}). Because mass evolution due to wind mass loss was not included, this could result in a moderate overestimate of the number of massive stripped stars.

In our stellar evolution models, we implemented a fixed step-like convective overshoot (see Sect. \ref{subsubsec:inputs}). Convective boundary mixing is an active field of research, and recent work suggests that the effect increases with stellar mass \citep[e.g.,][]{2021A&A...655A..29J, 2024ApJ...964..170J}. If this is true, it would cause the slope of the mass distribution to be shallower than what we predict here, meaning that we overestimate the number of lower mass stripped stars and underestimate the number of higher mass stripped stars. An example of how the mass distribution changes when adopting an overall lower core mass is shown by the dotted line in the top-left panel of Fig. \ref{fig:mdist}, which shows the mass distribution predicted by \cite{2024A&A...683A..37Y}.

The efficiencies of the three evolutionary pathways that we considered (see Sect. \ref{subsec:popsynth} and Fig. \ref{fig:cartoon}) also remain uncertain. Detailed binary evolution models predict that mass transfer can efficiently remove stellar envelopes for the rough parameters that we assume here, but the efficiency of common envelope ejection is substantially less certain \citep[see however, e.g.,][]{2012ApJ...759...52D}. From an observational perspective, the relative efficiencies of these evolutionary channels is largely unconstrained, since a statistical sample of observed stripped stars would be needed to perform suitable constraints. For this reason, we expect some uncertainties in the relative contributions of the three considered evolutionary channels (see also Appendix \ref{appendix:channels}).

Observing a population of stripped stars could help constrain these uncertainties.
This is one of the goals of the Ultraviolet Explorer mission \citep[UVEX, see][]{2021arXiv211115608K}, a UV satellite with a planned launch in 2030. In the meantime, we compare our predictions with nearby populations (see Sect. \ref{sec:nearby_pops}) and find reasonable agreement.

\subsection{Number estimates} \label{subsec:number_estimates}

\begin{figure}
    \centering
    \includegraphics[width=\columnwidth]{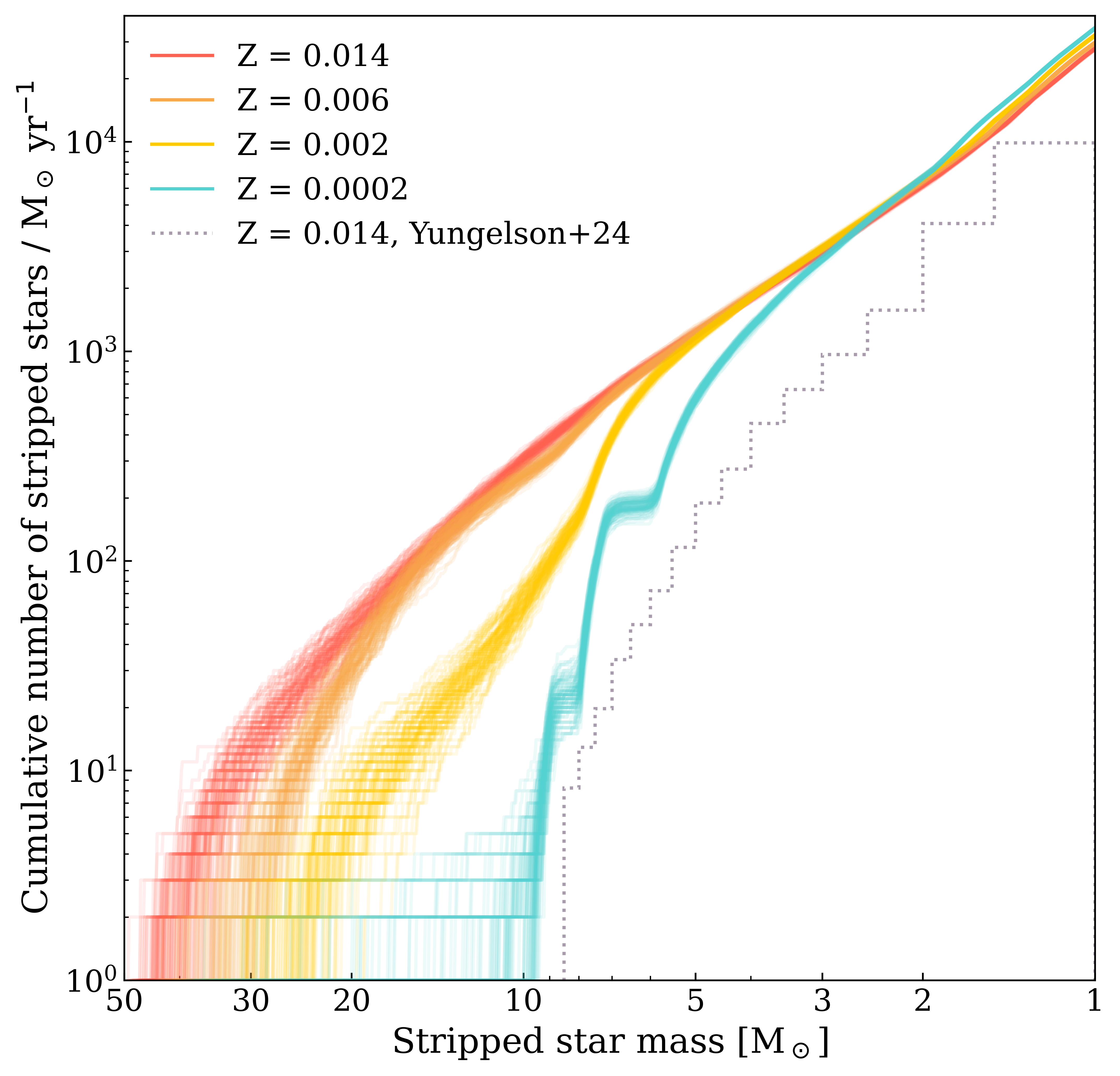}
    \caption{Cumulative distribution plot showing the number of hot stripped stars formed with mass greater than or equal to a given mass for all 100 runs at each metallicity. Lower metallicity environments form fewer high-mass stripped stars because of late expansion, but the expected total number of stripped stars is similar when considering a lower mass limit of $\sim 3 M_\odot$ or lower. 
    The different metallicity runs are marked with different colors, while we display the predictions for solar metallicity from \citet{2024A&A...683A..37Y} using a dotted gray line.}
    \label{fig:cumulative}
\end{figure}

In Fig. \ref{fig:cumulative}, we plot the cumulative distribution of stripped stars per unit constant star formation for all 100 runs from the population synthesis model and for each metallicity. In this cumulative distribution, we have inverted the horizontal axis and each point shows the number of stripped stars with mass equal to or greater than that mass.

\begin{table*}[ht!]
\caption{Estimated numbers of binary-stripped stars.}
\label{tab:number_estimate}
\centering
\begin{tabular}{c|cccc}
\hline\hline
M$_{\rm strip}$ & \multicolumn{4}{c}{Metallicity ($Z_*$)} \\
($M_\odot$) & $0.014$ & $0.006$ & $0.002$ & $0.0002$ \\
\hline
$1-50$ & $27788 \pm 173$ & $29594 \pm 180$ & $32126 \pm 182$ & $34766 \pm 190$ \\
$> M_{\rm drop}$ & & $5 \pm 2$ & $119 \pm 12$ & $213 \pm 13$ \\ 
$> M_{\rm WR}$\tablefootmark{a} & $560 \pm 22$ & $211 \pm 14$ & $11 \pm 4$ & \\ 
\hline
$1-1.4$ & $14733 \pm 133$ & $15533 \pm 119$ & $17168 \pm 140$ & $18207 \pm 122$ \\
$1.4-2.6$ & $9044 \pm 87$ & $9865 \pm 94$ & $10655 \pm 104$ & $12473 \pm 108$ \\
$2.6-5$ & $2579 \pm 46$ & $2743 \pm 50$ & $2913 \pm 61$ & $3204 \pm 62$ \\
$5-10$ & $912 \pm 30$ & $968 \pm 32$ & $1067 \pm 30$ & $589 \pm 24$ \\
$>10$ & $301 \pm 16$ & $253 \pm 14$ & $63 \pm 9$ & $3 \pm 2$ \\
\hline
\end{tabular}
\tablefoot{The estimates shown are for a population in equilibrium and in which stars form at a constant rate of 1 $M_\odot$/yr.
\tablefoottext{a}{The $M_{\rm WR}$ values are the $M_{\rm spec, He}^{\rm WR}$ values obtained by \citealt{2020A&A...634A..79S}: 7.5, 11, and 17 $M_\odot$ at $Z = 0.014$, 0.006, and 0.002, respectively.}
}
\end{table*}

For example, following the vertical grid line at 10 M$_\odot$, we can see that for the higher metallicities of $Z = 0.014$ and $Z = 0.006$, we expect around 300 stripped stars with masses of at least $10 M_\odot$ to exist, per $1 M_\odot$/yr of star formation (specifically, $301 \pm 16$ and $253 \pm 14$ stripped stars, respectively). However, for the same star formation rate and $Z = 0.002$, only $63 \pm 9$ stripped stars should have masses of at least 10 M$_\odot$, and at the even lower $Z = 0.0002$, we only expect $3 \pm 2$ stripped stars this massive. 
At lower masses, the differences are less prominent. If we consider the stripped stars that likely reach supernova \citep[$\gtrsim 2.6 M_\odot$,][]{2015MNRAS.451.2123T}, the metallicity difference is small and all populations predict $\sim 4,000$ such stripped stars to exist per $1 M_\odot$/yr. 
For clarity, we also display the expected numbers of stripped stars within certain given mass ranges and for the different metallicities in Table \ref{tab:number_estimate}.

Our expected number of stripped stars is higher than the predictions from \citet{2024A&A...683A..37Y} by a factor of a few up to an order of magnitude, depending on mass (see Table \ref{tab:number_estimate}). 
We display their predictions for the cumulative number of solar metallicity stripped stars in Fig. \ref{fig:cumulative} using a gray dotted line. 
Compared to our solar metallicity model, \citet{2024A&A...683A..37Y} only predict $\sim 100$ stripped stars with $M_{\rm strip} \sim 5-8 M_\odot$ to exist per $1 M_\odot$/yr, while our estimate is closer to $\sim 700$.

Classical WR stars are often thought to be helium stars in the helium core burning phase, potentially stripped through strong stellar winds or binary interaction.
\citet{2020A&A...634A..79S} compared the luminosity of WR stars with evolutionary models and found that helium stars with mass above roughly 7.5, 11, and 17 $M_\odot$ and in the Milky Way, LMC, and SMC, respectively, should display WR-type spectra (see their Sect. 3.2).
For a constant star formation rate of $1 M_\odot$/yr evaluated after 1 Gyr, we therefore expect $560 \pm 22$ binary-stripped stars of WR type in a solar metallicity population, $211 \pm 14$ in a $Z = 0.006$ population (similar to the LMC), and $11 \pm 4$ in a $Z = 0.002$ population (similar to the SMC). These numbers are also included in Table \ref{tab:number_estimate}.

\subsection{A helium star desert}\label{subsec:he_desert}

\begin{figure}
    \centering
    \includegraphics[width=\columnwidth]{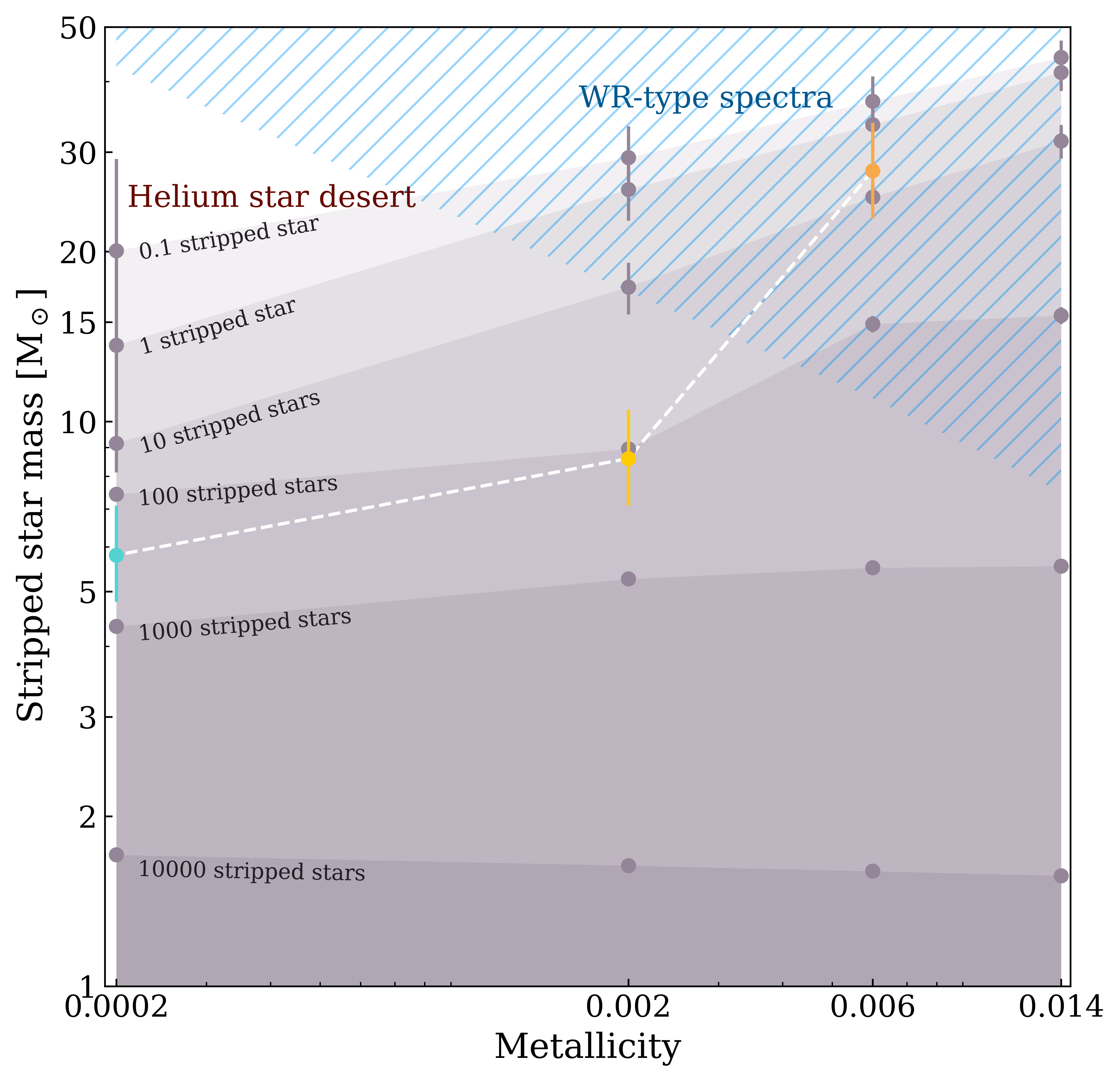}
    \caption{Contour plot depicting the number of hot stripped stars we expect for various metallicities above various masses, assuming a constant star formation rate of 1 $M_\odot$/yr. Each contour is labeled by the expected number of hot stripped stars with higher mass than the contour. At the low-mass end, the contours are close to horizontal, indicating that the formation efficiency of low-mass hot stripped stars is not affected by metallicity, but a metallicity dependence appears with increasing mass as higher-mass stripped stars are able to form at higher metallicities but are missing at lower metallicities because of late expansion. We use blue hatching to mark the general regime where WR-type spectra are expected to occur \citep{2020A&A...634A..79S}. The low-metallicity ($Z \lesssim 0.002$), high-mass ($M_{\rm strip} \sim 12-30 M_\odot$) region is practically devoid of both WR phenomena and hot binary-stripped stars, creating a ``desert'' of helium stars. 
    The locations of the drops identified in Fig. \ref{fig:mdist} are indicated in their corresponding colors and connected by a white dashed line.}
    \label{fig:contour}
\end{figure}

In Fig. \ref{fig:contour}, we illustrate the regions in mass and metallicity where we expect---and where we do not expect---hot stripped stars created through binary interaction (as described in the previous sections). Each contour in Fig. \ref{fig:contour} corresponds to a horizontal grid line from Fig. \ref{fig:cumulative} and shows, for each metallicity, the stripped star mass above which the labeled number of stripped stars are expected (assuming a constant star formation rate of $1 M_\odot$/yr).

For low stripped star masses, the expected number of stripped stars is largely insensitive to metallicity.
For a $1 M_\odot / {\rm yr}$ constant star formation rate, $\sim 10,000$ stripped stars with masses greater than $\sim 1.8 M_\odot$ are expected at all metallicities.
Similarly, $\sim 1,000$ stripped stars with masses greater than $\sim 4-5 M_\odot$ are also expected across all metallicities.

However, as already demonstrated in Figs. \ref{fig:mdist} and \ref{fig:cumulative}, the number of massive stripped stars is highly metallicity dependent.
We expect $\sim 100$ hot binary-stripped stars with mass above $15 M_\odot$ for solar metallicity, while not even one is expected for the lowest metallicity $Z=0.0002$ (Fig. \ref{fig:contour}).

For reference, we indicate the predicted mass ranges where the WR phenomenon is observed to occur in the Milky Way, LMC, and SMC in blue hatching in Fig. \ref{fig:contour} \citep[][Table 1]{2020A&A...634A..79S}. Because the data do not extend to metallicities lower than $Z=0.002$, the shaded area past this (to the left of $Z = 0.002$) is an extrapolation.
As the figure shows, binary-stripped stars could appear as WR stars at higher masses and metallicities, making them difficult to distinguish from the wind-stripped massive stars that are created at these higher masses.
However, this becomes increasingly rare with decreasing metallicity, and assuming that helium stars can only be created through envelope-stripping and wind-stripping, we expect a dearth of helium stars with masses $\sim 12-30 M_\odot$ and metallicity $\lesssim 0.002$, since neither binary evolution nor stellar winds should be efficient at producing these \citep[e.g.,][]{2013A&A...558A.103G, 2019A&A...627A..24G}.
We label this region of parameter space a ``helium star desert.''

Finally, for $Z = 0.006$, $Z = 0.002$, and $Z = 0.0002$, we plot the mass of the drop from Fig. \ref{fig:mdist} in the color corresponding to each metallicity (orange, yellow, and light blue, respectively).
The three points are connected by a white dashed line, which roughly outlines the helium star desert. 

\section{Expectations for nearby populations}\label{sec:nearby_pops}

\subsection{Predicted numbers of stripped stars} \label{subsec:nbr_nearby_pops}

The Milky Way, LMC, and SMC are the closest stellar populations in which we can observe individual stars. These populations have approximately solar (Milky Way), $\frac{1}{2}$ solar (LMC), and $\frac{1}{5}$ solar metallicity (SMC).
In the recent past, they have also had relatively constant star formation rates of 2 $M_\odot$/yr \citep[Milky Way,][]{2018ApJS..237...33X, 2022ApJ...941..162E}, 0.2 $M_\odot$/yr \citep[LMC,][]{2009AJ....138.1243H}, and 0.05 $M_\odot$/yr \citep[SMC,][]{2015MNRAS.449..639R}. Our number estimates for hot, binary-stripped stars can therefore be used to predict the existing number of helium stars in these populations.

We scaled our population predictions for $Z=0.014$, $Z=0.006$ and $Z=0.002$ (see Fig.~\ref{fig:cumulative}, Table \ref{tab:number_estimate} and Sect. \ref{subsec:number_estimates}) to the star formation rates of the Milky Way, LMC, and SMC, and thus estimated that they should contain $\sim 60,000$, $\sim 6,000$, and $\sim 1,600$ stripped stars with $M_{\rm strip} > 1 M_\odot$, respectively.
Similarly, $\sim 8,000$, $\sim 800$, and $\sim 200$ stripped stars should have sufficient mass to reach core collapse \citep[$M_{\rm strip} \gtrsim 2.6 M_\odot$,][]{2015MNRAS.451.2123T} in the Milky Way, LMC, and SMC, respectively.

\subsection{Comparison with the Milky Way supernova rate}

Stripped stars with masses above $\sim 2.6 M_\odot$ are expected to explode as supernovae \citep{2015MNRAS.451.2123T}. We can therefore compare the predicted supernova rate from our simulated populations to the observed rate of hydrogen-poor core-collapse supernovae (e.g. Type IIb/Ib/Ic). While H-poor core-collapse supernova can, in principle, be formed either from wind or binary stripping, numerous lines of evidence---from ejecta masses and intrinsic rates to the lack of progenitor detections in pre-explosion imaging---have been used to argue that binary stripped stars are the dominant formation mechanism \citep[e.g.][]{2011ApJ...741...97D,2016MNRAS.457..328L,2011MNRAS.412.1522S,2013MNRAS.436..774E}. 

To perform an order of magnitude estimate, we note that within volume-limited studies, approximately one third of all core-collapse supernovae are observed to be stripped-envelope supernovae \citep{2017ApJ...837..120G, 2017ApJ...837..121G}. Coupled with a Galactic supernova rate of about 2 per century \citep{2006Natur.439...45D}, this suggests that the Galactic stripped-envelope supernova rate is $\sim$0.0067 yr$^{-1}$.

To estimate a rate from our $Z = 0.014$ population, we output the duration of the stripped star phase, $t_{\rm strip}$, for each stripped star as determined from the evolutionary models. We then calculated the stripped envelope supernova (SESN) rate to be the sum of the inverses of the stripped star durations for each star with $M_{\rm strip} \geq 2.6 M_\odot$; that is,
\begin{equation}
    R_{\rm SESN} = \sum_{M_{\rm strip} \geq 2.6 M_\odot} \frac{1}{t_{\rm strip}}.
\end{equation}
Scaling according to the Milky Way star formation rate of $2 M_\odot$/yr, we found that $R_{\rm SESN} = 0.0074 \text{ yr}^{-1}$. In addition, while there is active debate about whether massive WR stars should explode as core-collapse supernovae or direct collapse to black holes \citep{2001ApJ...554..548F}, we also attempted to account for the contribution of WR stars. Since they are more massive, they only live $\sim 500,000$ years \citep{2014A&A...564A..30G}. There are $\sim 800$ WR stars in the Milky Way that are single and therefore not accounted for in our estimate (see Sect. \ref{subsec:WRestimate}). This means that we needed to add $800/500,000 = 0.0016$ yr$^{-1}$ to the rate of stripped-envelope supernovae, thus reaching 0.0090 yr$^{-1}$.

The reasonable match between our predicted and the observed stripped-envelope supernova rates provides support for our population predictions and for the mass at which helium stars can reach core collapse.
However, see Sect. \ref{subsubsec:uncertainties} for the primary sources of uncertainty that could be responsible for the slight mismatch.

\subsection{Investigation of late expansion using the SMC mass distribution}\label{subsec:SMCdrop}

As described in Sect. \ref{subsubsec:drop}, our models predict a drop in the SMC mass distribution at $M_{\rm strip} \sim 9 M_\odot$ because of its low metallicity, which is predicted to cause massive stars to expand late and therefore become stripped late.
Scaling by the adopted star formation rate of the SMC, we expect $\sim 5$ binary-stripped stars with masses above this drop. By contrast, the solar metallicity model, which does not show a drop in the mass distribution, predicts $\sim 20$ stripped stars with masses $\gtrsim 9 M_\odot$ when scaled by the SMC star formation rate.

The number of stripped stars with $M_{\rm strip} > 9\; M_\odot$ in the SMC can therefore be used as a test of the presence of a drop.
If the number of such massive stripped stars in the SMC is closer to 20 than 5, it could disprove the theorized late expansion at low metallicity. Alternatively, it could indicate that these stripped stars were created through a different formation scenario than envelope-stripping in binaries (see e.g., Sect. \ref{subsec:CHE}).

There is already a detected population of massive helium stars in the SMC---the WR stars. Our $Z=0.002$ models predict that envelope-stripping should produce 0.5 WR star in the SMC, assuming that stripped stars with $M_{\rm strip} > 17 \, M_\odot$ show WR type spectra. However, while the SMC has 12 WR stars \citep{2015A&A...581A..21H, 2016A&A...591A..22S}, 5 of them are in such close binary systems \citep[$\sim 20$ day orbital periods,][]{2003MNRAS.338..360F, 2024A&A...689A.157S} that they could have been produced by envelope-stripping in binaries \citep[cf.][]{1994A&A...288..475P, 2017MNRAS.464.2066S}. Although the numbers are small and stochastic effects play a role, it is intriguing that $\sim 3.5$ WR stars are expected for the same star formation rate but assuming the distribution with no drop.
This better match could suggest that stars with  metallicity comparable to that of the SMC do not experience late expansion, at least not above $\sim 40 M_\odot$. On the other hand, intriguingly, none of the WR binaries in the SMC reside in intermediate period orbits ($\sim 20-400$ days), which would be the likely outcome of binary stripping in models that currently fail to produce stripped stars due to late expansion. To fully test binary models and the phenomenon of late expansion, it would be ideal to map out the complete stripped star distribution in the SMC, at least down to masses below the expected drop at $\sim 9 M_\odot$.

Another interesting population to compare with are the luminous blue supergiants that are cooler than the main-sequence in the SMC. If late expansion does occur, the binaries that interact during or after helium-core burning should appear as such cool blue supergiants during the helium-core burning stage. With WR bolometric luminosities of $\log_{10} L/L_\odot > 5.6$, one could expect such cool blue supergiants to have similar bolometric luminosities. Comparing the mass distribution that contains a drop with the one that does not, we expect that if the SMC has a drop in its mass distribution, there should be at least 5 times more cool blue supergiants than WR stars. With 5 WR stars present in binaries, this corresponds to an expected 25 cool blue supergiants. However, as demonstrated by \cite{2021A&A...646A.106S}, the SMC contains only a handful, thus further strengthening the suggestion that SMC stars above $40 M_\odot$ do not experience late expansion.
We note, however, that \cite{2024A&A...689A.157S} also found a strong indication that these stars are subject to eruptive mass loss or significantly stronger stellar winds than commonly predicted, which may affect their radial expansion and further complicate the picture.

\subsection{Constraints from the Wolf-Rayet population}\label{subsec:WRestimate}

Adopting the mass thresholds for WR-type spectral morphology from \citet[][see also Table \ref{tab:number_estimate}]{2020A&A...634A..79S} and convolving with the adopted star formation rates, we estimate that there should be around $1100$, $45$ and $0.5$ (or $3.5$, depending on whether the drop is present, see Sect. \ref{subsec:SMCdrop}) WR stars produced from binary interaction in the Milky Way, LMC, and SMC, respectively.

The Milky Way has an observed sample of almost $700$\footnote{See the WR catalog by P.A.\ Crowther on \url{https://pacrowther.staff.shef.ac.uk/WRcat/}.}, which increases to $\sim 2,000$ after correcting for observational biases \citep{2024arXiv241004436S}. In contrast, the LMC and SMC WR samples are considered to be relatively complete, with 154 and 12 WR stars, respectively \citep{2014A&A...565A..27H,2015A&A...581A..21H,2016A&A...591A..22S,2017ApJ...841...20N,2019A&A...627A.151S}. \citet{2024A&A...692A.109D} observed the binary fraction of WN-type WR stars in the Milky Way to be $\sim 55$\% \citep[see also][]{2020A&A...641A..26D,2022A&A...664A..93D,2023A&A...674A..88D}, meaning that $\sim 1,200$ of the expected $\sim 2,000$ WR stars in the Milky Way should be WR binaries and the remaining $\sim 800$ should be single. The LMC has $44$ WR binary candidates \citep{2019A&A...627A.151S}, and the SMC has $5$ \citep{2016A&A...591A..22S, 2018A&A...616A.103S}. These observed numbers of WR binaries match well with our predictions. However, the large fraction of the WR stars at all metallicities that cannot be explained by binary stripping further underscores the need to better understand the mechanisms behind WR star formation \citep{2020A&A...634A..79S}.

\subsection{Comparison with stripped star detections in the Magellanic Clouds}

\citet{2023Sci...382.1287D} recently presented the results of an observational search for stripped stars in the Magellanic Clouds. Because candidates were identified through their excess UV emission compared to main-sequence stars, only systems with low extinction and lower-mass or dark companions could be identified. This means that the sample is incomplete, but it is still interesting to note that the most massive stripped star found in the SMC has $\sim 7-8 M_\odot$ \citep{2023ApJ...959..125G}, which is just below the expected drop at $\sim 9 M_\odot$.
On the other hand, the WR stars in the SMC are substantially brighter and likely more massive \citep[$\gtrsim 20\; M_\odot$,][]{2015A&A...581A..21H, 2016A&A...591A..22S} than these stripped star candidates, leaving a gap in mass that appears empty of stripped stars.
While higher mass stripped stars are more difficult to select via UV excess (due to their overlap with main sequence stars at higher luminosities), it is possible that this gap is the helium star desert (see Sect. \ref{subsec:he_desert}), and that the WR stars formed through other pathways, such as enhanced mass loss via wind stripping or chemically homogeneous evolution (Sect. \ref{subsec:CHE}).
A better understanding of these mechanisms will be necessary to confirm the origins of the SMC's WR stars.

A larger data set could help to confirm the existence of the helium star desert in other ways.
With complete observations of stripped stars down to $M_{\rm strip} \sim 3-5 M_\odot$, it would be possible to constrain the slope of the stripped star mass distribution (see Fig. \ref{fig:mdist}). While dependent on many factors, the slope would provide an additional constraint for uncertainties related to, for example, convective overshoot and the outcome of different envelope-stripping mechanisms. With an upcoming, newly reduced UV photometry catalog (Ludwig et al.,\ in prep.) based on data from the Swift Ultraviolet Survey of the Magellanic Clouds \citep[SUMaC,][]{2017MNRAS.466.4540H}, suitable data could soon be available. This will also be an interesting case for the UV mission UVEX, scheduled to launch in 2030, which will map the entire mass distribution of stripped stars \citep[][see science pillar 1]{2021arXiv211115608K}. However, we note that direct comparison to samples of hot stripped stars identified via UV excess will likely require further characterization of the expected contributions of stripped star plus compact object systems. While subdominant to the overall population (see Sect. \ref{subsec:popsynth}), these systems are more  efficiently identified via the UV excess method \citep{2023Sci...382.1287D}.

\section{Implications and potential observational tests}\label{sec:discussion}

\subsection{Partially stripped stars as the counterpart of hot stripped stars at low metallicity}\label{subsec:partial_strip}

In this study, we have considered the production of hot stripped stars through envelope-stripping in binaries that is initiated before central helium burning. However, at low metallicity, if a star is radiative and initiates mass transfer during or shortly before central helium burning, it could become a cool, partially stripped star or continue mass transfer until stellar death \citep{2022A&A...662A..56K}. 

These partially stripped stars could be more numerous than hot stripped stars if they originate from high-mass donor stars and low-metallicity environments. However, this hinges on the prediction of late expansion for massive stars at low metallicity. As described in Sect. \ref{subsec:nbr_nearby_pops}, a mass distribution with a drop predicts that the SMC will have only 5 hot binary-stripped stars with masses $>9 M_\odot$ and initial masses $\gtrsim 25M_\odot$, while the distribution without a drop predicts 20 such stars. This suggests that if the SMC does exhibit a drop, there should be $\sim 15$ stars with initial mass $\gtrsim 25 M_\odot$ that entered nuclear timescale mass transfer or that are partially stripped. Since they likely would quickly lose a large fraction, but not the entirety, of their hydrogen-rich envelopes, we expect that their masses should be higher (e.g., $\gtrsim 10 M_\odot$) than their hot counterparts \citep[see, e.g., Fig. 11 of][]{2022A&A...662A..56K}. 

Furthermore, these stars should be bright ($\log L/L_\odot \gtrsim 5$) and have temperatures that are similar to, or cooler than, the main sequence ($T_{\rm eff} \sim 10-30$kK).
Because of their lower temperatures, partially stripped stars should be significantly brighter in the optical wavelength range compared to hot stripped stars, which emit primarily in the UV \citep[$T_{\rm eff}\gtrsim 50$kK,][]{2018A&A...615A..78G}. This suggests that partially stripped stars are easily detectable but can masquerade as regular main-sequence stars or blue supergiants with somewhat lower surface gravity ($\log g \sim 2-4$) and higher surface helium abundance ($Y_{\rm surf} \sim 0.5$) than expected \citep{2022A&A...662A..56K}. However, while undergoing mass transfer, they could also be distinguished by appearing cooler than main-sequence stars and as eclipsing binaries and ellipsoidal variables \citep{2008ApJ...673L..59P}. 

In recent years, several intriguing observations have been presented of proposed $\sim 3-8 M_\odot$ partially stripped stars in the SMC and LMC \citep{2023A&A...674L..12R, 2024A&A...692A..90R}. These partially stripped stars appear as O- or B-type giants or supergiants and show surface enrichment in nitrogen and, in some cases, helium. An alternative explanation for these stars is that they are helium stars that have just undergone envelope-stripping and are observed while contracting before helium core burning starts. This evolutionary phase has been invoked to explain observations of several other stars \citep[e.g.,][see also \citealt{2024A&A...687A.215D}]{2020A&A...633L...5I,2020A&A...641A..43B,2022NatAs...6.1414I,2023MNRAS.525.5121V}.
Mapping out the full parameter space of partially stripped stars and such bloated stripped stars would help to constrain both late expansion and the outcome of binary interaction at low metallicity.

It is worth noting as well that models predict that stars undergoing partial envelope-stripping or mass transfer that initiates only after the end of core helium burning should have a carbon-oxygen core structure similar to that of single stars and would be significantly different from that of stars stripped fully earlier in their evolution \citep{2021A&A...645A...5S,2022A&A...662A..56K}. Those core differences could affect these stars' explodability and the formation of compact objects \citep{2021A&A...645A...5S,2021ApJ...916L...5V,2021A&A...656A..58L,2025A&A...695A..71L}.
 
\subsection{Implications for high-redshift galaxies and the production of ionizing photons} \label{subsec:ionizing}

Whether low-metallicity massive stars predominantly create hot stripped stars or cool partially stripped stars directly impacts the output of ionizing radiation from stellar populations. 
This is because of the difference in effective temperature between these two binary products \citep[cf.\ e.g.,][]{2023ApJ...959..125G,2024A&A...692A..90R}.
The ionizing output is of particular interest for high-redshift galaxies, both because ionizing radiation powered the reionization of the Universe, and because it is responsible for the nebular emission lines that are the main observable features of these distant populations. 
Metallicity is typically lower at high redshift than in most present-day galaxies, and most galaxies with redshift $z > 2$ have a stellar metallicity at least as low as the SMC \citep{2016ApJ...826..159S, 2018ApJ...868..117S, 2019MNRAS.487.2038C, 2020MNRAS.499.1652T, 2024MNRAS.532.3102S}.
Therefore, the helium star desert, if real, should impact metal-poor, high-redshift galaxies.

While we did not model the spectral energy distribution of massive stripped stars in this study, we used previous predicted ionizing emission rates to estimate the influence of the helium star desert on the hydrogen- and helium-ionizing output from a stellar population. We did this for a population with metallicity $Z=0.002$ and constant star formation rate of $1 M_\odot$/yr, adopting the mass distributions presented in Sect. \ref{subsec:mdist}. 

\begin{figure}
\centering
\includegraphics[width=\columnwidth]{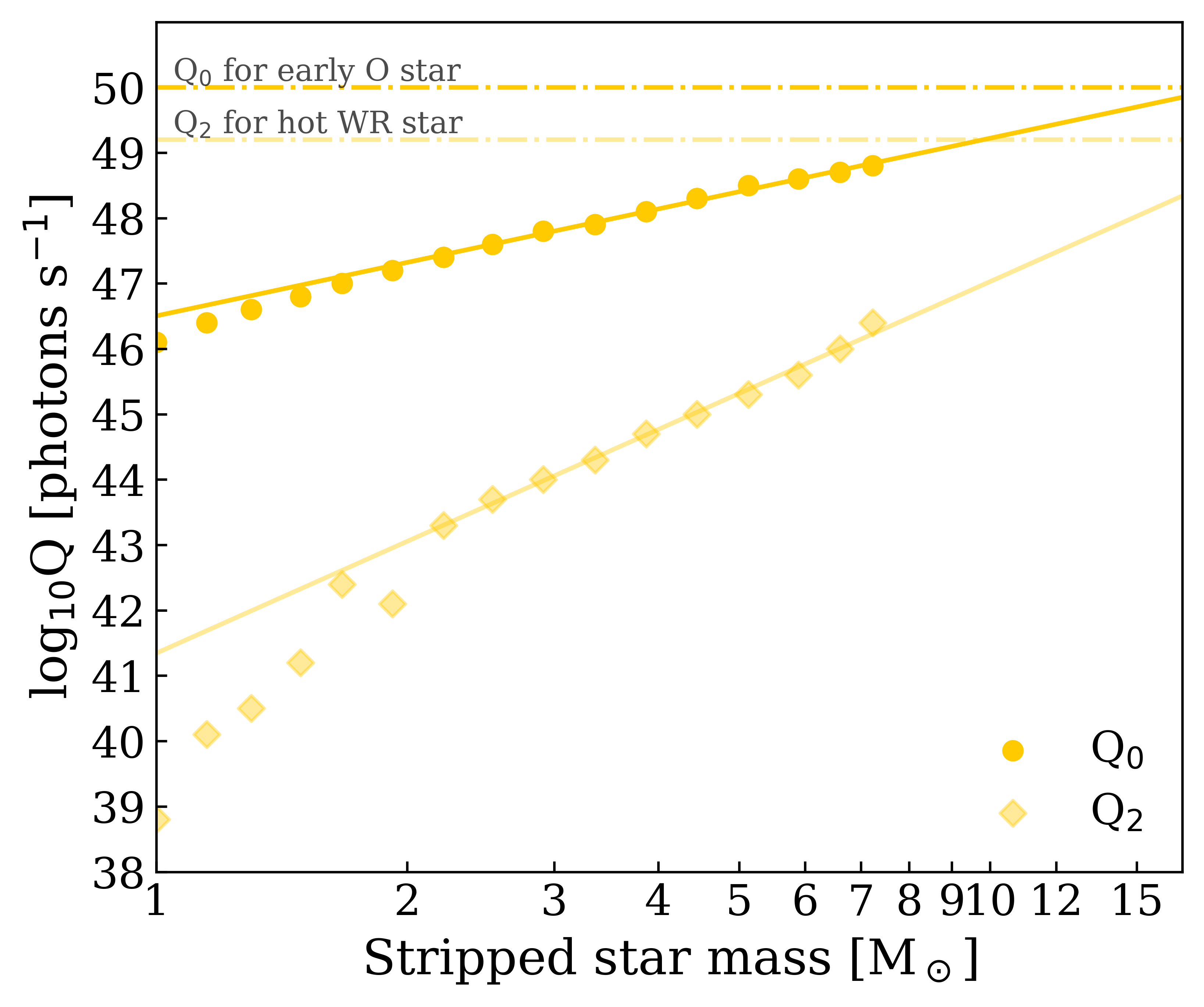}
\caption{Estimates of hydrogen- and fully helium-ionizing photon emission rates of individual stripped stars at $Z=0.002$ from \citet[][$Q_0$ and $Q_2$, respectively]{2018A&A...615A..78G}. To estimate the ionizing emission from massive and hot stripped stars in a $Z=0.002$ population, we fit a line to the stars with $M_{\rm strip} > 2 \, M_\odot$. We then extrapolated this function up to stripped star masses of 17 $M_\odot$ to represent the emission from stripped stars that do not appear as WR type. For reference, we mark the $Q_0$ and $Q_2$ for an early O-type star and a hot WR star, respectively, according to models from \citet{2002MNRAS.337.1309S}.}
\label{fig:Q_fit}
\end{figure}

We used the ionizing emission rate estimates from \citet{2019A&A...629A.134G}, adopting $10^{52}$ photons per second as the total hydrogen-ionizing emission rate ($Q_0$) from hot stripped stars with masses $\leq 8.5 M_\odot$. In this population, the authors used the same assumptions for the binary population synthesis and the same underlying evolutionary models as we did here, but they only included stripped stars with masses up to 8.5 $M_\odot$ (corresponding to initial masses up to  $\sim 20 M_\odot$). They interpolated over their spectral models to estimate the total ionizing emission from stripped stars across this mass range. 

To account for the contribution from higher mass stripped stars, we fit a straight line to $Q_0$ as function of stripped star mass using the predictions for individual stripped stars with mass $>2 M_\odot$ by \citet{2018A&A...615A..78G}. We then extrapolated this function out to $M_{\rm strip} = 17 M_\odot$ to estimate the ionizing emission rates of the stripped stars with masses above 8.5 $M_\odot$. We excluded stripped stars with mass $> 17 M_\odot$, as these may have WR-type spectra \citep{2020A&A...634A..79S}, which can have cooler photospheres located in the stellar wind and opaque stellar winds that block ionizing emission. This mass boundary is also below the mass where \citet{2020MNRAS.499..873S} estimate that wind mass loss rates of $Z = 0.002$ stars significantly increase. 
Fig. \ref{fig:Q_fit} shows the model predictions and the extrapolation of the fit. 
We also display predictions for two other ionizing stars - an early O-type star and a hot WR star \citep{2002MNRAS.337.1309S}. The extrapolation falls below these numbers, but detailed evolutionary and spectral models would be necessary to provide a more accurate estimate. 

\begin{figure}
\centering
\includegraphics[width=\columnwidth]{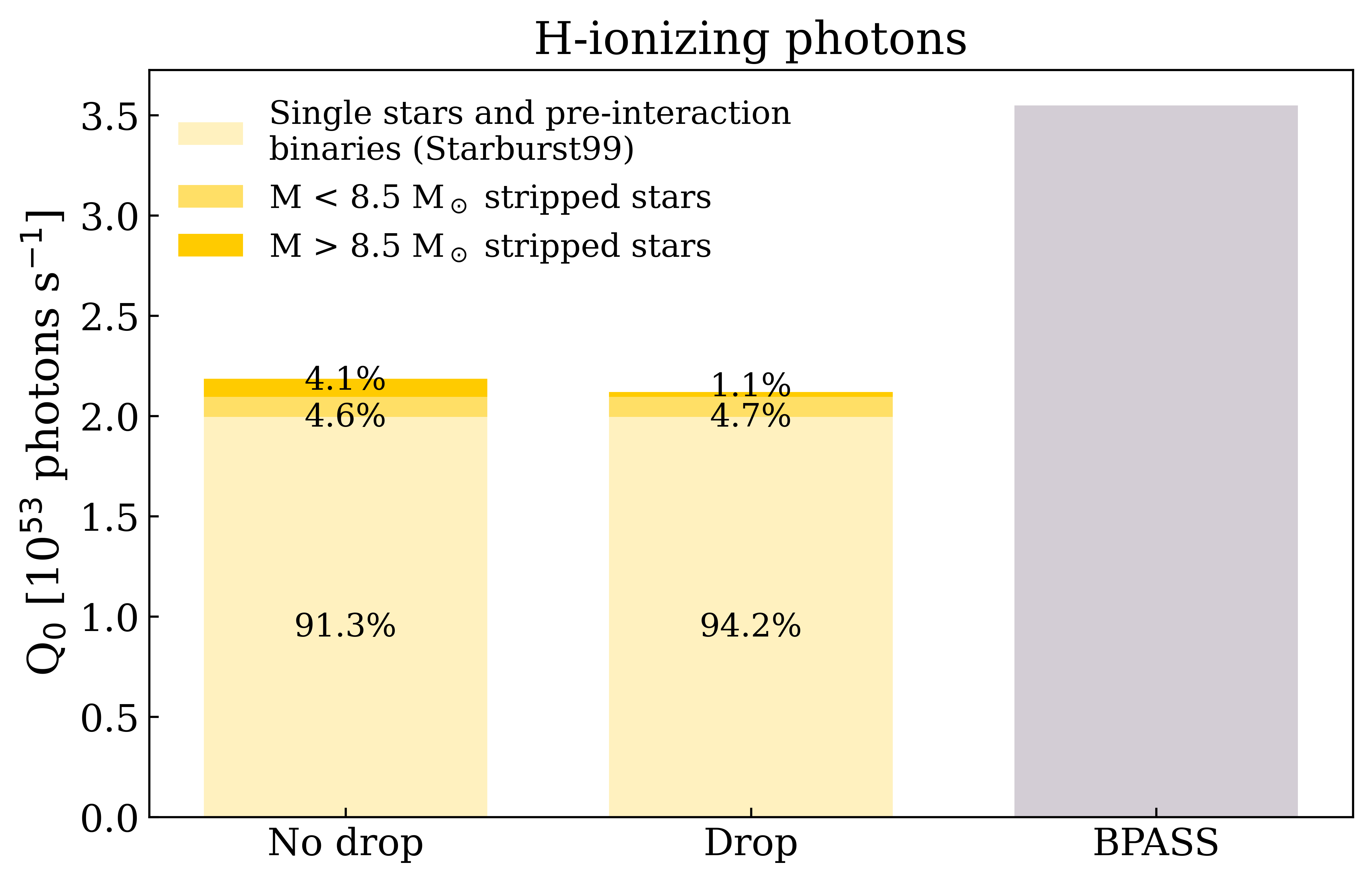}
\includegraphics[width=\columnwidth]{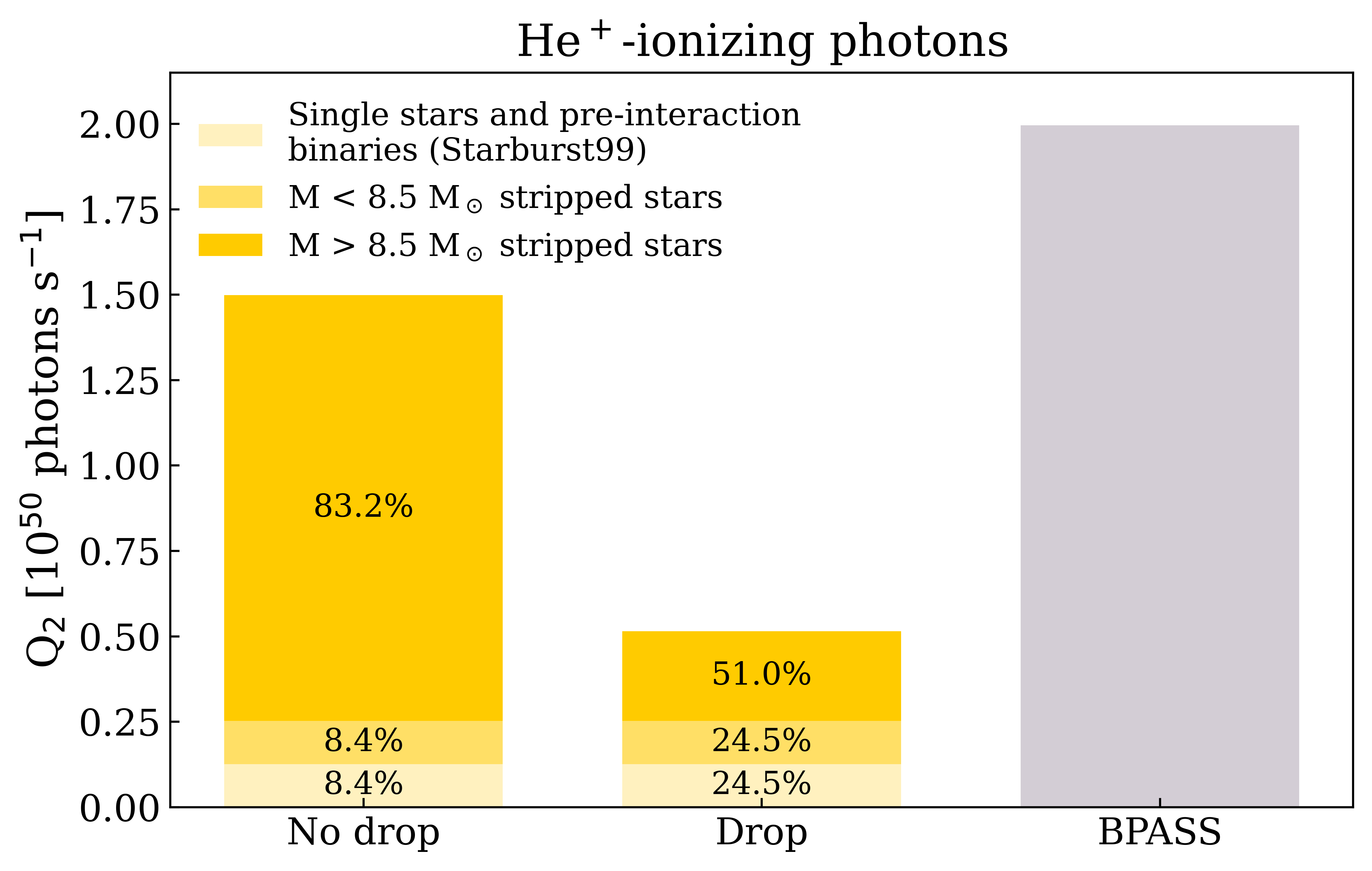}
\caption{Rough estimates for hydrogen- ($Q_0$, top) and helium-ionizing ($Q_2$, bottom) emission rates. The total rates are shown as a sum of rates for single stars, low-mass stripped stars whose rates are calculated via interpolation ($M < 8.5 M_\odot$), and higher-mass stripped stars whose rates are calculated via extrapolation ($M > 8.5 M_\odot$). We did this calculation separately in the case that there is no drop (left) and in the case that there is a drop (middle), and we show the estimate from BPASS for comparison (right).}
\label{fig:Q}
\end{figure}

Using this extrapolation to assign ionizing emission rates to the stripped stars in our $Z = 0.002$ mass distribution, we predict that the stripped stars with $M_{\rm strip} > 8.5\, M_\odot$ should contribute $10^{51.4}$ hydrogen-ionizing photons per second in a population in which stars form at a constant rate of $1 \, M_\odot$/yr.
This constitutes a $\sim 25 \%$ boost in the contribution from stripped stars compared to \citet[][see their Table~A.3]{2019A&A...629A.134G}. 
If we instead use the mass distribution from the $Z=0.014$ population to emulate a low-$Z$ distribution without a drop in the mass distribution, we estimate that massive stripped stars would contribute $10^{52.0}$ hydrogen-ionizing photons per second for a constant star formation rate of $1 \, M_\odot$/yr. This would double the contribution from stripped stars compared to \cite{2019A&A...629A.134G}.
However, in either case, stripped stars are not a dominant source of H-ionizing photon production from a full stellar population.
Adopting the ionizing contribution from single stars from Starburst99 ($Q_0 = 10^{53.3}$ photons s$^{-1}$, \citealt{1999ApJS..123....3L}), we estimate that the total contribution from stripped stars of all masses to the hydrogen-ionizing emission rate from a low-$Z$ population should be between $10^{52.1}$ s$^{-1}$ and $10^{52.3}$ s$^{-1}$, or 6.5 to 9.1\%, depending on whether or not the mass distribution exhibits a drop, respectively.
These results are summarized visually in the top panel of Fig. \ref{fig:Q}. 

The emission rate of photons that can fully ionize helium ($Q_2$) is an uncertain property for all stars, because such hard ionizing emission originates from the steep Wien part of the spectral energy distribution, which is highly dependent on density and recombination in the stellar atmospheres \citep[see e.g.,][]{2022arXiv221105424S}. With wind mass loss-associated uncertainties directly influencing the atmosphere's density \citep{2017A&A...608A..11G}, it is clear that the He$^+$-ionizing regime is in urgent need of observational constraints. 

Stripped stars with transparent atmospheres, that is, ones that do not show WR type spectra, could be expected to emit a substantial amount of He$^+$-ionizing emission.
We therefore made a rough estimate of the contribution of massive stripped stars to the helium-ionizing budget of a stellar population. We took the same approach as described previously in this section for the hydrogen-ionizing emission estimate (see Fig.~\ref{fig:Q_fit}). Again, we extrapolated up to $M_{\rm strip} = 17 M_\odot$, which we estimated has $Q_2 = 48.3$ s$^{-1}$. This agrees with the estimate for the hottest WN star (WR stars with nitrogen-dominated spectra) models from \citet{2002MNRAS.337.1309S} and the estimate for massive helium stars from \citet{2020MNRAS.499..873S}.
By summing the He$^+$-ionizing emission for individual stars, we estimated the contribution from a population of massive stripped stars at $Z = 0.002$ to be $Q_2 = 10^{49.4}$ photons per second for a constant star formation rate of $1 \, M_\odot$/yr. If, instead, late expansion does not occur at low metallicity, and the mass distribution does not have a drop (i.e., we adopted the $Z=0.014$ mass distribution), the contribution could be as high as $Q_2 = 10^{50.1}$ photons per second.

This exercise suggests that massive stripped stars without WR-type spectra could be the most prominent source of He$^+$-ionizing emission from a stellar population. However, it is a rough estimate and observational confirmation would be required.
The $Q_2$ estimate for main sequence and WR stars from Starburst99 is $10^{49.1}$ photons s$^{-1}$ and for stripped stars with $M_{\rm strip} < 8.5\, M_\odot$ it is $10^{49.1}$ photons s$^{-1}$ \citep{2019A&A...629A.134G}.
Therefore, if there is no drop, the He$^+$-ionizing budget could increase by 500\%, reaching $10^{50.2}$ photons s$^{-1}$.
Even if there is a drop, the addition of massive stripped stars would still double the total budget of He$^+$-ionizing photons, reaching a total of $10^{49.7}$ photons s$^{-1}$ (Fig. \ref{fig:Q}, bottom panel).
These values are similar to predictions from version 2.0 of the BPASS binary spectral population synthesis code \citep{2016MNRAS.456..485S}, included for comparison in Fig. \ref{fig:Q}. However, we note that the main ionizing source likely differs between our two estimates, since we did not include rapidly rotating, chemically homogeneously evolving accretor stars, while in  BPASS they are prominent sources of ionizing emission. 

Matching the observed He II recombination line emission is a long-standing problem for models of stellar populations in low-metallicity galaxies \citep[e.g.,][]{2000ApJ...531..776G, 2012MNRAS.421.1043S, 2013A&A...556A..68C, 2017MNRAS.472.2608S, 2019MNRAS.488.3492S, 2020A&A...636A..47S}. Low-metallicity galaxies exhibit strong emission in ionized helium recombination lines, suggesting that more He$^+$-ionizing photons are produced than spectral population synthesis models predict \citep[e.g.,][]{2018ApJ...859..164B, 2019A&A...624A..89N, 2022ApJ...938...16O}.
The sources of this He$^+$-ionizing emission remain a puzzle.
However, stellar population synthesis models including effects from binaries (e.g., BPASS), have been shown to be the only ones able to in some cases match both the H-ionizing and higher-energy He$^+$-ionizing emission \citep{2016ApJ...826..159S}.
Stripped stars, which do not seem to significantly modify the H-ionizing production (see the top panel of Fig. \ref{fig:Q}) but do enhance harder ionizing photon emission (bottom panel of Fig.\ \ref{fig:Q}), are a promising solution to this problem without necessitating large differences at longer UV wavelengths.
Our rough estimate for the contribution from massive stripped stars is therefore intriguing and emphasizes the importance of better modeling these stars, both in terms of evolutionary modeling and spectral modeling, which were approximated in our study.

Furthermore, observations of massive ($\gtrsim 8.5 M_\odot$) stripped stars at low-metallicity ($Z\sim 0.002$) that do not exhibit WR-type spectra are still largely missing (see, however, star 1 in the SMC presented in \citealt{2023Sci...382.1287D} and \citealt{2023ApJ...959..125G}). A more representative sample will be essential to constrain the ionizing emission of massive stripped stars and understand their contribution to stellar populations' spectra.  

\subsection{Implications for the formation of gravitational wave progenitors}

Understanding the formation pathways of black hole and neutron star binary systems is crucial to accurately predicting the mass distributions and merger rates of compact objects.
In the standard isolated binary evolution scenario leading to a binary compact object merger, two envelope-stripping phases are expected to occur, one of which is required in order to bring the two stars in the system sufficiently close for a merger to occur within a Hubble time \citep{2016Natur.534..512B, 2016A&A...596A..58K, 2017NatCo...814906S,2017ApJ...846..170T}.

The progenitors of massive ($>$20 M$_\odot$) black holes are thought to originate in low metallicity environments \citep[$Z < 0.002$,][]{2010ApJ...714.1217B, 2018MNRAS.480.2011G, 2018A&A...619A..77K, 2019MNRAS.482.5012C, 2019MNRAS.490.3740N, 2022MNRAS.516.5737B}. 
This suggests that these progenitors were either the massive hot stripped stars we modeled in this work or the partially stripped or mass-transferring cool objects discussed in Sect. \ref{subsec:partial_strip}. While late expansion at low metallicity is included in rapid binary population synthesis codes \citep{1998MNRAS.298..525P,2000MNRAS.315..543H}, the complex mass transfer phase associated with massive, low-metallicity stars merits more in-depth study and likely an updated treatment for population synthesis to ensure that the evolution in populations agrees with what is found in detailed evolutionary models \citep[cf.\ e.g.,][]{2022A&A...662A..56K,2024A&A...685A..58E}. 
This could be particularly relevant to explore in context of the proposed evolutionary channel where stable mass transfer can bring two stars sufficiently close that a common envelope is not required for the future compact object system to merge within a Hubble time (\citealt{2017MNRAS.468.5020I, 2017MNRAS.471.4256V, 2021A&A...650A.107M}, see also \citealt{2021A&A...647A.153B}).

The late expansion of massive stars in binaries also determines whether mass transfer is initiated right after the end of the main sequence (i.e., Hertzprung gap donors) or only later during helium burning (i.e., core-He burning donors). Those two types of stars are commonly differentiated in rapid binary evolution codes \citep[following][]{2002MNRAS.329..897H}. In population studies of gravitational wave sources, common envelope evolution has typically been limited to core-He burning donors \citep[e.g.,][]{2012ApJ...759...52D, 2016Natur.534..512B, 2020PASA...37...38V, 2021MNRAS.508.5028B}, with Hertzprung gap giants assumed to always lead to mergers and the formation of Thorne-\.Zytkow objects \citep[cf.][]{2023MNRAS.524.1692F, 2024arXiv241002896O}. As a result, the phenomenon of late expansion has an effect on the predicted formation rates of gravitational wave sources as a function of metallicity, contributing to the preference for low metallicity \citep[though in the current models it is not the leading-order effect, see e.g.,][]{2025ApJ...979..209V}. More recently, it has been demonstrated that the outcome of common envelope evolution is primarily determined by the type of the outer envelope (radiative vs.\ convective), rather than the evolutionary stage of the donor (Hertzprung gap vs.\ core-He burning) \citep{2016A&A...596A..58K, 2021A&A...645A..54K, 2025A&A...693A.137R}. It is still unclear, however, how the late expansion of massive stars affects the stable mass transfer channel. \cite{2022A&A...662A..56K} found that mass transfer from a core-He burning donor is characterized by a lower mass exchange rate compared to that from a Hertzprung gap giant. This suggests an increased mass transfer stability (Klencki et al.\ in prep.), which is a crucial factor favoring the formation of close-orbit binary black hole systems that merge within a Hubble time \citep{2021A&A...650A.107M, 2021A&A...651A.100O, 2022ApJ...931...17V, 2024A&A...681A..31P}.

\subsection{Rapidly rotating stars and the helium star desert}\label{subsec:CHE}

In this study, we have considered only helium stars formed through one of two pathways: those formed through the binary envelope-stripping that we also model in this work and those formed through strong stellar winds. Our models predict that neither formation scenario should be efficient at producing massive ($M_{\rm strip} \sim 10-30 M_\odot$) helium stars at low metallicities ($Z \lesssim 0.002$, see the helium star desert in Fig. \ref{fig:contour}).

However, at low metallicity, where line-driven winds are expected to be weak, rapid rotation has been predicted to be able to induce such strong interior mixing that massive stars \citep[$\gtrsim 20 \, M_\odot$ at $Z=0.002$,][]{2011A&A...530A.115B} should be able to evolve chemically homogeneously  \citep[see also][]{1987A&A...178..159M,2009A&A...495..257M, 2015A&A...581A..21H, 2018A&A...611A..75S}. As a result, during the main-sequence evolution the stars slowly convert themselves into helium stars, becoming hotter and bluer than the zero-age main-sequence \citep{2009A&A...497..243D, 2015A&A...581A..15S, 2019A&A...623A...8K}. The observational signatures of these stars include surface helium and nitrogen enrichment along with rapid rotation. Chemically homogeneously evolving stars have, however, not been conclusively observed. For example, \citet{2019ApJ...880..115A, 2021A&A...651A..96A} found no chemical enrichment in massive, rapidly spinning overcontact binaries. Additionally, \citet{2018A&A...611A..75S} found that structure models of chemically homogeneously evolving stars were a poor match for the SMC WR stars, whose rotation rates are too slow for stars of this type. But, if they exist, chemically homogeneously evolving stars could populate the helium star desert. 

It could be possible to look for these objects and test if they exist using our stripped star mass distribution. If more than ten helium stars more massive than $\sim 10 M_\odot$ are found in very metal-poor environments with constant star formation rates of $1 M_\odot$/yr (e.g.,  $Z=0.0002$), it is not expected from either binary envelope-stripping or wind-stripping. Such objects would thus provide excellent candidates for exploring chemically homogeneous evolution.

\section{Summary and conclusions}\label{sec:conclusions}

In this work, we explored the effect of metallicity on the formation of hot stripped stars through envelope-stripping in binary stars.
To do this, we used a Monte Carlo binary population synthesis code created specifically for binary-stripped stars (see Sect. \ref{subsec:popsynth}) and based on detailed single and binary stellar evolution models (see Sect. \ref{subsec:mesa}). 
We summarize our findings below:

\begin{enumerate}
    \item We estimated the number of hot stripped stars that should exist in a population as a function of mass (see Sect. \ref{subsec:number_estimates}, Table~\ref{tab:number_estimate}, and Fig.~\ref{fig:cumulative}). Independent of metallicity, we found that $\sim 4,000$ stripped stars per $1 \ M_\odot / {\rm yr}$ of constant star formation should be present and sufficiently massive to reach core collapse ($> 2.6 \ M_\odot$, \citealt{2015MNRAS.451.2123T}). Similarly, we found that $\sim 30,000$ stripped stars with mass $>1 M_\odot$ should exist. 

    \item Scaling the number estimates with appropriate star formation rates, we estimated the number of hot stripped stars in the Milky Way, LMC, and SMC (see Sect. \ref{sec:nearby_pops}). We found that there should be around 8,000 (60,000), 800 (6,000), and 200 (1,600) stripped stars in the Milky Way, LMC, and SMC, respectively, and with mass above 2.6$M_\odot$ (1$M_\odot$). 

    \item We studied the shape of the mass distribution for hot stripped stars (see Sect. \ref{subsec:mdist}). At the low-mass end, the slope of the present-day mass function ($\alpha = -2.01$ to $-1.87$) is similar for all four metallicities that we considered, and it is also shallower than the slope of the adopted initial mass function ($\alpha = -2.3$). This slope is sensitive to, for example, the assumed binary fraction and the adopted prescription for main sequence core overshoot. We discuss how, if measured from observations, this slope could potentially be used to help constrain these uncertain properties.
    
    \item We found that low-metallicity environments are expected to host many fewer massive hot stripped stars than high-metallicity environments. The reason is that, at low metallicity, massive stars are predicted to expand predominantly only after central helium burning has started, delaying binary interaction. The resulting lack of massive stripped stars can be seen in our computed mass function as a drop at the high-mass end of the distribution (see Fig.~\ref{fig:mdist}).
    For example, we estimated that only $\sim 5$ hot binary-stripped stars with $M_{\rm strip} > 9 \ M_\odot$ should exist in the SMC (Sect. \ref{subsec:SMCdrop}). Of these, we only expect $\sim 1$ to be sufficiently massive to exhibit a WR-type spectrum. Our results thus suggest that binary interaction is not efficient at forming WR stars at low metallicity. 
    
    \item This lack of massive hot stripped stars at low metallicity can be described as a ``helium star desert'' --- a triangular region of mass-metallicity space with $Z \lesssim 0.002$ and ranging from $M_{\rm strip} \sim 15-30 \ M_\odot$ (see Fig.~\ref{fig:contour} and Sect. \ref{subsec:he_desert}), where our models predict that helium stars neither form efficiently through binary interaction nor through wind mass loss.
    If an unexpected amount ($\gtrsim 10$ per 1$M_\odot$/yr at $Z=0.002$) of helium stars are found in this region of mass-metallicity space, it is possible that they are rapidly rotating stars that evolve chemically homogeneously (see Sect. \ref{subsec:CHE}). However, it could also mean that the predicted late expansion at low metallicity does not, in fact, occur in nature, and that additional work is required to constrain and model the post-main sequence evolution of massive stars across metallicities.
\end{enumerate}

Our work is complementary to earlier studies of the binary stripped star populations \citep{2019A&A...629A.134G, 2020ApJ...904...56G, 2020A&A...634A.126W, 2021ApJ...908...67S, 2024A&A...683A..37Y, 2024ApJ...975L..20W}. However, our focus on the effect of metallicity on the shape of the mass distribution is new and has various interesting implications for adjacent astrophysical research. 

For example, the predicted lack of massive hot stripped stars at low metallicity could limit their contribution to the hard ionizing radiation observed in high-$z$ galaxies and their local analogs, as well as their contribution to cosmic reionization \citep[cf.][]{2020A&A...634A.134G}. We made a rough estimate of the ionizing contribution from massive stripped stars (see Sect. \ref{subsec:ionizing}). We found that while they are non-negligible contributors of H-ionizing photons, stripped stars could potentially dominate the output of He$^+$-ionizing photons from galaxies, even with the significant drop in their mass distribution. To better understand their roles, more careful spectral modeling and better understanding of wind driving from hot stripped stars are needed \citep[such as][]{2017A&A...607L...8V, 2020MNRAS.499..873S}, as are observations that can confirm or refute theoretical predictions.

In addition, if it is difficult for massive stripped stars to form at low metallicity, this could also affect the efficiency of the conventionally assumed isolated binary evolution pathway to the formation of compact object binaries. Here, we did not explore the impact of late expansion on compact object formation in depth, since it would require a larger exploration of detailed binary evolution models. 

As the stripped star population is beginning to be revealed by observations \citep{2023Sci...382.1287D, 2023ApJ...959..125G, 2023MNRAS.525.5121V, 2023A&A...674L..12R, 2024A&A...692A..90R}, our predictions should soon become testable. 
Further, new and ongoing observations of high-$z$ galaxies and their local analogs will allow for comparison with models with various treatments of massive stars and binaries, further constraining massive binary evolution.

Our study considered only the masses and numbers of stripped stars in a constant star-forming population. Future work will expand these population models to also include predictions for the companion stars, final binaries orbits, and spectral contributions. Such populations would be particularly useful as a guide for observational campaigns to search for populations of stripped stars in upcoming wide-field photometric and spectroscopic surveys.

\begin{acknowledgements}
We thank the anonymous referee for providing a constructive report.

We thank Tomer Shenar and Selma de Mink for the interesting discussions that helped us improve the content of Sect. \ref{sec:nearby_pops}.
Thank you to Jorick Vink and Andreas Sander for helpful discussions about wind driving.

BHA thanks the Caltech Summer Undergraduate Research Fellowship (SURF) program and Peter Adams for supporting this project in memory of Alain Porter and Arthur R. Adams.
BHA thanks Gwen Rudie for organizing the Carnegie Astrophysics Summer Student Internship (CASSI) program and all the staﬀ at Carnegie Observatories who help to support this program.

BHA also thanks Laura Jaliff, Sal Wanying Fu, Ivanna Escala, Johanna Teske, Tony Piro, Brian Lorenz, and Peter Senchyna for their mentorship during this project.

Computing resources used for this work were made possible by a grant from the Ahmanson Foundation. We thank the Observatories of the Carnegie Institution for Science for support, including Chris Burns for help with computations.

This work used computing resources provided by Northwestern University and the Center for Interdisciplinary Exploration and Research in Astrophysics (CIERA). This research was supported in part through the computational resources and staff contributions provided for the Quest high performance computing facility at Northwestern University which is jointly supported by the Office of the Provost, the Office for Research, and Northwestern University Information Technology.

MRD acknowledges support from the NSERC through grant RGPIN-2019-06186, the Canada Research Chairs Program, and the Dunlap Institute at the University of Toronto.

BHA is supported by the National Science Foundation Graduate Research Fellowship under Grant No. DGE-2234667.
\end{acknowledgements}

\bibliographystyle{aa}
\bibliography{ref}

\begin{thebibliography}{189}
\expandafter\ifx\csname natexlab\endcsname\relax\def\natexlab#1{#1}\fi

\bibitem[{{Abbott} {et~al.}(2023){Abbott}, {Abbott}, {Acernese}, {Ackley}, {Adams}, {Adhikari}, {Adhikari}, {Adya}, {Affeldt}, {Agarwal}, {Agathos}, {Agatsuma}, {Aggarwal}, {Aguiar}, {Aiello}, {Ain}, {Ajith}, {Akcay}, {Akutsu}, {Albanesi}, {Allocca}, {Altin}, {Amato}, {Anand}, {Anand}, {Ananyeva}, {Anderson}, {Anderson}, {Ando}, {Andrade}, {Andres}, {Andri{\'c}}, {Angelova}, {Ansoldi}, {Antelis}, {Antier}, {Appert}, {Arai}, {Arai}, {Arai}, {Araki}, {Araya}, {Araya}, {Areeda}, {Ar{\`e}ne}, {Aritomi}, {Arnaud}, {Arogeti}, {Aronson}, {Arun}, {Asada}, {Asali}, {Ashton}, {Aso}, {Assiduo}, {Aston}, {Astone}, {Aubin}, {Austin}, {Babak}, {Badaracco}, {Bader}, {Badger}, {Bae}, {Bae}, {Baer}, {Bagnasco}, {Bai}, {Baiotti}, {Baird}, {Bajpai}, {Ball}, {Ballardin}, {Ballmer}, {Balsamo}, {Baltus}, {Banagiri}, {Bankar}, {Barayoga}, {Barbieri}, {Barish}, {Barker}, {Barneo}, {Barone}, {Barr}, {Barsotti}, {Barsuglia}, {Barta}, {Bartlett}, {Barton}, {Bartos}, {Bassiri}, {Basti}, {Bawaj}, {Bayley}, {Baylor}, {Bazzan},
  {B{\'e}csy}, {Bedakihale}, {Bejger}, {Belahcene}, {Benedetto}, {Beniwal}, {Bennett}, {Bentley}, {Benyaala}, {Bergamin}, {Berger}, {Bernuzzi}, {Berry}, {Bersanetti}, {Bertolini}, {Betzwieser}, {Beveridge}, {Bhandare}, {Bhardwaj}, {Bhattacharjee}, {Bhaumik}, {Bilenko}, {Billingsley}, {Bini}, {Birney}, {Birnholtz}, {Biscans}, {Bischi}, {Biscoveanu}, {Bisht}, {Biswas}, {Bitossi}, {Bizouard}, {Blackburn}, {Blair}, {Blair}, {Blair}, {Bobba}, {Bode}, {Boer}, {Bogaert}, {Boldrini}, {Bonavena}, {Bondu}, {Bonilla}, {Bonnand}, {Booker}, {Boom}, {Bork}, {Boschi}, {Bose}, {Bose}, {Bossilkov}, {Boudart}, {Bouffanais}, {Bozzi}, {Bradaschia}, {Brady}, {Bramley}, {Branch}, {Branchesi}, {Brandt}, {Brau}, {Breschi}, {Briant}, {Briggs}, {Brillet}, {Brinkmann}, {Brockill}, {Brooks}, {Brooks}, {Brown}, {Brunett}, {Bruno}, {Bruntz}, {Bryant}, {Bulik}, {Bulten}, {Buonanno}, {Buscicchio}, {Buskulic}, {Buy}, {Byer}, {Davies}, {Cadonati}, {Cagnoli}, {Cahillane}, {Bustillo}, {Callaghan}, {Callister}, {Calloni}, {Cameron}, {Camp},
  {Canepa}, {Canevarolo}, {Cannavacciuolo}, {Cannon}, {Cao}, {Cao}, {Capocasa}, {Capote}, {Carapella}, {Carbognani}, {Carlin}, {Carney}, {Carpinelli}, {Carrillo}, {Carullo}, {Carver}, {Diaz}, {Casentini}, {Castaldi}, {Caudill}, {Cavagli{\`a}}, {Cavalier}, {Cavalieri}, {Ceasar}, {Cella}, {Cerd{\'a}-Dur{\'a}n}, {Cesarini}, {Chaibi}, {Chakravarti}, {Subrahmanya}, {Champion}, {Chan}, {Chan}, {Chan}, {Chan}, {Chan}, {Chandra}, {Chanial}, {Chao}, {Chapman-Bird}, {Charlton}, {Chase}, {Chassande-Mottin}, {Chatterjee}, {Chatterjee}, {Chatterjee}, {Chaturvedi}, {Chaty}, {Chatziioannou}, {Chen}, {Chen}, {Chen}, {Chen}, {Chen}, {Chen}, {Chen}, {Chen}, {Cheng}, {Cheong}, {Cheung}, {Chia}, {Chiadini}, {Chiang}, {Chiarini}, {Chierici}, {Chincarini}, {Chiofalo}, {Chiummo}, {Cho}, {Cho}, {Choudhary}, {Choudhary}, {Christensen}, {Chu}, {Chu}, {Chu}, {Chua}, {Chung}, {Ciani}, {Ciecielag}, {Cie{\'s}lar}, {Cifaldi}, {Ciobanu}, {Ciolfi}, {Cipriano}, {Cirone}, {Clara}, {Clark}, {Clark}, {Clarke}, {Clearwater}, {Clesse}, {Cleva},
  {Coccia}, {Codazzo}, {Cohadon}, {Cohen}, {Cohen}, {Colleoni}, {Collette}, {Colombo}, {Colpi}, {Compton}, {Constancio}, {Conti}, {Cooper}, {Corban}, {Corbitt}, {Cordero-Carri{\'o}n}, {Corezzi}, {Corley}, {Cornish}, {Corre}, {Corsi}, {Cortese}, {Costa}, {Cotesta}, {Coughlin}, {Coulon}, {Countryman}, {Cousins}, {Couvares}, {Coward}, {Cowart}, {Coyne}, {Coyne}, {Creighton}, {Creighton}, {Criswell}, {Croquette}, {Crowder}, {Cudell}, {Cullen}, {Cumming}, {Cummings}, {Cunningham}, {Cuoco}, {Cury{\l}o}, {Dabadie}, {Canton}, {Dall'Osso}, {D{\'a}lya}, {Dana}, {Daneshgaranbajastani}, {D'Angelo}, {Danila}, {Danilishin}, {D'Antonio}, {Danzmann}, {Darsow-Fromm}, {Dasgupta}, {Datrier}, {Dattilo}, {Dave}, {Davier}, {Davis}, {Davis}, {Daw}, {de Alarc{\'o}n}, {Dean}, {Debra}, {Deenadayalan}, {Degallaix}, {de Laurentis}, {Del{\'e}glise}, {Del Favero}, {de Lillo}, {de Lillo}, {Del Pozzo}, {Demarchi}, {de Matteis}, {D'Emilio}, {Demos}, {Dent}, {Depasse}, {de Pietri}, {De Rosa}, {de Rossi}, {Desalvo}, {de Simone}, {Dhurandhar},
  {D{\'\i}az}, {Diaz-Ortiz}, {Didio}, {Dietrich}, {di Fiore}, {di Fronzo}, {di Giorgio}, {di Giovanni}, {di Giovanni}, {di Girolamo}, {di Lieto}, {Ding}, {di Pace}, {di Palma}, {di Renzo}, {Divakarla}, {Dmitriev}, {Doctor}, {D'Onofrio}, {Donovan}, {Dooley}, {Doravari}, {Dorrington}, {Drago}, {Driggers}, {Drori}, {Ducoin}, {Dupej}, {Durante}, {D'Urso}, {Duverne}, {Dwyer}, {Eassa}, {Easter}, {Ebersold}, {Eckhardt}, {Eddolls}, {Edelman}, {Edo}, {Edy}, {Effler}, {Eguchi}, {Eichholz}, {Eikenberry}, {Eisenmann}, {Eisenstein}, {Ejlli}, {Engelby}, {Enomoto}, {Errico}, {Essick}, {Estell{\'e}s}, {Estevez}, {Etienne}, {Etzel}, {Evans}, {Evans}, {Ewing}, {Fafone}, {Fair}, {Fairhurst}, {Farah}, {Farinon}, {Farr}, {Farr}, {Farrow}, {Fauchon-Jones}, {Favaro}, {Favata}, {Fays}, {Fazio}, {Feicht}, {Fejer}, {Fenyvesi}, {Ferguson}, {Fernandez-Galiana}, {Ferrante}, {Ferreira}, {Fidecaro}, {Figura}, {Fiori}, {Fishbach}, {Fisher}, {Fittipaldi}, {Fiumara}, {Flaminio}, {Floden}, {Fong}, {Font}, {Fornal}, {Forsyth}, {Franke},
  {Frasca}, {Frasconi}, {Frederick}, {Freed}, {Frei}, {Freise}, {Frey}, {Fritschel}, {Frolov}, {Fronz{\'e}}, {Fujii}, {Fujikawa}, {Fukunaga}, {Fukushima}, {Fulda}, {Fyffe}, {Gabbard}, {Gabella}, {Gadre}, {Gair}, {Gais}, {Galaudage}, {Gamba}, {Ganapathy}, {Ganguly}, {Gao}, {Gaonkar}, {Garaventa}, {Garc{\'\i}a}, {Garc{\'\i}a-N{\'u}{\~n}ez}, {Garc{\'\i}a-Quir{\'o}s}, {Garufi}, {Gateley}, {Gaudio}, {Gayathri}, {Ge}, {Gemme}, {Gennai}, {George}, {George}, {Gerberding}, {Gergely}, {Gewecke}, {Ghonge}, {Ghosh}, {Ghosh}, {Ghosh}, {Ghosh}, {Giacomazzo}, {Giacoppo}, {Giaime}, {Giardina}, {Gibson}, {Gier}, {Giesler}, {Giri}, {Gissi}, {Glanzer}, {Gleckl}, {Godwin}, {Goetz}, {Goetz}, {Gohlke}, {Golomb}, {Goncharov}, {Gonz{\'a}lez}, {Gopakumar}, {Gosselin}, {Gouaty}, {Gould}, {Grace}, {Grado}, {Granata}, {Granata}, {Grant}, {Gras}, {Grassia}, {Gray}, {Gray}, {Greco}, {Green}, {Green}, {Gretarsson}, {Gretarsson}, {Griffith}, {Griffiths}, {Griggs}, {Grignani}, {Grimaldi}, {Grimm}, {Grote}, {Grunewald}, {Gruning}, {Guerra},
  {Guidi}, {Guimaraes}, {Guix{\'e}}, {Gulati}, {Guo}, {Guo}, {Gupta}, {Gupta}, {Gupta}, {Gustafson}, {Gustafson}, {Guzman}, {Ha}, {Haegel}, {Hagiwara}, {Haino}, {Halim}, {Hall}, {Hamilton}, {Hammond}, {Han}, {Haney}, {Hanks}, {Hanna}, {Hannam}, {Hannuksela}, {Hansen}, {Hansen}, {Hanson}, {Harder}, {Hardwick}, {Haris}, {Harms}, {Harry}, {Harry}, {Hartwig}, {Hasegawa}, {Haskell}, {Hasskew}, {Haster}, {Hattori}, {Haughian}, {Hayakawa}, {Hayama}, {Hayes}, {Healy}, {Heidmann}, {Heidt}, {Heintze}, {Heinze}, {Heinzel}, {Heitmann}, {Hellman}, {Hello}, {Helmling-Cornell}, {Hemming}, {Hendry}, {Heng}, {Hennes}, {Hennig}, {Hennig}, {Hernandez}, {Hernandez Vivanco}, {Heurs}, {Hild}, {Hill}, {Himemoto}, {Hines}, {Hiranuma}, {Hirata}, {Hirose}, {Hochheim}, {Hofman}, {Hohmann}, {Holcomb}, {Holland}, {Holley-Bockelmann}, {Hollows}, {Holmes}, {Holt}, {Holz}, {Hong}, {Hopkins}, {Hough}, {Hourihane}, {Howell}, {Hoy}, {Hoyland}, {Hreibi}, {Hsieh}, {Hsu}, {Huang}, {Huang}, {Huang}, {Huang}, {Huang}, {Huang}, {H{\"u}bner},
  {Huddart}, {Hughey}, {Hui}, {Hui}, {Husa}, {Huttner}, {Huxford}, {Huynh-Dinh}, {Ide}, {Idzkowski}, {Iess}, {Ikenoue}, {Imam}, {Inayoshi}, {Ingram}, {Inoue}, {Ioka}, {Isi}, {Isleif}, {Ito}, {Itoh}, {Iyer}, {Izumi}, {Jaberianhamedan}, {Jacqmin}, {Jadhav}, {Jadhav}, {James}, {Jan}, {Jani}, {Janquart}, {Janssens}, {Janthalur}, {Jaranowski}, {Jariwala}, {Jaume}, {Jenkins}, {Jenner}, {Jeon}, {Jeunon}, {Jia}, {Jin}, {Johns}, {Johnson-McDaniel}, {Jones}, {Jones}, {Jones}, {Jones}, {Jones}, {Jonker}, {Ju}, {Jung}, {Jung}, {Junker}, {Juste}, {Kaihotsu}, {Kajita}, {Kakizaki}, {Kalaghatgi}, {Kalogera}, {Kamai}, {Kamiizumi}, {Kanda}, {Kandhasamy}, {Kang}, {Kanner}, {Kao}, {Kapadia}, {Kapasi}, {Karat}, {Karathanasis}, {Karki}, {Kashyap}, {Kasprzack}, {Kastaun}, {Katsanevas}, {Katsavounidis}, {Katzman}, {Kaur}, {Kawabe}, {Kawaguchi}, {Kawai}, {Kawasaki}, {K{\'e}f{\'e}lian}, {Keitel}, {Key}, {Khadka}, {Khalili}, {Khan}, {Khazanov}, {Khetan}, {Khursheed}, {Kijbunchoo}, {Kim}, {Kim}, {Kim}, {Kim}, {Kim}, {Kim}, {Kimball},
  {Kimura}, {Kinley-Hanlon}, {Kirchhoff}, {Kissel}, {Kita}, {Kitazawa}, {Kleybolte}, {Klimenko}, {Knee}, {Knowles}, {Knyazev}, {Koch}, {Koekoek}, {Kojima}, {Kokeyama}, {Koley}, {Kolitsidou}, {Kolstein}, {Komori}, {Kondrashov}, {Kong}, {Kontos}, {Koper}, {Korobko}, {Kotake}, {Kovalam}, {Kozak}, {Kozakai}, {Kozu}, {Kringel}, {Krishnendu}, {Kr{\'o}lak}, {Kuehn}, {Kuei}, {Kuijer}, {Kulkarni}, {Kumar}, {Kumar}, {Kumar}, {Kumar}, {Kume}, {Kuns}, {Kuo}, {Kuo}, {Kuromiya}, {Kuroyanagi}, {Kusayanagi}, {Kuwahara}, {Kwak}, {Lagabbe}, {Laghi}, {Lalande}, {Lam}, {Lamberts}, {Landry}, {Lane}, {Lang}, {Lange}, {Lantz}, {La Rosa}, {Lartaux-Vollard}, {Lasky}, {Laxen}, {Lazzarini}, {Lazzaro}, {Leaci}, {Leavey}, {Lecoeuche}, {Lee}, {Lee}, {Lee}, {Lee}, {Lee}, {Lee}, {Lehmann}, {Lema{\^\i}tre}, {Leonardi}, {Leroy}, {Letendre}, {Levesque}, {Levin}, {Leviton}, {Leyde}, {Li}, {Li}, {Li}, {Li}, {Li}, {Li}, {Lin}, {Lin}, {Lin}, {Lin}, {Lin}, {Linde}, {Linker}, {Linley}, {Littenberg}, {Liu}, {Liu}, {Liu}, {Liu}, {Llamas},
  {Llorens-Monteagudo}, {Lo}, {Lockwood}, {Loh}, {London}, {Longo}, {Lopez}, {Portilla}, {Lorenzini}, {Loriette}, {Lormand}, {Losurdo}, {Lott}, {Lough}, {Lousto}, {Lovelace}, {Lucaccioni}, {L{\"u}ck}, {Lumaca}, {Lundgren}, {Luo}, {Lynam}, {Macas}, {Macinnis}, {MacLeod}, {MacMillan}, {Macquet}, {Hernandez}, {Magazz{\`u}}, {Magee}, {Maggiore}, {Magnozzi}, {Mahesh}, {Majorana}, {Makarem}, {Maksimovic}, {Maliakal}, {Malik}, {Man}, {Mandic}, {Mangano}, {Mango}, {Mansell}, {Manske}, {Mantovani}, {Mapelli}, {Marchesoni}, {Marchio}, {Marion}, {Mark}, {M{\'a}rka}, {M{\'a}rka}, {Markakis}, {Markosyan}, {Markowitz}, {Maros}, {Marquina}, {Marsat}, {Martelli}, {Martin}, {Martin}, {Martinez}, {Martinez}, {Martinez}, {Martinovic}, {Martynov}, {Marx}, {Masalehdan}, {Mason}, {Massera}, {Masserot}, {Massinger}, {Masso-Reid}, {Mastrogiovanni}, {Matas}, {Mateu-Lucena}, {Matichard}, {Matiushechkina}, {Mavalvala}, {McCann}, {McCarthy}, {McClelland}, {McClincy}, {McCormick}, {McCuller}, {McGhee}, {McGuire}, {McIsaac}, {McIver},
  {McRae}, {McWilliams}, {Meacher}, {Mehmet}, {Mehta}, {Meijer}, {Melatos}, {Melchor}, {Mendell}, {Menendez-Vazquez}, {Menoni}, {Mercer}, {Mereni}, {Merfeld}, {Merilh}, {Merritt}, {Merzougui}, {Meshkov}, {Messenger}, {Messick}, {Meyers}, {Meylahn}, {Mhaske}, {Miani}, {Miao}, {Michaloliakos}, {Michel}, {Michimura}, {Middleton}, {Milano}, {Miller}, {Miller}, {Miller}, {Millhouse}, {Mills}, {Milotti}, {Minazzoli}, {Minenkov}, {Mio}, {Mir}, {Miravet-Ten{\'e}s}, {Mishra}, {Mishra}, {Mistry}, {Mitra}, {Mitrofanov}, {Mitselmakher}, {Mittleman}, {Miyakawa}, {Miyamoto}, {Miyazaki}, {Miyo}, {Miyoki}, {Mo}, {Modafferi}, {Moguel}, {Mogushi}, {Mohapatra}, {Mohite}, {Molina}, {Molina-Ruiz}, {Mondin}, {Montani}, {Moore}, {Moraru}, {Morawski}, {More}, {Moreno}, {Moreno}, {Mori}, {Morisaki}, {Moriwaki}, {Morr{\'a}s}, {Mours}, {Mow-Lowry}, {Mozzon}, {Muciaccia}, {Mukherjee}, {Mukherjee}, {Mukherjee}, {Mukherjee}, {Mukherjee}, {Mukund}, {Mullavey}, {Munch}, {Mu{\~n}iz}, {Murray}, {Musenich}, {Muusse}, {Nadji}, {Nagano},
  {Nagano}, {Nagar}, {Nakamura}, {Nakano}, {Nakano}, {Nakashima}, {Nakayama}, {Napolano}, {Nardecchia}, {Narikawa}, {Naticchioni}, {Nayak}, {Nayak}, {Negishi}, {Neil}, {Neilson}, {Nelemans}, {Nelson}, {Nery}, {Neubauer}, {Neunzert}, {Ng}, {Ng}, {Nguyen}, {Nguyen}, {Nguyen}, {Quynh}, {Ni}, {Nichols}, {Nishizawa}, {Nissanke}, {Nitoglia}, {Nocera}, {Norman}, {North}, {Nozaki}, {Siles}, {Nuttall}, {Oberling}, {O'Brien}, {Obuchi}, {O'Dell}, {Oelker}, {Ogaki}, {Oganesyan}, {Oh}, {Oh}, {Oh}, {Ohashi}, {Ohishi}, {Ohkawa}, {Ohme}, {Ohta}, {Okada}, {Okutani}, {Okutomi}, {Olivetto}, {Oohara}, {Ooi}, {Oram}, {O'Reilly}, {Ormiston}, {Ormsby}, {Ortega}, {O'Shaughnessy}, {O'Shea}, {Oshino}, {Ossokine}, {Osthelder}, {Otabe}, {Ottaway}, {Overmier}, {Pace}, {Pagano}, {Page}, {Pagliaroli}, {Pai}, {Pai}, {Palamos}, {Palashov}, {Palomba}, {Pan}, {Pan}, {Panda}, {Pang}, {Pang}, {Pankow}, {Pannarale}, {Pant}, {Panther}, {Paoletti}, {Paoli}, {Paolone}, {Parisi}, {Park}, {Park}, {Parker}, {Pascucci}, {Pasqualetti}, {Passaquieti},
  {Passuello}, {Patel}, {Pathak}, {Patricelli}, {Patron}, {Paul}, {Payne}, {Pedraza}, {Pegoraro}, {Pele}, {Arellano}, {Penn}, {Perego}, {Pereira}, {Pereira}, {Perez}, {P{\'e}rigois}, {Perkins}, {Perreca}, {Perri{\`e}s}, {Petermann}, {Petterson}, {Pfeiffer}, {Pham}, {Phukon}, {Piccinni}, {Pichot}, {Piendibene}, {Piergiovanni}, {Pierini}, {Pierro}, {Pillant}, {Pillas}, {Pilo}, {Pinard}, {Pinto}, {Pinto}, {Piotrzkowski}, {Piotrzkowski}, {Pirello}, {Pitkin}, {Placidi}, {Planas}, {Plastino}, {Pluchar}, {Poggiani}, {Polini}, {Pong}, {Ponrathnam}, {Popolizio}, {Porter}, {Poulton}, {Powell}, {Pracchia}, {Pradier}, {Prajapati}, {Prasai}, {Prasanna}, {Pratten}, {Principe}, {Prodi}, {Prokhorov}, {Prosposito}, {Prudenzi}, {Puecher}, {Punturo}, {Puosi}, {Puppo}, {P{\"u}rrer}, {Qi}, {Quetschke}, {Quitzow-James}, {Qutob}, {Raab}, {Raaijmakers}, {Radkins}, {Radulesco}, {Raffai}, {Rail}, {Raja}, {Rajan}, {Ramirez}, {Ramirez}, {Ramos-Buades}, {Rana}, {Rapagnani}, {Rapol}, {Ray}, {Raymond}, {Raza}, {Razzano}, {Read}, {Rees},
  {Regimbau}, {Rei}, {Reid}, {Reid}, {Reitze}, {Relton}, {Renzini}, {Rettegno}, {Reza}, {Rezac}, {Ricci}, {Richards}, {Richardson}, {Richardson}, {Riemenschneider}, {Riles}, {Rinaldi}, {Rink}, {Rizzo}, {Robertson}, {Robie}, {Robinet}, {Rocchi}, {Rodriguez}, {Rolland}, {Rollins}, {Romanelli}, {Romano}, {Romel}, {Romero-Rodr{\'\i}guez}, {Romero-Shaw}, {Romie}, {Ronchini}, {Rosa}, {Rose}, {Rosi{\'n}ska}, {Ross}, {Rowan}, {Rowlinson}, {Roy}, {Roy}, {Roy}, {Rozza}, {Ruggi}, {Ruiz-Rocha}, {Ryan}, {Sachdev}, {Sadecki}, {Sadiq}, {Sago}, {Saito}, {Saito}, {Sakai}, {Sakai}, {Sakellariadou}, {Sakuno}, {Salafia}, {Salconi}, {Saleem}, {Salemi}, {Samajdar}, {Sanchez}, {Sanchez}, {Sanchez}, {Sanchis-Gual}, {Sanders}, {Sanuy}, {Saravanan}, {Sarin}, {Sassolas}, {Satari}, {Sathyaprakash}, {Sato}, {Sato}, {Sauter}, {Savage}, {Sawada}, {Sawant}, {Sawant}, {Sayah}, {Schaetzl}, {Scheel}, {Scheuer}, {Schiworski}, {Schmidt}, {Schmidt}, {Schnabel}, {Schneewind}, {Schofield}, {Sch{\"o}nbeck}, {Schulte}, {Schutz}, {Schwartz}, {Scott},
  {Scott}, {Seglar-Arroyo}, {Sekiguchi}, {Sekiguchi}, {Sellers}, {Sengupta}, {Sentenac}, {Seo}, {Sequino}, {Sergeev}, {Setyawati}, {Shaffer}, {Shahriar}, {Shams}, {Shao}, {Sharma}, {Sharma}, {Shawhan}, {Shcheblanov}, {Shibagaki}, {Shikauchi}, {Shimizu}, {Shimoda}, {Shimode}, {Shinkai}, {Shishido}, {Shoda}, {Shoemaker}, {Shoemaker}, {Shyamsundar}, {Sieniawska}, {Sigg}, {Singer}, {Singh}, {Singh}, {Singha}, {Sintes}, {Sipala}, {Skliris}, {Slagmolen}, {Slaven-Blair}, {Smetana}, {Smith}, {Smith}, {Soldateschi}, {Somala}, {Somiya}, {Son}, {Soni}, {Soni}, {Sordini}, {Sorrentino}, {Sorrentino}, {Sotani}, {Soulard}, {Souradeep}, {Sowell}, {Spagnuolo}, {Spencer}, {Spera}, {Srinivasan}, {Srivastava}, {Srivastava}, {Staats}, {Stachie}, {Steer}, {Steinhoff}, {Steinlechner}, {Steinlechner}, {Stevenson}, {Stops}, {Stover}, {Strain}, {Strang}, {Stratta}, {Strunk}, {Sturani}, {Stuver}, {Sudhagar}, {Sudhir}, {Sugimoto}, {Suh}, {Sullivan}, {Sullivan}, {Summerscales}, {Sun}, {Sun}, {Sunil}, {Sur}, {Suresh}, {Sutton}, {Suzuki},
  {Suzuki}, {Swinkels}, {Szczepa{\'n}czyk}, {Szewczyk}, {Tacca}, {Tagoshi}, {Tait}, {Takahashi}, {Takahashi}, {Takamori}, {Takano}, {Takeda}, {Takeda}, {Talbot}, {Talbot}, {Tanaka}, {Tanaka}, {Tanaka}, {Tanaka}, {Tanaka}, {Tanasijczuk}, {Tanioka}, {Tanner}, {Tao}, {Tao}, {Mart{\'\i}n}, {Taranto}, {Tasson}, {Telada}, {Tenorio}, {Terhune}, {Terkowski}, {Thirugnanasambandam}, {Thomas}, {Thomas}, {Thomas}, {Thompson}, {Thondapu}, {Thorne}, {Thrane}, {Tiwari}, {Tiwari}, {Tiwari}, {Toivonen}, {Toland}, {Tolley}, {Tomaru}, {Tomigami}, {Tomura}, {Tonelli}, {Torres-Forn{\'e}}, {Torrie}, {E Melo}, {T{\"o}yr{\"a}}, {Trapananti}, {Travasso}, {Traylor}, {Trevor}, {Tringali}, {Tripathee}, {Troiano}, {Trovato}, {Trozzo}, {Trudeau}, {Tsai}, {Tsai}, {Tsang}, {Tsang}, {Tsao}, {Tse}, {Tso}, {Tsubono}, {Tsuchida}, {Tsukada}, {Tsuna}, {Tsutsui}, {Tsuzuki}, {Turbang}, {Turconi}, {Tuyenbayev}, {Ubhi}, {Uchikata}, {Uchiyama}, {Udall}, {Ueda}, {Uehara}, {Ueno}, {Ueshima}, {Unnikrishnan}, {Uraguchi}, {Urban}, {Ushiba}, {Utina},
  {Vahlbruch}, {Vajente}, {Vajpeyi}, {Valdes}, {Valentini}, {Valsan}, {van Bakel}, {van Beuzekom}, {van den Brand}, {van den Broeck}, {Vander-Hyde}, {van der Schaaf}, {van Heijningen}, {Vanosky}, {van Putten}, {van Remortel}, {Vardaro}, {Vargas}, {Varma}, {Vas{\'u}th}, {Vecchio}, {Vedovato}, {Veitch}, {Veitch}, {Venneberg}, {Venugopalan}, {Verkindt}, {Verma}, {Verma}, {Veske}, {Vetrano}, {Vicer{\'e}}, {Vidyant}, {Viets}, {Vijaykumar}, {Villa-Ortega}, {Vinet}, {Virtuoso}, {Vitale}, {Vo}, {Vocca}, {von Reis}, {von Wrangel}, {Vorvick}, {Vyatchanin}, {Wade}, {Wade}, {Wagner}, {Walet}, {Walker}, {Wallace}, {Wallace}, {Walsh}, {Wang}, {Wang}, {Wang}, {Ward}, {Warner}, {Was}, {Washimi}, {Washington}, {Watchi}, {Weaver}, {Webster}, {Weinert}, {Weinstein}, {Weiss}, {Weller}, {Weller}, {Wellmann}, {Wen}, {We{\ss}els}, {Wette}, {Whelan}, {White}, {Whiting}, {Whittle}, {Wilken}, {Williams}, {Williams}, {Williams}, {Williamson}, {Willis}, {Willke}, {Wilson}, {Winkler}, {Wipf}, {Wlodarczyk}, {Woan}, {Woehler}, {Wofford},
  {Wong}, {Wu}, {Wu}, {Wu}, {Wu}, {Wysocki}, {Xiao}, {Xu}, {Yamada}, {Yamamoto}, {Yamamoto}, {Yamamoto}, {Yamamoto}, {Yamashita}, {Yamazaki}, {Yang}, {Yang}, {Yang}, {Yang}, {Yang}, {Yap}, {Yeeles}, {Yelikar}, {Ying}, {Yokogawa}, {Yokoyama}, {Yokozawa}, {Yoo}, {Yoshioka}, {Yu}, {Yu}, {Yuzurihara}, {Zadro{\.z}ny}, {Zanolin}, {Zeidler}, {Zelenova}, {Zendri}, {Zevin}, {Zhan}, {Zhang}, {Zhang}, {Zhang}, {Zhang}, {Zhang}, {Zhao}, {Zhao}, {Zhao}, {Zhao}, {Zheng}, {Zhou}, {Zhou}, {Zhu}, {Zhu}, {Zimmerman}, {Zlochower}, {Zucker}, {Zweizig}, {Ligo Scientific Collaboration}, {VIRGO Collaboration}, \& {Kagra Collaboration}}]{2023PhRvX..13d1039A}
{Abbott}, R., {Abbott}, T.~D., {Acernese}, F., {et~al.} 2023, Physical Review X, 13, 041039

\bibitem[{{Abdul-Masih} {et~al.}(2021){Abdul-Masih}, {Sana}, {Hawcroft}, {Almeida}, {Brands}, {de Mink}, {Justham}, {Langer}, {Mahy}, {Marchant}, {Menon}, {Puls}, \& {Sundqvist}}]{2021A&A...651A..96A}
{Abdul-Masih}, M., {Sana}, H., {Hawcroft}, C., {et~al.} 2021, \aap, 651, A96

\bibitem[{{Abdul-Masih} {et~al.}(2019){Abdul-Masih}, {Sana}, {Sundqvist}, {Mahy}, {Menon}, {Almeida}, {De Koter}, {de Mink}, {Justham}, {Langer}, {Puls}, {Shenar}, \& {Tramper}}]{2019ApJ...880..115A}
{Abdul-Masih}, M., {Sana}, H., {Sundqvist}, J., {et~al.} 2019, \apj, 880, 115

\bibitem[{{Agrawal} {et~al.}(2023){Agrawal}, {Hurley}, {Stevenson}, {Rodriguez}, {Sz{\'e}csi}, \& {Kemp}}]{2023MNRAS.525..933A}
{Agrawal}, P., {Hurley}, J., {Stevenson}, S., {et~al.} 2023, \mnras, 525, 933

\bibitem[{{Agrawal} {et~al.}(2020){Agrawal}, {Hurley}, {Stevenson}, {Sz{\'e}csi}, \& {Flynn}}]{2020MNRAS.497.4549A}
{Agrawal}, P., {Hurley}, J., {Stevenson}, S., {Sz{\'e}csi}, D., \& {Flynn}, C. 2020, \mnras, 497, 4549

\bibitem[{{Bavera} {et~al.}(2021){Bavera}, {Fragos}, {Zevin}, {Berry}, {Marchant}, {Andrews}, {Coughlin}, {Dotter}, {Kovlakas}, {Misra}, {Serra-Perez}, {Qin}, {Rocha}, {Rom{\'a}n-Garza}, {Tran}, \& {Zapartas}}]{2021A&A...647A.153B}
{Bavera}, S.~S., {Fragos}, T., {Zevin}, M., {et~al.} 2021, \aap, 647, A153

\bibitem[{{Belczynski} {et~al.}(2010){Belczynski}, {Bulik}, {Fryer}, {Ruiter}, {Valsecchi}, {Vink}, \& {Hurley}}]{2010ApJ...714.1217B}
{Belczynski}, K., {Bulik}, T., {Fryer}, C.~L., {et~al.} 2010, \apj, 714, 1217

\bibitem[{{Belczynski} {et~al.}(2016){Belczynski}, {Holz}, {Bulik}, \& {O'Shaughnessy}}]{2016Natur.534..512B}
{Belczynski}, K., {Holz}, D.~E., {Bulik}, T., \& {O'Shaughnessy}, R. 2016, \nat, 534, 512

\bibitem[{{Berg} {et~al.}(2019){Berg}, {Chisholm}, {Erb}, {Pogge}, {Henry}, \& {Olivier}}]{2019ApJ...878L...3B}
{Berg}, D.~A., {Chisholm}, J., {Erb}, D.~K., {et~al.} 2019, \apjl, 878, L3

\bibitem[{{Berg} {et~al.}(2018){Berg}, {Erb}, {Auger}, {Pettini}, \& {Brammer}}]{2018ApJ...859..164B}
{Berg}, D.~A., {Erb}, D.~K., {Auger}, M.~W., {Pettini}, M., \& {Brammer}, G.~B. 2018, \apj, 859, 164

\bibitem[{{Berg} {et~al.}(2022){Berg}, {James}, {King}, {McDonald}, {Chen}, {Chisholm}, {Heckman}, {Martin}, {Stark}, {Aloisi}, {Amor{\'\i}n}, {Arellano-C{\'o}rdova}, {Bayliss}, {Bordoloi}, {Brinchmann}, {Charlot}, {Chevallard}, {Clark}, {Erb}, {Feltre}, {Gronke}, {Hayes}, {Henry}, {Hernandez}, {Jaskot}, {Jones}, {Kewley}, {Kumari}, {Leitherer}, {Llerena}, {Maseda}, {Mingozzi}, {Nanayakkara}, {Ouchi}, {Plat}, {Pogge}, {Ravindranath}, {Rigby}, {Sanders}, {Scarlata}, {Senchyna}, {Skillman}, {Steidel}, {Strom}, {Sugahara}, {Wilkins}, {Wofford}, {Xu}, \& {Classy Team}}]{2022ApJS..261...31B}
{Berg}, D.~A., {James}, B.~L., {King}, T., {et~al.} 2022, \apjs, 261, 31

\bibitem[{{Bilinski} {et~al.}(2024){Bilinski}, {Smith}, {Williams}, {Smith}, {Leonard}, {Hoffman}, {Andrews}, \& {Milne}}]{2024MNRAS.529.1104B}
{Bilinski}, C., {Smith}, N., {Williams}, G.~G., {et~al.} 2024, \mnras, 529, 1104

\bibitem[{{Bodensteiner} {et~al.}(2020){Bodensteiner}, {Shenar}, {Mahy}, {Fabry}, {Marchant}, {Abdul-Masih}, {Banyard}, {Bowman}, {Dsilva}, {Frost}, {Hawcroft}, {Reggiani}, \& {Sana}}]{2020A&A...641A..43B}
{Bodensteiner}, J., {Shenar}, T., {Mahy}, L., {et~al.} 2020, \aap, 641, A43

\bibitem[{{Broekgaarden} {et~al.}(2021){Broekgaarden}, {Berger}, {Neijssel}, {Vigna-G{\'o}mez}, {Chattopadhyay}, {Stevenson}, {Chruslinska}, {Justham}, {de Mink}, \& {Mandel}}]{2021MNRAS.508.5028B}
{Broekgaarden}, F.~S., {Berger}, E., {Neijssel}, C.~J., {et~al.} 2021, \mnras, 508, 5028

\bibitem[{{Broekgaarden} {et~al.}(2022){Broekgaarden}, {Berger}, {Stevenson}, {Justham}, {Mandel}, {Chru{\'s}li{\'n}ska}, {van Son}, {Wagg}, {Vigna-G{\'o}mez}, {de Mink}, {Chattopadhyay}, \& {Neijssel}}]{2022MNRAS.516.5737B}
{Broekgaarden}, F.~S., {Berger}, E., {Stevenson}, S., {et~al.} 2022, \mnras, 516, 5737

\bibitem[{{Brott} {et~al.}(2011){Brott}, {de Mink}, {Cantiello}, {Langer}, {de Koter}, {Evans}, {Hunter}, {Trundle}, \& {Vink}}]{2011A&A...530A.115B}
{Brott}, I., {de Mink}, S.~E., {Cantiello}, M., {et~al.} 2011, \aap, 530, A115

\bibitem[{{Bunker} {et~al.}(2023){Bunker}, {Saxena}, {Cameron}, {Willott}, {Curtis-Lake}, {Jakobsen}, {Carniani}, {Smit}, {Maiolino}, {Witstok}, {Curti}, {D'Eugenio}, {Jones}, {Ferruit}, {Arribas}, {Charlot}, {Chevallard}, {Giardino}, {de Graaff}, {Looser}, {L{\"u}tzgendorf}, {Maseda}, {Rawle}, {Rix}, {Del Pino}, {Alberts}, {Egami}, {Eisenstein}, {Endsley}, {Hainline}, {Hausen}, {Johnson}, {Rieke}, {Rieke}, {Robertson}, {Shivaei}, {Stark}, {Sun}, {Tacchella}, {Tang}, {Williams}, {Willmer}, {Baker}, {Baum}, {Bhatawdekar}, {Bowler}, {Boyett}, {Chen}, {Circosta}, {Helton}, {Ji}, {Kumari}, {Lyu}, {Nelson}, {Parlanti}, {Perna}, {Sandles}, {Scholtz}, {Suess}, {Topping}, {{\"U}bler}, {Wallace}, \& {Whitler}}]{2023A&A...677A..88B}
{Bunker}, A.~J., {Saxena}, A., {Cameron}, A.~J., {et~al.} 2023, \aap, 677, A88

\bibitem[{{Cassata} {et~al.}(2013){Cassata}, {Le F{\`e}vre}, {Charlot}, {Contini}, {Cucciati}, {Garilli}, {Zamorani}, {Adami}, {Bardelli}, {Le Brun}, {Lemaux}, {Maccagni}, {Pollo}, {Pozzetti}, {Tresse}, {Vergani}, {Zanichelli}, \& {Zucca}}]{2013A&A...556A..68C}
{Cassata}, P., {Le F{\`e}vre}, O., {Charlot}, S., {et~al.} 2013, \aap, 556, A68

\bibitem[{{Cheng} {et~al.}(2024){Cheng}, {Goldberg}, {Cantiello}, {Bauer}, {Renzo}, \& {Conroy}}]{2024ApJ...974..270C}
{Cheng}, S.~J., {Goldberg}, J.~A., {Cantiello}, M., {et~al.} 2024, \apj, 974, 270

\bibitem[{{Chruslinska} {et~al.}(2019){Chruslinska}, {Nelemans}, \& {Belczynski}}]{2019MNRAS.482.5012C}
{Chruslinska}, M., {Nelemans}, G., \& {Belczynski}, K. 2019, \mnras, 482, 5012

\bibitem[{{Crowther}(2007)}]{2007ARA&A..45..177C}
{Crowther}, P.~A. 2007, \araa, 45, 177

\bibitem[{{Crowther} {et~al.}(2016){Crowther}, {Caballero-Nieves}, {Bostroem}, {Ma{\'\i}z Apell{\'a}niz}, {Schneider}, {Walborn}, {Angus}, {Brott}, {Bonanos}, {de Koter}, {de Mink}, {Evans}, {Gr{\"a}fener}, {Herrero}, {Howarth}, {Langer}, {Lennon}, {Puls}, {Sana}, \& {Vink}}]{2016MNRAS.458..624C}
{Crowther}, P.~A., {Caballero-Nieves}, S.~M., {Bostroem}, K.~A., {et~al.} 2016, \mnras, 458, 624

\bibitem[{{Cullen} {et~al.}(2019){Cullen}, {McLure}, {Dunlop}, {Khochfar}, {Dav{\'e}}, {Amor{\'\i}n}, {Bolzonella}, {Carnall}, {Castellano}, {Cimatti}, {Cirasuolo}, {Cresci}, {Fynbo}, {Fontanot}, {Gargiulo}, {Garilli}, {Guaita}, {Hathi}, {Hibon}, {Mannucci}, {Marchi}, {McLeod}, {Pentericci}, {Pozzetti}, {Shapley}, {Talia}, \& {Zamorani}}]{2019MNRAS.487.2038C}
{Cullen}, F., {McLure}, R.~J., {Dunlop}, J.~S., {et~al.} 2019, \mnras, 487, 2038

\bibitem[{{de Mink} {et~al.}(2009){de Mink}, {Cantiello}, {Langer}, {Pols}, {Brott}, \& {Yoon}}]{2009A&A...497..243D}
{de Mink}, S.~E., {Cantiello}, M., {Langer}, N., {et~al.} 2009, \aap, 497, 243

\bibitem[{{de Mink} {et~al.}(2007){de Mink}, {Pols}, \& {Hilditch}}]{2007A&A...467.1181D}
{de Mink}, S.~E., {Pols}, O.~R., \& {Hilditch}, R.~W. 2007, \aap, 467, 1181

\bibitem[{{Deshmukh} {et~al.}(2024){Deshmukh}, {Sana}, {M{\'e}rand}, {Bordier}, {Langer}, {Bodensteiner}, {Dsilva}, {Frost}, {Gosset}, {Le Bouquin}, {Lefever}, {Mahy}, {Patrick}, {Reggiani}, {Sander}, {Shenar}, {Tramper}, {Villase{\~n}or}, \& {Waisberg}}]{2024A&A...692A.109D}
{Deshmukh}, K., {Sana}, H., {M{\'e}rand}, A., {et~al.} 2024, \aap, 692, A109

\bibitem[{{Dewi} \& {Tauris}(2000)}]{2000A&A...360.1043D}
{Dewi}, J.~D.~M. \& {Tauris}, T.~M. 2000, \aap, 360, 1043

\bibitem[{{Diehl} {et~al.}(2006){Diehl}, {Halloin}, {Kretschmer}, {Lichti}, {Sch{\"o}nfelder}, {Strong}, {von Kienlin}, {Wang}, {Jean}, {Kn{\"o}dlseder}, {Roques}, {Weidenspointner}, {Schanne}, {Hartmann}, {Winkler}, \& {Wunderer}}]{2006Natur.439...45D}
{Diehl}, R., {Halloin}, H., {Kretschmer}, K., {et~al.} 2006, \nat, 439, 45

\bibitem[{{Dominik} {et~al.}(2012){Dominik}, {Belczynski}, {Fryer}, {Holz}, {Berti}, {Bulik}, {Mandel}, \& {O'Shaughnessy}}]{2012ApJ...759...52D}
{Dominik}, M., {Belczynski}, K., {Fryer}, C., {et~al.} 2012, \apj, 759, 52

\bibitem[{{Drout} {et~al.}(2023){Drout}, {G{\"o}tberg}, {Ludwig}, {Groh}, {de Mink}, {O'Grady}, \& {Smith}}]{2023Sci...382.1287D}
{Drout}, M.~R., {G{\"o}tberg}, Y., {Ludwig}, B.~A., {et~al.} 2023, Science, 382, 1287

\bibitem[{{Drout} {et~al.}(2011){Drout}, {Soderberg}, {Gal-Yam}, {Cenko}, {Fox}, {Leonard}, {Sand}, {Moon}, {Arcavi}, \& {Green}}]{2011ApJ...741...97D}
{Drout}, M.~R., {Soderberg}, A.~M., {Gal-Yam}, A., {et~al.} 2011, \apj, 741, 97

\bibitem[{{Dsilva} {et~al.}(2020){Dsilva}, {Shenar}, {Sana}, \& {Marchant}}]{2020A&A...641A..26D}
{Dsilva}, K., {Shenar}, T., {Sana}, H., \& {Marchant}, P. 2020, \aap, 641, A26

\bibitem[{{Dsilva} {et~al.}(2022){Dsilva}, {Shenar}, {Sana}, \& {Marchant}}]{2022A&A...664A..93D}
{Dsilva}, K., {Shenar}, T., {Sana}, H., \& {Marchant}, P. 2022, \aap, 664, A93

\bibitem[{{Dsilva} {et~al.}(2023){Dsilva}, {Shenar}, {Sana}, \& {Marchant}}]{2023A&A...674A..88D}
{Dsilva}, K., {Shenar}, T., {Sana}, H., \& {Marchant}, P. 2023, \aap, 674, A88

\bibitem[{{Dutta} \& {Klencki}(2024)}]{2024A&A...687A.215D}
{Dutta}, D. \& {Klencki}, J. 2024, \aap, 687, A215

\bibitem[{{Eggleton}(1983)}]{1983ApJ...268..368E}
{Eggleton}, P.~P. 1983, \apj, 268, 368

\bibitem[{{Eldridge} {et~al.}(2013){Eldridge}, {Fraser}, {Smartt}, {Maund}, \& {Crockett}}]{2013MNRAS.436..774E}
{Eldridge}, J.~J., {Fraser}, M., {Smartt}, S.~J., {Maund}, J.~R., \& {Crockett}, R.~M. 2013, \mnras, 436, 774

\bibitem[{{Eldridge} {et~al.}(2017){Eldridge}, {Stanway}, {Xiao}, {McClelland}, {Taylor}, {Ng}, {Greis}, \& {Bray}}]{2017PASA...34...58E}
{Eldridge}, J.~J., {Stanway}, E.~R., {Xiao}, L., {et~al.} 2017, \pasa, 34, e058

\bibitem[{{Elia} {et~al.}(2022){Elia}, {Molinari}, {Schisano}, {Soler}, {Merello}, {Russeil}, {Veneziani}, {Zavagno}, {Noriega-Crespo}, {Olmi}, {Benedettini}, {Hennebelle}, {Klessen}, {Leurini}, {Paladini}, {Pezzuto}, {Traficante}, {Eden}, {Martin}, {Sormani}, {Coletta}, {Colman}, {Plume}, {Maruccia}, {Mininni}, \& {Liu}}]{2022ApJ...941..162E}
{Elia}, D., {Molinari}, S., {Schisano}, E., {et~al.} 2022, \apj, 941, 162

\bibitem[{{Ercolino} {et~al.}(2024){Ercolino}, {Jin}, {Langer}, \& {Dessart}}]{2024A&A...685A..58E}
{Ercolino}, A., {Jin}, H., {Langer}, N., \& {Dessart}, L. 2024, \aap, 685, A58

\bibitem[{{Farmer} {et~al.}(2023){Farmer}, {Renzo}, {G{\"o}tberg}, {Bellinger}, {Justham}, \& {de Mink}}]{2023MNRAS.524.1692F}
{Farmer}, R., {Renzo}, M., {G{\"o}tberg}, Y., {et~al.} 2023, \mnras, 524, 1692

\bibitem[{{Farrell} {et~al.}(2022){Farrell}, {Groh}, {Meynet}, \& {Eldridge}}]{2022MNRAS.512.4116F}
{Farrell}, E., {Groh}, J.~H., {Meynet}, G., \& {Eldridge}, J.~J. 2022, \mnras, 512, 4116

\bibitem[{{Foellmi} {et~al.}(2003){Foellmi}, {Moffat}, \& {Guerrero}}]{2003MNRAS.338..360F}
{Foellmi}, C., {Moffat}, A.~F.~J., \& {Guerrero}, M.~A. 2003, \mnras, 338, 360

\bibitem[{{Fragos} {et~al.}(2023){Fragos}, {Andrews}, {Bavera}, {Berry}, {Coughlin}, {Dotter}, {Giri}, {Kalogera}, {Katsaggelos}, {Kovlakas}, {Lalvani}, {Misra}, {Srivastava}, {Qin}, {Rocha}, {Rom{\'a}n-Garza}, {Serra}, {Stahle}, {Sun}, {Teng}, {Trajcevski}, {Tran}, {Xing}, {Zapartas}, \& {Zevin}}]{2023ApJS..264...45F}
{Fragos}, T., {Andrews}, J.~J., {Bavera}, S.~S., {et~al.} 2023, \apjs, 264, 45

\bibitem[{{Fryer} \& {Kalogera}(2001)}]{2001ApJ...554..548F}
{Fryer}, C.~L. \& {Kalogera}, V. 2001, \apj, 554, 548

\bibitem[{{Georgy} {et~al.}(2013){Georgy}, {Ekstr{\"o}m}, {Eggenberger}, {Meynet}, {Haemmerl{\'e}}, {Maeder}, {Granada}, {Groh}, {Hirschi}, {Mowlavi}, {Yusof}, {Charbonnel}, {Decressin}, \& {Barblan}}]{2013A&A...558A.103G}
{Georgy}, C., {Ekstr{\"o}m}, S., {Eggenberger}, P., {et~al.} 2013, \aap, 558, A103

\bibitem[{{Giacobbo} \& {Mapelli}(2018)}]{2018MNRAS.480.2011G}
{Giacobbo}, N. \& {Mapelli}, M. 2018, \mnras, 480, 2011

\bibitem[{{Gilkis} {et~al.}(2021){Gilkis}, {Shenar}, {Ramachandran}, {Jermyn}, {Mahy}, {Oskinova}, {Arcavi}, \& {Sana}}]{2021MNRAS.503.1884G}
{Gilkis}, A., {Shenar}, T., {Ramachandran}, V., {et~al.} 2021, \mnras, 503, 1884

\bibitem[{{Gilkis} {et~al.}(2019){Gilkis}, {Vink}, {Eldridge}, \& {Tout}}]{2019MNRAS.486.4451G}
{Gilkis}, A., {Vink}, J.~S., {Eldridge}, J.~J., \& {Tout}, C.~A. 2019, \mnras, 486, 4451

\bibitem[{{G{\"o}tberg} {et~al.}(2017){G{\"o}tberg}, {de Mink}, \& {Groh}}]{2017A&A...608A..11G}
{G{\"o}tberg}, Y., {de Mink}, S.~E., \& {Groh}, J.~H. 2017, \aap, 608, A11

\bibitem[{{G{\"o}tberg} {et~al.}(2018){G{\"o}tberg}, {de Mink}, {Groh}, {Kupfer}, {Crowther}, {Zapartas}, \& {Renzo}}]{2018A&A...615A..78G}
{G{\"o}tberg}, Y., {de Mink}, S.~E., {Groh}, J.~H., {et~al.} 2018, \aap, 615, A78

\bibitem[{{G{\"o}tberg} {et~al.}(2019){G{\"o}tberg}, {de Mink}, {Groh}, {Leitherer}, \& {Norman}}]{2019A&A...629A.134G}
{G{\"o}tberg}, Y., {de Mink}, S.~E., {Groh}, J.~H., {Leitherer}, C., \& {Norman}, C. 2019, \aap, 629, A134

\bibitem[{{G{\"o}tberg} {et~al.}(2020{\natexlab{a}}){G{\"o}tberg}, {de Mink}, {McQuinn}, {Zapartas}, {Groh}, \& {Norman}}]{2020A&A...634A.134G}
{G{\"o}tberg}, Y., {de Mink}, S.~E., {McQuinn}, M., {et~al.} 2020{\natexlab{a}}, \aap, 634, A134

\bibitem[{{G{\"o}tberg} {et~al.}(2023){G{\"o}tberg}, {Drout}, {Ji}, {Groh}, {Ludwig}, {Crowther}, {Smith}, {de Koter}, \& {de Mink}}]{2023ApJ...959..125G}
{G{\"o}tberg}, Y., {Drout}, M.~R., {Ji}, A.~P., {et~al.} 2023, \apj, 959, 125

\bibitem[{{G{\"o}tberg} {et~al.}(2020{\natexlab{b}}){G{\"o}tberg}, {Korol}, {Lamberts}, {Kupfer}, {Breivik}, {Ludwig}, \& {Drout}}]{2020ApJ...904...56G}
{G{\"o}tberg}, Y., {Korol}, V., {Lamberts}, A., {et~al.} 2020{\natexlab{b}}, \apj, 904, 56

\bibitem[{{Graur} {et~al.}(2017{\natexlab{a}}){Graur}, {Bianco}, {Huang}, {Modjaz}, {Shivvers}, {Filippenko}, {Li}, \& {Eldridge}}]{2017ApJ...837..120G}
{Graur}, O., {Bianco}, F.~B., {Huang}, S., {et~al.} 2017{\natexlab{a}}, \apj, 837, 120

\bibitem[{{Graur} {et~al.}(2017{\natexlab{b}}){Graur}, {Bianco}, {Modjaz}, {Shivvers}, {Filippenko}, {Li}, \& {Smith}}]{2017ApJ...837..121G}
{Graur}, O., {Bianco}, F.~B., {Modjaz}, M., {et~al.} 2017{\natexlab{b}}, \apj, 837, 121

\bibitem[{{Grevesse} \& {Sauval}(1998)}]{1998SSRv...85..161G}
{Grevesse}, N. \& {Sauval}, A.~J. 1998, \ssr, 85, 161

\bibitem[{{Groh} {et~al.}(2019){Groh}, {Ekstr{\"o}m}, {Georgy}, {Meynet}, {Choplin}, {Eggenberger}, {Hirschi}, {Maeder}, {Murphy}, {Boian}, \& {Farrell}}]{2019A&A...627A..24G}
{Groh}, J.~H., {Ekstr{\"o}m}, S., {Georgy}, C., {et~al.} 2019, \aap, 627, A24

\bibitem[{{Groh} {et~al.}(2014){Groh}, {Meynet}, {Ekstr{\"o}m}, \& {Georgy}}]{2014A&A...564A..30G}
{Groh}, J.~H., {Meynet}, G., {Ekstr{\"o}m}, S., \& {Georgy}, C. 2014, \aap, 564, A30

\bibitem[{{Groh} {et~al.}(2008){Groh}, {Oliveira}, \& {Steiner}}]{2008A&A...485..245G}
{Groh}, J.~H., {Oliveira}, A.~S., \& {Steiner}, J.~E. 2008, \aap, 485, 245

\bibitem[{{Guseva} {et~al.}(2000){Guseva}, {Izotov}, \& {Thuan}}]{2000ApJ...531..776G}
{Guseva}, N.~G., {Izotov}, Y.~I., \& {Thuan}, T.~X. 2000, \apj, 531, 776

\bibitem[{{Hagen} {et~al.}(2017){Hagen}, {Siegel}, {Hoversten}, {Gronwall}, {Immler}, \& {Hagen}}]{2017MNRAS.466.4540H}
{Hagen}, L. M.~Z., {Siegel}, M.~H., {Hoversten}, E.~A., {et~al.} 2017, \mnras, 466, 4540

\bibitem[{{Hainich} {et~al.}(2015){Hainich}, {Pasemann}, {Todt}, {Shenar}, {Sander}, \& {Hamann}}]{2015A&A...581A..21H}
{Hainich}, R., {Pasemann}, D., {Todt}, H., {et~al.} 2015, \aap, 581, A21

\bibitem[{{Hainich} {et~al.}(2014){Hainich}, {R{\"u}hling}, {Todt}, {Oskinova}, {Liermann}, {Gr{\"a}fener}, {Foellmi}, {Schnurr}, \& {Hamann}}]{2014A&A...565A..27H}
{Hainich}, R., {R{\"u}hling}, U., {Todt}, H., {et~al.} 2014, \aap, 565, A27

\bibitem[{{Han} {et~al.}(2003){Han}, {Podsiadlowski}, {Maxted}, \& {Marsh}}]{2003MNRAS.341..669H}
{Han}, Z., {Podsiadlowski}, P., {Maxted}, P.~F.~L., \& {Marsh}, T.~R. 2003, \mnras, 341, 669

\bibitem[{{Han} {et~al.}(2002){Han}, {Podsiadlowski}, {Maxted}, {Marsh}, \& {Ivanova}}]{2002MNRAS.336..449H}
{Han}, Z., {Podsiadlowski}, P., {Maxted}, P.~F.~L., {Marsh}, T.~R., \& {Ivanova}, N. 2002, \mnras, 336, 449

\bibitem[{{Harris} \& {Zaritsky}(2009)}]{2009AJ....138.1243H}
{Harris}, J. \& {Zaritsky}, D. 2009, \aj, 138, 1243

\bibitem[{{Hurley} {et~al.}(2000){Hurley}, {Pols}, \& {Tout}}]{2000MNRAS.315..543H}
{Hurley}, J.~R., {Pols}, O.~R., \& {Tout}, C.~A. 2000, \mnras, 315, 543

\bibitem[{{Hurley} {et~al.}(2002){Hurley}, {Tout}, \& {Pols}}]{2002MNRAS.329..897H}
{Hurley}, J.~R., {Tout}, C.~A., \& {Pols}, O.~R. 2002, \mnras, 329, 897

\bibitem[{{Inayoshi} {et~al.}(2017){Inayoshi}, {Hirai}, {Kinugawa}, \& {Hotokezaka}}]{2017MNRAS.468.5020I}
{Inayoshi}, K., {Hirai}, R., {Kinugawa}, T., \& {Hotokezaka}, K. 2017, \mnras, 468, 5020

\bibitem[{{Irrgang} {et~al.}(2020){Irrgang}, {Geier}, {Kreuzer}, {Pelisoli}, \& {Heber}}]{2020A&A...633L...5I}
{Irrgang}, A., {Geier}, S., {Kreuzer}, S., {Pelisoli}, I., \& {Heber}, U. 2020, \aap, 633, L5

\bibitem[{{Irrgang} {et~al.}(2022){Irrgang}, {Przybilla}, \& {Meynet}}]{2022NatAs...6.1414I}
{Irrgang}, A., {Przybilla}, N., \& {Meynet}, G. 2022, Nature Astronomy, 6, 1414

\bibitem[{{Ivanova}(2011)}]{2011ApJ...730...76I}
{Ivanova}, N. 2011, \apj, 730, 76

\bibitem[{{Jermyn} {et~al.}(2023){Jermyn}, {Bauer}, {Schwab}, {Farmer}, {Ball}, {Bellinger}, {Dotter}, {Joyce}, {Marchant}, {Mombarg}, {Wolf}, {Sunny Wong}, {Cinquegrana}, {Farrell}, {Smolec}, {Thoul}, {Cantiello}, {Herwig}, {Toloza}, {Bildsten}, {Townsend}, \& {Timmes}}]{2023ApJS..265...15J}
{Jermyn}, A.~S., {Bauer}, E.~B., {Schwab}, J., {et~al.} 2023, \apjs, 265, 15

\bibitem[{{Johnston}(2021)}]{2021A&A...655A..29J}
{Johnston}, C. 2021, \aap, 655, A29

\bibitem[{{Johnston} {et~al.}(2024){Johnston}, {Michielsen}, {Anders}, {Renzo}, {Cantiello}, {Marchant}, {Goldberg}, {Townsend}, {Sabhahit}, \& {Jermyn}}]{2024ApJ...964..170J}
{Johnston}, C., {Michielsen}, M., {Anders}, E.~H., {et~al.} 2024, \apj, 964, 170

\bibitem[{{Kiminki} \& {Kobulnicky}(2012)}]{2012ApJ...751....4K}
{Kiminki}, D.~C. \& {Kobulnicky}, H.~A. 2012, \apj, 751, 4

\bibitem[{{Kippenhahn} \& {Weigert}(1967)}]{1967ZA.....65..251K}
{Kippenhahn}, R. \& {Weigert}, A. 1967, \zap, 65, 251

\bibitem[{{Klencki} {et~al.}(2022){Klencki}, {Istrate}, {Nelemans}, \& {Pols}}]{2022A&A...662A..56K}
{Klencki}, J., {Istrate}, A., {Nelemans}, G., \& {Pols}, O. 2022, \aap, 662, A56

\bibitem[{{Klencki} {et~al.}(2018){Klencki}, {Moe}, {Gladysz}, {Chruslinska}, {Holz}, \& {Belczynski}}]{2018A&A...619A..77K}
{Klencki}, J., {Moe}, M., {Gladysz}, W., {et~al.} 2018, \aap, 619, A77

\bibitem[{{Klencki} {et~al.}(2021){Klencki}, {Nelemans}, {Istrate}, \& {Chruslinska}}]{2021A&A...645A..54K}
{Klencki}, J., {Nelemans}, G., {Istrate}, A.~G., \& {Chruslinska}, M. 2021, \aap, 645, A54

\bibitem[{{Klencki} {et~al.}(2020){Klencki}, {Nelemans}, {Istrate}, \& {Pols}}]{2020A&A...638A..55K}
{Klencki}, J., {Nelemans}, G., {Istrate}, A.~G., \& {Pols}, O. 2020, \aap, 638, A55

\bibitem[{{Kroupa}(2001)}]{2001MNRAS.322..231K}
{Kroupa}, P. 2001, \mnras, 322, 231

\bibitem[{{Kruckow} {et~al.}(2018){Kruckow}, {Tauris}, {Langer}, {Kramer}, \& {Izzard}}]{2018MNRAS.481.1908K}
{Kruckow}, M.~U., {Tauris}, T.~M., {Langer}, N., {Kramer}, M., \& {Izzard}, R.~G. 2018, \mnras, 481, 1908

\bibitem[{{Kruckow} {et~al.}(2016){Kruckow}, {Tauris}, {Langer}, {Sz{\'e}csi}, {Marchant}, \& {Podsiadlowski}}]{2016A&A...596A..58K}
{Kruckow}, M.~U., {Tauris}, T.~M., {Langer}, N., {et~al.} 2016, \aap, 596, A58

\bibitem[{{Kub{\'a}tov{\'a}} {et~al.}(2019){Kub{\'a}tov{\'a}}, {Sz{\'e}csi}, {Sander}, {Kub{\'a}t}, {Tramper}, {Krti{\v{c}}ka}, {Kehrig}, {Hamann}, {Hainich}, \& {Shenar}}]{2019A&A...623A...8K}
{Kub{\'a}tov{\'a}}, B., {Sz{\'e}csi}, D., {Sander}, A.~A.~C., {et~al.} 2019, \aap, 623, A8

\bibitem[{{Kulkarni} {et~al.}(2021){Kulkarni}, {Harrison}, {Grefenstette}, {Earnshaw}, {Andreoni}, {Berg}, {Bloom}, {Cenko}, {Chornock}, {Christiansen}, {Coughlin}, {Wuollet Criswell}, {Darvish}, {Das}, {De}, {Dessart}, {Dixon}, {Dorsman}, {El-Badry}, {Evans}, {Ford}, {Fremling}, {Gansicke}, {Gezari}, {Goetberg}, {Green}, {Graham}, {Heida}, {Ho}, {Jaodand}, {Johns-Krull}, {Kasliwal}, {Lazzarini}, {Lu}, {Margutti}, {Martin}, {Masters}, {McKernan}, {Naze}, {Nissanke}, {Parazin}, {Perley}, {Phinney}, {Piro}, {Raaijmakers}, {Rauw}, {Rodriguez}, {Sana}, {Senchyna}, {Singer}, {Spake}, {Stassun}, {Stern}, {Teplitz}, {Weisz}, \& {Yao}}]{2021arXiv211115608K}
{Kulkarni}, S.~R., {Harrison}, F.~A., {Grefenstette}, B.~W., {et~al.} 2021, arXiv e-prints, arXiv:2111.15608

\bibitem[{{Langer}(1991)}]{1991A&A...252..669L}
{Langer}, N. 1991, \aap, 252, 669

\bibitem[{{Langer} {et~al.}(1985){Langer}, {El Eid}, \& {Fricke}}]{1985A&A...145..179L}
{Langer}, N., {El Eid}, M.~F., \& {Fricke}, K.~J. 1985, \aap, 145, 179

\bibitem[{{Laplace} {et~al.}(2020){Laplace}, {G{\"o}tberg}, {de Mink}, {Justham}, \& {Farmer}}]{2020A&A...637A...6L}
{Laplace}, E., {G{\"o}tberg}, Y., {de Mink}, S.~E., {Justham}, S., \& {Farmer}, R. 2020, \aap, 637, A6

\bibitem[{{Laplace} {et~al.}(2021){Laplace}, {Justham}, {Renzo}, {G{\"o}tberg}, {Farmer}, {Vartanyan}, \& {de Mink}}]{2021A&A...656A..58L}
{Laplace}, E., {Justham}, S., {Renzo}, M., {et~al.} 2021, \aap, 656, A58

\bibitem[{{Laplace} {et~al.}(2025){Laplace}, {Schneider}, \& {Podsiadlowski}}]{2025A&A...695A..71L}
{Laplace}, E., {Schneider}, F.~R.~N., \& {Podsiadlowski}, P. 2025, \aap, 695, A71

\bibitem[{{Leitherer} {et~al.}(1999){Leitherer}, {Schaerer}, {Goldader}, {Delgado}, {Robert}, {Kune}, {de Mello}, {Devost}, \& {Heckman}}]{1999ApJS..123....3L}
{Leitherer}, C., {Schaerer}, D., {Goldader}, J.~D., {et~al.} 1999, \apjs, 123, 3

\bibitem[{{Lyman} {et~al.}(2016){Lyman}, {Bersier}, {James}, {Mazzali}, {Eldridge}, {Fraser}, \& {Pian}}]{2016MNRAS.457..328L}
{Lyman}, J.~D., {Bersier}, D., {James}, P.~A., {et~al.} 2016, \mnras, 457, 328

\bibitem[{{Ma} {et~al.}(2016){Ma}, {Hopkins}, {Kasen}, {Quataert}, {Faucher-Gigu{\`e}re}, {Kere{\v{s}}}, {Murray}, \& {Strom}}]{2016MNRAS.459.3614M}
{Ma}, X., {Hopkins}, P.~F., {Kasen}, D., {et~al.} 2016, \mnras, 459, 3614

\bibitem[{{Maeder}(1987)}]{1987A&A...178..159M}
{Maeder}, A. 1987, \aap, 178, 159

\bibitem[{{Mandel} \& {Broekgaarden}(2022)}]{2022LRR....25....1M}
{Mandel}, I. \& {Broekgaarden}, F.~S. 2022, Living Reviews in Relativity, 25, 1

\bibitem[{{Marchant} {et~al.}(2021){Marchant}, {Pappas}, {Gallegos-Garcia}, {Berry}, {Taam}, {Kalogera}, \& {Podsiadlowski}}]{2021A&A...650A.107M}
{Marchant}, P., {Pappas}, K. M.~W., {Gallegos-Garcia}, M., {et~al.} 2021, \aap, 650, A107

\bibitem[{{Marigo} {et~al.}(2001){Marigo}, {Girardi}, {Chiosi}, \& {Wood}}]{2001A&A...371..152M}
{Marigo}, P., {Girardi}, L., {Chiosi}, C., \& {Wood}, P.~R. 2001, \aap, 371, 152

\bibitem[{{Martins} {et~al.}(2009){Martins}, {Hillier}, {Bouret}, {Depagne}, {Foellmi}, {Marchenko}, \& {Moffat}}]{2009A&A...495..257M}
{Martins}, F., {Hillier}, D.~J., {Bouret}, J.~C., {et~al.} 2009, \aap, 495, 257

\bibitem[{{Massey}(2003)}]{2003ARA&A..41...15M}
{Massey}, P. 2003, \araa, 41, 15

\bibitem[{{Moe} \& {Di Stefano}(2017)}]{2017ApJS..230...15M}
{Moe}, M. \& {Di Stefano}, R. 2017, \apjs, 230, 15

\bibitem[{{Nanayakkara} {et~al.}(2019){Nanayakkara}, {Brinchmann}, {Boogaard}, {Bouwens}, {Cantalupo}, {Feltre}, {Kollatschny}, {Marino}, {Maseda}, {Matthee}, {Paalvast}, {Richard}, \& {Verhamme}}]{2019A&A...624A..89N}
{Nanayakkara}, T., {Brinchmann}, J., {Boogaard}, L., {et~al.} 2019, \aap, 624, A89

\bibitem[{{Neijssel} {et~al.}(2019){Neijssel}, {Vigna-G{\'o}mez}, {Stevenson}, {Barrett}, {Gaebel}, {Broekgaarden}, {de Mink}, {Sz{\'e}csi}, {Vinciguerra}, \& {Mandel}}]{2019MNRAS.490.3740N}
{Neijssel}, C.~J., {Vigna-G{\'o}mez}, A., {Stevenson}, S., {et~al.} 2019, \mnras, 490, 3740

\bibitem[{{Neugent} {et~al.}(2017){Neugent}, {Massey}, {Hillier}, \& {Morrell}}]{2017ApJ...841...20N}
{Neugent}, K.~F., {Massey}, P., {Hillier}, D.~J., \& {Morrell}, N. 2017, \apj, 841, 20

\bibitem[{{Nugis} \& {Lamers}(2000)}]{2000A&A...360..227N}
{Nugis}, T. \& {Lamers}, H.~J.~G.~L.~M. 2000, \aap, 360, 227

\bibitem[{{O'Grady} {et~al.}(2024){O'Grady}, {Moriya}, {Renzo}, \& {Vigna-G{\'o}mez}}]{2024arXiv241002896O}
{O'Grady}, A. J.~G., {Moriya}, T.~J., {Renzo}, M., \& {Vigna-G{\'o}mez}, A. 2024, arXiv e-prints, arXiv:2410.02896

\bibitem[{{Olejak} {et~al.}(2021){Olejak}, {Belczynski}, \& {Ivanova}}]{2021A&A...651A.100O}
{Olejak}, A., {Belczynski}, K., \& {Ivanova}, N. 2021, \aap, 651, A100

\bibitem[{{Olivier} {et~al.}(2022){Olivier}, {Berg}, {Chisholm}, {Erb}, {Pogge}, \& {Skillman}}]{2022ApJ...938...16O}
{Olivier}, G.~M., {Berg}, D.~A., {Chisholm}, J., {et~al.} 2022, \apj, 938, 16

\bibitem[{{{\"O}pik}(1924)}]{1924PTarO..25f...1O}
{{\"O}pik}, E. 1924, Publications of the Tartu Astrofizica Observatory, 25, 1

\bibitem[{{Paczy{\'n}ski}(1967)}]{1967AcA....17..355P}
{Paczy{\'n}ski}, B. 1967, \actaa, 17, 355

\bibitem[{{Paxton} {et~al.}(2011){Paxton}, {Bildsten}, {Dotter}, {Herwig}, {Lesaffre}, \& {Timmes}}]{2011ApJS..192....3P}
{Paxton}, B., {Bildsten}, L., {Dotter}, A., {et~al.} 2011, \apjs, 192, 3

\bibitem[{{Paxton} {et~al.}(2013){Paxton}, {Cantiello}, {Arras}, {Bildsten}, {Brown}, {Dotter}, {Mankovich}, {Montgomery}, {Stello}, {Timmes}, \& {Townsend}}]{2013ApJS..208....4P}
{Paxton}, B., {Cantiello}, M., {Arras}, P., {et~al.} 2013, \apjs, 208, 4

\bibitem[{{Paxton} {et~al.}(2015){Paxton}, {Marchant}, {Schwab}, {Bauer}, {Bildsten}, {Cantiello}, {Dessart}, {Farmer}, {Hu}, {Langer}, {Townsend}, {Townsley}, \& {Timmes}}]{2015ApJS..220...15P}
{Paxton}, B., {Marchant}, P., {Schwab}, J., {et~al.} 2015, \apjs, 220, 15

\bibitem[{{Paxton} {et~al.}(2018){Paxton}, {Schwab}, {Bauer}, {Bildsten}, {Blinnikov}, {Duffell}, {Farmer}, {Goldberg}, {Marchant}, {Sorokina}, {Thoul}, {Townsend}, \& {Timmes}}]{2018ApJS..234...34P}
{Paxton}, B., {Schwab}, J., {Bauer}, E.~B., {et~al.} 2018, \apjs, 234, 34

\bibitem[{{Paxton} {et~al.}(2019){Paxton}, {Smolec}, {Schwab}, {Gautschy}, {Bildsten}, {Cantiello}, {Dotter}, {Farmer}, {Goldberg}, {Jermyn}, {Kanbur}, {Marchant}, {Thoul}, {Townsend}, {Wolf}, {Zhang}, \& {Timmes}}]{2019ApJS..243...10P}
{Paxton}, B., {Smolec}, R., {Schwab}, J., {et~al.} 2019, \apjs, 243, 10

\bibitem[{{Pettini} {et~al.}(2002){Pettini}, {Ellison}, {Bergeron}, \& {Petitjean}}]{2002A&A...391...21P}
{Pettini}, M., {Ellison}, S.~L., {Bergeron}, J., \& {Petitjean}, P. 2002, \aap, 391, 21

\bibitem[{{Pettini} {et~al.}(2001){Pettini}, {Shapley}, {Steidel}, {Cuby}, {Dickinson}, {Moorwood}, {Adelberger}, \& {Giavalisco}}]{2001ApJ...554..981P}
{Pettini}, M., {Shapley}, A.~E., {Steidel}, C.~C., {et~al.} 2001, \apj, 554, 981

\bibitem[{{Picco} {et~al.}(2024){Picco}, {Marchant}, {Sana}, \& {Nelemans}}]{2024A&A...681A..31P}
{Picco}, A., {Marchant}, P., {Sana}, H., \& {Nelemans}, G. 2024, \aap, 681, A31

\bibitem[{{Pols} \& {Marinus}(1994)}]{1994A&A...288..475P}
{Pols}, O.~R. \& {Marinus}, M. 1994, \aap, 288, 475

\bibitem[{{Pols} {et~al.}(1998){Pols}, {Schr{\"o}der}, {Hurley}, {Tout}, \& {Eggleton}}]{1998MNRAS.298..525P}
{Pols}, O.~R., {Schr{\"o}der}, K.-P., {Hurley}, J.~R., {Tout}, C.~A., \& {Eggleton}, P.~P. 1998, \mnras, 298, 525

\bibitem[{{Prieto} {et~al.}(2008){Prieto}, {Stanek}, {Kochanek}, {Weisz}, {Baruffolo}, {Bechtold}, {Burwitz}, {De Santis}, {Gallozzi}, {Garnavich}, {Giallongo}, {Hill}, {Pogge}, {Ragazzoni}, {Speziali}, {Thompson}, \& {Wagner}}]{2008ApJ...673L..59P}
{Prieto}, J.~L., {Stanek}, K.~Z., {Kochanek}, C.~S., {et~al.} 2008, \apjl, 673, L59

\bibitem[{{Ramachandran} {et~al.}(2023){Ramachandran}, {Klencki}, {Sander}, {Pauli}, {Shenar}, {Oskinova}, \& {Hamann}}]{2023A&A...674L..12R}
{Ramachandran}, V., {Klencki}, J., {Sander}, A.~A.~C., {et~al.} 2023, \aap, 674, L12

\bibitem[{{Ramachandran} {et~al.}(2024){Ramachandran}, {Sander}, {Pauli}, {Klencki}, {Backs}, {Tramper}, {Bernini-Peron}, {Crowther}, {Hamann}, {Ignace}, {Kuiper}, {Oey}, {Oskinova}, {Shenar}, {Todt}, {Vink}, {Wang}, {Wofford}, \& {the XShootU Collaboration}}]{2024A&A...692A..90R}
{Ramachandran}, V., {Sander}, A.~A.~C., {Pauli}, D., {et~al.} 2024, \aap, 692, A90

\bibitem[{{Renzo} {et~al.}(2019){Renzo}, {Zapartas}, {de Mink}, {G{\"o}tberg}, {Justham}, {Farmer}, {Izzard}, {Toonen}, \& {Sana}}]{2019A&A...624A..66R}
{Renzo}, M., {Zapartas}, E., {de Mink}, S.~E., {et~al.} 2019, \aap, 624, A66

\bibitem[{{Rogers} {et~al.}(2024){Rogers}, {Strom}, {Rudie}, {Trainor}, {Raptis}, \& {von Raesfeld}}]{2024ApJ...964L..12R}
{Rogers}, N. S.~J., {Strom}, A.~L., {Rudie}, G.~C., {et~al.} 2024, \apjl, 964, L12

\bibitem[{{Romagnolo} {et~al.}(2025){Romagnolo}, {Klencki}, {Vigna-G{\'o}mez}, \& {Belczynski}}]{2025A&A...693A.137R}
{Romagnolo}, A., {Klencki}, J., {Vigna-G{\'o}mez}, A., \& {Belczynski}, K. 2025, \aap, 693, A137

\bibitem[{{Rubele} {et~al.}(2015){Rubele}, {Girardi}, {Kerber}, {Cioni}, {Piatti}, {Zaggia}, {Bekki}, {Bressan}, {Clementini}, {de Grijs}, {Emerson}, {Groenewegen}, {Ivanov}, {Marconi}, {Marigo}, {Moretti}, {Ripepi}, {Subramanian}, {Tatton}, \& {van Loon}}]{2015MNRAS.449..639R}
{Rubele}, S., {Girardi}, L., {Kerber}, L., {et~al.} 2015, \mnras, 449, 639

\bibitem[{{Sana} {et~al.}(2012){Sana}, {de Mink}, {de Koter}, {Langer}, {Evans}, {Gieles}, {Gosset}, {Izzard}, {Le Bouquin}, \& {Schneider}}]{2012Sci...337..444S}
{Sana}, H., {de Mink}, S.~E., {de Koter}, A., {et~al.} 2012, Science, 337, 444

\bibitem[{{Sander}(2022)}]{2022arXiv221105424S}
{Sander}, A. A.~C. 2022, arXiv e-prints, arXiv:2211.05424

\bibitem[{{Sander} \& {Vink}(2020)}]{2020MNRAS.499..873S}
{Sander}, A. A.~C. \& {Vink}, J.~S. 2020, \mnras, 499, 873

\bibitem[{{Saxena} {et~al.}(2020){Saxena}, {Pentericci}, {Mirabelli}, {Schaerer}, {Schneider}, {Cullen}, {Amorin}, {Bolzonella}, {Bongiorno}, {Carnall}, {Castellano}, {Cucciati}, {Fontana}, {Fynbo}, {Garilli}, {Gargiulo}, {Guaita}, {Hathi}, {Hutchison}, {Koekemoer}, {Marchi}, {McLeod}, {McLure}, {Papovich}, {Pozzetti}, {Talia}, \& {Zamorani}}]{2020A&A...636A..47S}
{Saxena}, A., {Pentericci}, L., {Mirabelli}, M., {et~al.} 2020, \aap, 636, A47

\bibitem[{{Schaerer} {et~al.}(2019){Schaerer}, {Fragos}, \& {Izotov}}]{2019A&A...622L..10S}
{Schaerer}, D., {Fragos}, T., \& {Izotov}, Y.~I. 2019, \aap, 622, L10

\bibitem[{{Schneider} {et~al.}(2014){Schneider}, {Izzard}, {de Mink}, {Langer}, {Stolte}, {de Koter}, {Gvaramadze}, {Hu{\ss}mann}, {Liermann}, \& {Sana}}]{2014ApJ...780..117S}
{Schneider}, F.~R.~N., {Izzard}, R.~G., {de Mink}, S.~E., {et~al.} 2014, \apj, 780, 117

\bibitem[{{Schneider} {et~al.}(2021){Schneider}, {Podsiadlowski}, \& {M{\"u}ller}}]{2021A&A...645A...5S}
{Schneider}, F.~R.~N., {Podsiadlowski}, P., \& {M{\"u}ller}, B. 2021, \aap, 645, A5

\bibitem[{{Schootemeijer} \& {Langer}(2018)}]{2018A&A...611A..75S}
{Schootemeijer}, A. \& {Langer}, N. 2018, \aap, 611, A75

\bibitem[{{Schootemeijer} {et~al.}(2019){Schootemeijer}, {Langer}, {Grin}, \& {Wang}}]{2019A&A...625A.132S}
{Schootemeijer}, A., {Langer}, N., {Grin}, N.~J., \& {Wang}, C. 2019, \aap, 625, A132

\bibitem[{{Schootemeijer} {et~al.}(2021){Schootemeijer}, {Langer}, {Lennon}, {Evans}, {Crowther}, {Geen}, {Howarth}, {de Koter}, {Menten}, \& {Vink}}]{2021A&A...646A.106S}
{Schootemeijer}, A., {Langer}, N., {Lennon}, D., {et~al.} 2021, \aap, 646, A106

\bibitem[{{Schootemeijer} {et~al.}(2024){Schootemeijer}, {Shenar}, {Langer}, {Grin}, {Sana}, {Gr{\"a}fener}, {Sch{\"u}rmann}, {Wang}, \& {Xu}}]{2024A&A...689A.157S}
{Schootemeijer}, A., {Shenar}, T., {Langer}, N., {et~al.} 2024, \aap, 689, A157

\bibitem[{{Sch{\"u}rmann} {et~al.}(2024){Sch{\"u}rmann}, {Langer}, {Kramer}, {Marchant}, {Wang}, \& {Sen}}]{2024A&A...690A.282S}
{Sch{\"u}rmann}, C., {Langer}, N., {Kramer}, J.~A., {et~al.} 2024, \aap, 690, A282

\bibitem[{{Sen} {et~al.}(2023){Sen}, {Langer}, {Pauli}, {Gr{\"a}fener}, {Schootemeijer}, {Sana}, {Shenar}, {Mahy}, \& {Wang}}]{2023A&A...672A.198S}
{Sen}, K., {Langer}, N., {Pauli}, D., {et~al.} 2023, \aap, 672, A198

\bibitem[{{Senchyna} {et~al.}(2019){Senchyna}, {Stark}, {Chevallard}, {Charlot}, {Jones}, \& {Vidal-Garc{\'\i}a}}]{2019MNRAS.488.3492S}
{Senchyna}, P., {Stark}, D.~P., {Chevallard}, J., {et~al.} 2019, \mnras, 488, 3492

\bibitem[{{Senchyna} {et~al.}(2020){Senchyna}, {Stark}, {Mirocha}, {Reines}, {Charlot}, {Jones}, \& {Mulchaey}}]{2020MNRAS.494..941S}
{Senchyna}, P., {Stark}, D.~P., {Mirocha}, J., {et~al.} 2020, \mnras, 494, 941

\bibitem[{{Senchyna} {et~al.}(2017){Senchyna}, {Stark}, {Vidal-Garc{\'\i}a}, {Chevallard}, {Charlot}, {Mainali}, {Jones}, {Wofford}, {Feltre}, \& {Gutkin}}]{2017MNRAS.472.2608S}
{Senchyna}, P., {Stark}, D.~P., {Vidal-Garc{\'\i}a}, A., {et~al.} 2017, \mnras, 472, 2608

\bibitem[{{Shao} \& {Li}(2021)}]{2021ApJ...908...67S}
{Shao}, Y. \& {Li}, X.-D. 2021, \apj, 908, 67

\bibitem[{{Shapley} {et~al.}(2003){Shapley}, {Steidel}, {Pettini}, \& {Adelberger}}]{2003ApJ...588...65S}
{Shapley}, A.~E., {Steidel}, C.~C., {Pettini}, M., \& {Adelberger}, K.~L. 2003, \apj, 588, 65

\bibitem[{{Shara} {et~al.}(2017){Shara}, {Crawford}, {Vanbeveren}, {Moffat}, {Zurek}, \& {Crause}}]{2017MNRAS.464.2066S}
{Shara}, M.~M., {Crawford}, S.~M., {Vanbeveren}, D., {et~al.} 2017, \mnras, 464, 2066

\bibitem[{{Shara} {et~al.}(2020){Shara}, {Crawford}, {Vanbeveren}, {Moffat}, {Zurek}, \& {Crause}}]{2020MNRAS.492.4430S}
{Shara}, M.~M., {Crawford}, S.~M., {Vanbeveren}, D., {et~al.} 2020, \mnras, 492, 4430

\bibitem[{{Shenar}(2024)}]{2024arXiv241004436S}
{Shenar}, T. 2024, arXiv e-prints, arXiv:2410.04436

\bibitem[{{Shenar} {et~al.}(2020){Shenar}, {Gilkis}, {Vink}, {Sana}, \& {Sander}}]{2020A&A...634A..79S}
{Shenar}, T., {Gilkis}, A., {Vink}, J.~S., {Sana}, H., \& {Sander}, A.~A.~C. 2020, \aap, 634, A79

\bibitem[{{Shenar} {et~al.}(2018){Shenar}, {Hainich}, {Todt}, {Moffat}, {Sander}, {Oskinova}, {Ramachandran}, {Munoz}, {Pablo}, {Sana}, \& {Hamann}}]{2018A&A...616A.103S}
{Shenar}, T., {Hainich}, R., {Todt}, H., {et~al.} 2018, \aap, 616, A103

\bibitem[{{Shenar} {et~al.}(2016){Shenar}, {Hainich}, {Todt}, {Sander}, {Hamann}, {Moffat}, {Eldridge}, {Pablo}, {Oskinova}, \& {Richardson}}]{2016A&A...591A..22S}
{Shenar}, T., {Hainich}, R., {Todt}, H., {et~al.} 2016, \aap, 591, A22

\bibitem[{{Shenar} {et~al.}(2019){Shenar}, {Sablowski}, {Hainich}, {Todt}, {Moffat}, {Oskinova}, {Ramachandran}, {Sana}, {Sander}, {Schnurr}, {St-Louis}, {Vanbeveren}, {G{\"o}tberg}, \& {Hamann}}]{2019A&A...627A.151S}
{Shenar}, T., {Sablowski}, D.~P., {Hainich}, R., {et~al.} 2019, \aap, 627, A151

\bibitem[{{Shenar} {et~al.}(2023){Shenar}, {Wade}, {Marchant}, {Bagnulo}, {Bodensteiner}, {Bowman}, {Gilkis}, {Langer}, {Nicolas-Chen{\'e}}, {Oskinova}, {Van Reeth}, {Sana}, {St-Louis}, {de Oliveira}, {Todt}, \& {Toonen}}]{2023Sci...381..761S}
{Shenar}, T., {Wade}, G.~A., {Marchant}, P., {et~al.} 2023, Science, 381, 761

\bibitem[{{Shirazi} \& {Brinchmann}(2012)}]{2012MNRAS.421.1043S}
{Shirazi}, M. \& {Brinchmann}, J. 2012, \mnras, 421, 1043

\bibitem[{{Smith} {et~al.}(2002){Smith}, {Norris}, \& {Crowther}}]{2002MNRAS.337.1309S}
{Smith}, L.~J., {Norris}, R. P.~F., \& {Crowther}, P.~A. 2002, \mnras, 337, 1309

\bibitem[{{Smith}(2014)}]{2014ARA&A..52..487S}
{Smith}, N. 2014, \araa, 52, 487

\bibitem[{{Smith} {et~al.}(2011){Smith}, {Li}, {Filippenko}, \& {Chornock}}]{2011MNRAS.412.1522S}
{Smith}, N., {Li}, W., {Filippenko}, A.~V., \& {Chornock}, R. 2011, \mnras, 412, 1522

\bibitem[{{Stanton} {et~al.}(2024){Stanton}, {Cullen}, {McLure}, {Shapley}, {Arellano-C{\'o}rdova}, {Begley}, {Amor{\'\i}n}, {Barrufet}, {Calabr{\`o}}, {Carnall}, {Cirasuolo}, {Dunlop}, {Donnan}, {Hamadouche}, {Liu}, {McLeod}, {Pentericci}, {Pozzetti}, {Sanders}, {Scholte}, \& {Topping}}]{2024MNRAS.532.3102S}
{Stanton}, T.~M., {Cullen}, F., {McLure}, R.~J., {et~al.} 2024, \mnras, 532, 3102

\bibitem[{{Stanway} {et~al.}(2016){Stanway}, {Eldridge}, \& {Becker}}]{2016MNRAS.456..485S}
{Stanway}, E.~R., {Eldridge}, J.~J., \& {Becker}, G.~D. 2016, \mnras, 456, 485

\bibitem[{{Steidel} {et~al.}(2016){Steidel}, {Strom}, {Pettini}, {Rudie}, {Reddy}, \& {Trainor}}]{2016ApJ...826..159S}
{Steidel}, C.~C., {Strom}, A.~L., {Pettini}, M., {et~al.} 2016, \apj, 826, 159

\bibitem[{{Stevenson} {et~al.}(2017){Stevenson}, {Vigna-G{\'o}mez}, {Mandel}, {Barrett}, {Neijssel}, {Perkins}, \& {de Mink}}]{2017NatCo...814906S}
{Stevenson}, S., {Vigna-G{\'o}mez}, A., {Mandel}, I., {et~al.} 2017, Nature Communications, 8, 14906

\bibitem[{{Strom} {et~al.}(2018){Strom}, {Steidel}, {Rudie}, {Trainor}, \& {Pettini}}]{2018ApJ...868..117S}
{Strom}, A.~L., {Steidel}, C.~C., {Rudie}, G.~C., {Trainor}, R.~F., \& {Pettini}, M. 2018, \apj, 868, 117

\bibitem[{{Sz{\'e}csi} {et~al.}(2015){Sz{\'e}csi}, {Langer}, {Yoon}, {Sanyal}, {de Mink}, {Evans}, \& {Dermine}}]{2015A&A...581A..15S}
{Sz{\'e}csi}, D., {Langer}, N., {Yoon}, S.-C., {et~al.} 2015, \aap, 581, A15

\bibitem[{{Tang} {et~al.}(2014){Tang}, {Bressan}, {Rosenfield}, {Slemer}, {Marigo}, {Girardi}, \& {Bianchi}}]{2014MNRAS.445.4287T}
{Tang}, J., {Bressan}, A., {Rosenfield}, P., {et~al.} 2014, \mnras, 445, 4287

\bibitem[{{Tauris} \& {Dewi}(2001)}]{2001A&A...369..170T}
{Tauris}, T.~M. \& {Dewi}, J.~D.~M. 2001, \aap, 369, 170

\bibitem[{{Tauris} {et~al.}(2017){Tauris}, {Kramer}, {Freire}, {Wex}, {Janka}, {Langer}, {Podsiadlowski}, {Bozzo}, {Chaty}, {Kruckow}, {van den Heuvel}, {Antoniadis}, {Breton}, \& {Champion}}]{2017ApJ...846..170T}
{Tauris}, T.~M., {Kramer}, M., {Freire}, P.~C.~C., {et~al.} 2017, \apj, 846, 170

\bibitem[{{Tauris} {et~al.}(2015){Tauris}, {Langer}, \& {Podsiadlowski}}]{2015MNRAS.451.2123T}
{Tauris}, T.~M., {Langer}, N., \& {Podsiadlowski}, P. 2015, \mnras, 451, 2123

\bibitem[{{Topping} {et~al.}(2020){Topping}, {Shapley}, {Reddy}, {Sanders}, {Coil}, {Kriek}, {Mobasher}, \& {Siana}}]{2020MNRAS.499.1652T}
{Topping}, M.~W., {Shapley}, A.~E., {Reddy}, N.~A., {et~al.} 2020, \mnras, 499, 1652

\bibitem[{{Topping} {et~al.}(2024){Topping}, {Stark}, {Senchyna}, {Plat}, {Zitrin}, {Endsley}, {Charlot}, {Furtak}, {Maseda}, {Smit}, {Mainali}, {Chevallard}, {Molyneux}, \& {Rigby}}]{2024MNRAS.529.3301T}
{Topping}, M.~W., {Stark}, D.~P., {Senchyna}, P., {et~al.} 2024, \mnras, 529, 3301

\bibitem[{{van den Heuvel} {et~al.}(2017){van den Heuvel}, {Portegies Zwart}, \& {de Mink}}]{2017MNRAS.471.4256V}
{van den Heuvel}, E.~P.~J., {Portegies Zwart}, S.~F., \& {de Mink}, S.~E. 2017, \mnras, 471, 4256

\bibitem[{{van Son} {et~al.}(2022){van Son}, {de Mink}, {Callister}, {Justham}, {Renzo}, {Wagg}, {Broekgaarden}, {Kummer}, {Pakmor}, \& {Mandel}}]{2022ApJ...931...17V}
{van Son}, L.~A.~C., {de Mink}, S.~E., {Callister}, T., {et~al.} 2022, \apj, 931, 17

\bibitem[{{van Son} {et~al.}(2025){van Son}, {Roy}, {Mandel}, {Farr}, {Lam}, {Merritt}, {Broekgaarden}, {Sander}, \& {Andrews}}]{2025ApJ...979..209V}
{van Son}, L.~A.~C., {Roy}, S.~K., {Mandel}, I., {et~al.} 2025, \apj, 979, 209

\bibitem[{{Vanbeveren} {et~al.}(1998){Vanbeveren}, {De Donder}, {Van Bever}, {Van Rensbergen}, \& {De Loore}}]{1998NewA....3..443V}
{Vanbeveren}, D., {De Donder}, E., {Van Bever}, J., {Van Rensbergen}, W., \& {De Loore}, C. 1998, \na, 3, 443

\bibitem[{{Vartanyan} {et~al.}(2021){Vartanyan}, {Laplace}, {Renzo}, {G{\"o}tberg}, {Burrows}, \& {de Mink}}]{2021ApJ...916L...5V}
{Vartanyan}, D., {Laplace}, E., {Renzo}, M., {et~al.} 2021, \apjl, 916, L5

\bibitem[{{Vigna-G{\'o}mez} {et~al.}(2020){Vigna-G{\'o}mez}, {MacLeod}, {Neijssel}, {Broekgaarden}, {Justham}, {Howitt}, {de Mink}, {Vinciguerra}, \& {Mandel}}]{2020PASA...37...38V}
{Vigna-G{\'o}mez}, A., {MacLeod}, M., {Neijssel}, C.~J., {et~al.} 2020, \pasa, 37, e038

\bibitem[{{Vigna-G{\'o}mez} {et~al.}(2022){Vigna-G{\'o}mez}, {Wassink}, {Klencki}, {Istrate}, {Nelemans}, \& {Mandel}}]{2022MNRAS.511.2326V}
{Vigna-G{\'o}mez}, A., {Wassink}, M., {Klencki}, J., {et~al.} 2022, \mnras, 511, 2326

\bibitem[{{Villase{\~n}or} {et~al.}(2023){Villase{\~n}or}, {Lennon}, {Picco}, {Shenar}, {Marchant}, {Langer}, {Dufton}, {Nardini}, {Evans}, {Bodensteiner}, {de Mink}, {G{\"o}tberg}, {Soszy{\'n}ski}, {Taylor}, \& {Sana}}]{2023MNRAS.525.5121V}
{Villase{\~n}or}, J.~I., {Lennon}, D.~J., {Picco}, A., {et~al.} 2023, \mnras, 525, 5121

\bibitem[{{Vink}(2017)}]{2017A&A...607L...8V}
{Vink}, J.~S. 2017, \aap, 607, L8

\bibitem[{{Wang} {et~al.}(2024){Wang}, {Bodensteiner}, {Xu}, {de Mink}, {Langer}, {Laplace}, {Vigna-G{\'o}mez}, {Justham}, {Klencki}, {Olejak}, {Valli}, \& {Schootemeijer}}]{2024ApJ...975L..20W}
{Wang}, C., {Bodensteiner}, J., {Xu}, X.-T., {et~al.} 2024, \apjl, 975, L20

\bibitem[{{Webbink}(1984)}]{1984ApJ...277..355W}
{Webbink}, R.~F. 1984, \apj, 277, 355

\bibitem[{{Wu} {et~al.}(2020){Wu}, {Chen}, {Chen}, {Li}, \& {Han}}]{2020A&A...634A.126W}
{Wu}, Y., {Chen}, X., {Chen}, H., {Li}, Z., \& {Han}, Z. 2020, \aap, 634, A126

\bibitem[{{Xiang} {et~al.}(2018){Xiang}, {Shi}, {Liu}, {Yuan}, {Chen}, {Huang}, {Wang}, {Wu}, {Tian}, {Huo}, {Zhang}, \& {Zhang}}]{2018ApJS..237...33X}
{Xiang}, M., {Shi}, J., {Liu}, X., {et~al.} 2018, \apjs, 237, 33

\bibitem[{{Yoon} {et~al.}(2017){Yoon}, {Dessart}, \& {Clocchiatti}}]{2017ApJ...840...10Y}
{Yoon}, S.-C., {Dessart}, L., \& {Clocchiatti}, A. 2017, \apj, 840, 10

\bibitem[{{Yoon} {et~al.}(2010){Yoon}, {Woosley}, \& {Langer}}]{2010ApJ...725..940Y}
{Yoon}, S.~C., {Woosley}, S.~E., \& {Langer}, N. 2010, \apj, 725, 940

\bibitem[{{Yungelson} {et~al.}(2024){Yungelson}, {Kuranov}, {Postnov}, {Kuranova}, {Oskinova}, \& {Hamann}}]{2024A&A...683A..37Y}
{Yungelson}, L., {Kuranov}, A., {Postnov}, K., {et~al.} 2024, \aap, 683, A37

\bibitem[{{Zapartas} {et~al.}(2019){Zapartas}, {de Mink}, {Justham}, {Smith}, {de Koter}, {Renzo}, {Arcavi}, {Farmer}, {G{\"o}tberg}, \& {Toonen}}]{2019A&A...631A...5Z}
{Zapartas}, E., {de Mink}, S.~E., {Justham}, S., {et~al.} 2019, \aap, 631, A5

\bibitem[{{Zapartas} {et~al.}(2017){Zapartas}, {de Mink}, {Van Dyk}, {Fox}, {Smith}, {Bostroem}, {de Koter}, {Filippenko}, {Izzard}, {Kelly}, {Neijssel}, {Renzo}, \& {Ryder}}]{2017ApJ...842..125Z}
{Zapartas}, E., {de Mink}, S.~E., {Van Dyk}, S.~D., {et~al.} 2017, \apj, 842, 125

\end{thebibliography}

\begin{appendix}
\onecolumn
\section{Stripped stars formed through different channels}\label{appendix:channels}

As outlined in Sect. \ref{subsec:popsynth} and visualized in Fig.~\ref{fig:cartoon}, there are several evolutionary pathways that lead to stripped star formation through envelope-stripping in binaries. Here, we consider mass transfer initiated during the main-sequence evolution, mass transfer initiated during the Hertzsprung gap, and successful common envelope ejection that was initiated during the Hertzsprung gap evolution of the donor. 

Which envelope-stripping mechanism a star undergoes depends on how much it expands during the different evolutionary phases and its initial orbital period. 
This means that the importance of the different envelope-stripping mechanisms to the population as a whole varies as function of stellar mass, because the amount of expansion during the different evolutionary phases is dependent on the star's mass (see e.g.,  Fig.~\ref{fig:R_at_He}). 

\begin{figure*}[h!]
    \centering
    \includegraphics[width=\textwidth]{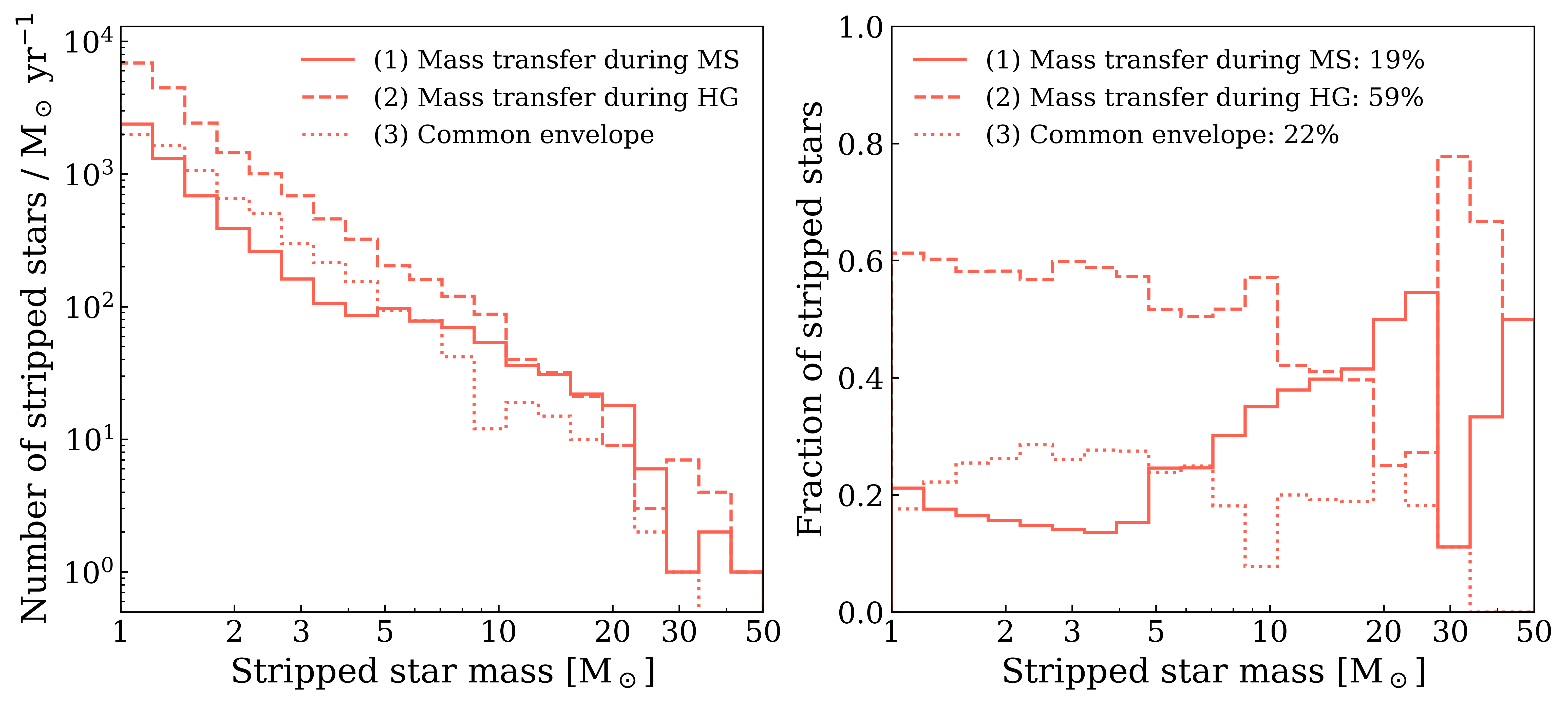}
    \caption{Stripped stars formed through each of the 3 channels considered, as raw numbers (left) and as a fraction of the total (right).
    We display the distributions for one of the solar metallicity model runs, although the metallicity distributions in general show similar trends. Common envelope evolution accounts for $\sim 20 \%$ of stripped stars across most mass bins and in total. Stable mass transfer during the Hertzsprung gap is the most common channel for $M \lesssim 15 M_\odot$, and stable mass transfer during the main sequence becomes more common at the very high-mass end. The stripped star masses of those formed through the latter channel may have been slightly overestimated.}
    \label{fig:channels}
\end{figure*}

In this appendix, we use our population synthesis models to explore the contributions of the different envelope-stripping channels. We display the fractional contribution as function of mass and for solar metallicity in the right panel of Fig.~\ref{fig:channels}, while the left panel shows the total contribution to the mass function. 

On average, mass transfer initiated during the main-sequence evolution (also known as Case A) and successful common envelope ejection initiated during the Hertzsprung gap passage of the donor (here referred to as Case B CEE) each contribute about 20\%, while mass transfer initiated during the Hertzsprung gap (also known as Case B) contributes about 60\%. Looking closer at the mass dependence, we find that for $M_{\rm strip} < 5 M_\odot$, Case B contributes 60\%, while Case A contributes 10-20\% and Case B CEE contributes 20-30\%. For $M_{\rm strip} \sim 10-20 M_\odot$, Case A and B both contribute 30-50\%, while the contribution of Case B CEE remains around 20\%. Case A becomes the dominant channel for envelope-stripping in binaries for $M_{\rm strip} > 15 M_\odot$, reaching 60\%, while Case B reduces to about 30\% contribution and Case B CEE drops from 10\% to no contribution at all. We expect that Case A should become even more important for massive stripped stars at lower metallicity, given the late expansion.   

The binary evolutionary models that we use as a base to model the population of stripped stars are created for Case B type mass transfer, initiated rather early during the Hertzsprung gap passage (Sect. \ref{subsec:mesa}, \citealt{2018A&A...615A..78G}). This relies on the assumption that stripped stars created through different envelope-stripping mechanisms should look similar to those produced through Case B mass transfer \citep[see however][]{2017ApJ...840...10Y, 2019MNRAS.486.4451G}. While this is a simplification, it is likely more accurate than the assumption that all stripped stars can be represented by pure helium stars, as is the case in most binary population synthesis codes. 
Stripped stars created through common envelope ejection are thought to have been more deeply stripped than stripped stars produced through mass transfer \citep[e.g.,][]{2011ApJ...730...76I, 2017A&A...608A..11G, 2022MNRAS.511.2326V}. 
Stripped stars produced through Case A mass transfer are thought to have lower masses than those produced through Case B, since the helium core is not fully defined until the main-sequence is complete.  Case A mass transfer therefore causes immediate recession of the central convective region, thus redefining the core \citep[e.g.,][]{2010ApJ...725..940Y}. 
While these predictions for the dependence of stripped star mass on the different envelope-stripping mechanisms have yet to be observationally constrained, we note that the lower masses predicted for Case A stripped stars would steepen the slope of the mass distribution, further enhancing the dearth of massive helium stars.

\section{Properties derived from MESA stellar evolution models}

\begin{table*}[ht!]
\caption{Initial mass, stripped star mass calculated two ways, main sequence lifetime, and stripped star lifetime for $Z=0.014$ models.}
\label{tab:mesa_properties_014}
\centering
\begin{tabular}{ccccc}
\hline\hline
$M_{\rm init} \ (M_\odot)$ & $M_{\rm He} \ (M_\odot)$\tablefootmark{a} & $M_{\rm strip} \ (M_\odot)$\tablefootmark{b} & $t_{\rm MS}$ (yr)\tablefootmark{a} & $t_{\rm strip}$ (yr)\tablefootmark{b} \\
\hline
2.00 & 0.22 & 0.35 & $1.17\times 10^{9}$ & -- \\
2.21 & 0.26 & 0.38 & $8.96\times 10^{8}$ & $4.14\times 10^{8}$ \\
2.44 & 0.29 & 0.44 & $6.84\times 10^{8}$ & -- \\
2.70 & 0.33 & 0.51 & $5.22\times 10^{8}$ & $1.33\times 10^{8}$ \\
2.99 & 0.38 & 0.58 & $4.00\times 10^{8}$ & -- \\
3.30 & 0.44 & 0.67 & $3.07\times 10^{8}$ & $6.39\times 10^{7}$ \\
3.65 & 0.50 & 0.74 & $2.37\times 10^{8}$ & $4.20\times 10^{7}$ \\
4.04 & 0.57 & 0.85 & $1.84\times 10^{8}$ & $2.81\times 10^{7}$ \\
4.46 & 0.65 & 0.97 & $1.44\times 10^{8}$ & $2.00\times 10^{7}$ \\
4.93 & 0.76 & 1.12 & $1.13\times 10^{8}$ & $1.65\times 10^{7}$ \\
5.45 & 0.88 & 1.28 & $8.94\times 10^{7}$ & $1.06\times 10^{7}$ \\
6.03 & 1.01 & 1.43 & $7.14\times 10^{7}$ & $8.04\times 10^{6}$ \\
6.66 & 1.18 & 1.64 & $5.74\times 10^{7}$ & $6.42\times 10^{6}$ \\
7.37 & 1.38 & 1.89 & $4.66\times 10^{7}$ & $4.41\times 10^{6}$ \\
8.15 & 1.61 & 2.18 & $3.81\times 10^{7}$ & $3.40\times 10^{6}$ \\
9.00 & 1.88 & 2.50 & $3.14\times 10^{7}$ & $2.70\times 10^{6}$ \\
9.95 & 2.20 & 2.88 & $2.62\times 10^{7}$ & $2.16\times 10^{6}$ \\
11.00 & 2.59 & 3.34 & $2.19\times 10^{7}$ & $1.74\times 10^{6}$ \\
12.17 & 3.04 & 3.86 & $1.85\times 10^{7}$ & $1.43\times 10^{6}$ \\
13.45 & 3.58 & 4.46 & $1.58\times 10^{7}$ & $1.20\times 10^{6}$ \\
14.87 & 4.22 & 5.14 & $1.37\times 10^{7}$ & $1.03\times 10^{6}$ \\
16.44 & 4.93 & 5.90 & $1.19\times 10^{7}$ & $8.93\times 10^{5}$ \\
18.17 & 5.73 & 6.74 & $1.05\times 10^{7}$ & $7.91\times 10^{5}$ \\
20.09 & 6.68 & -- & $9.33\times 10^{6}$ & -- \\
22.21 & 7.73 & -- & $8.34\times 10^{6}$ & -- \\
24.55 & 8.91 & -- & $7.53\times 10^{6}$ & -- \\
27.13 & 10.20 & -- & $6.83\times 10^{6}$ & -- \\
30.00 & 11.71 & -- & $6.25\times 10^{6}$ & -- \\
33.16 & 13.37 & -- & $5.75\times 10^{6}$ & -- \\
36.65 & 15.21 & -- & $5.31\times 10^{6}$ & -- \\
40.52 & 17.22 & -- & $4.94\times 10^{6}$ & -- \\
44.79 & 19.51 & -- & $4.62\times 10^{6}$ & -- \\
49.51 & 22.08 & -- & $4.33\times 10^{6}$ & -- \\
54.73 & 24.93 & -- & $4.08\times 10^{6}$ & -- \\
60.50 & 28.00 & -- & $3.86\times 10^{6}$ & -- \\
66.89 & 31.66 & -- & $3.66\times 10^{6}$ & -- \\
73.94 & 35.64 & -- & $3.48\times 10^{6}$ & -- \\
81.74 & 40.02 & -- & $3.32\times 10^{6}$ & -- \\
90.38 & 44.64 & -- & $3.19\times 10^{6}$ & -- \\
99.91 & 47.56 & -- & $3.06\times 10^{6}$ & -- \\
\hline
\end{tabular}
\tablefoot{For each of our MESA models, the initial mass, the mass of the helium core of a single star model at terminal age main sequence, the mass of the stripped star in a binary model when the central helium mass fraction reaches 0.5, the main sequence lifetime, and the stripped star lifetime. The second and third columns are the same values as plotted in Fig. \ref{fig:Minit_Mstrip_relation}.
\tablefoottext{a}{From the single star models.}
\tablefoottext{b}{From the binary models.}}
\end{table*}

\begin{table*}[ht!]
\caption{Same as Table \ref{tab:mesa_properties_014} for $Z=0.006$ models.}
\label{tab:mesa_properties_006}
\centering
\begin{tabular}{ccccc}
\hline\hline
$M_{\rm init} \ (M_\odot)$ & $M_{\rm He} \ (M_\odot)$ & $M_{\rm strip} \ (M_\odot)$ & $t_{\rm MS}$ (yr) & $t_{\rm strip}$ (yr) \\
\hline
2.00 & 0.24 & 0.37 & $9.83\times 10^{8}$ & -- \\
2.21 & 0.27 & 0.42 & $7.60\times 10^{8}$ & -- \\
2.44 & 0.31 & 0.48 & $5.88\times 10^{8}$ & -- \\
2.70 & 0.35 & 0.54 & $4.55\times 10^{8}$ & -- \\
2.99 & 0.40 & 0.62 & $3.53\times 10^{8}$ & $7.82\times 10^{7}$ \\
3.30 & 0.46 & 0.71 & $2.76\times 10^{8}$ & $4.84\times 10^{7}$ \\
3.65 & 0.52 & 0.81 & $2.16\times 10^{8}$ & $3.33\times 10^{7}$ \\
4.04 & 0.60 & 0.92 & $1.70\times 10^{8}$ & $2.37\times 10^{7}$ \\
4.46 & 0.68 & 1.06 & $1.35\times 10^{8}$ & $1.66\times 10^{7}$ \\
4.93 & 0.79 & 1.21 & $1.07\times 10^{8}$ & $1.41\times 10^{7}$ \\
5.45 & 0.91 & 1.38 & $8.60\times 10^{7}$ & $8.91\times 10^{6}$ \\
6.03 & 1.05 & 1.55 & $6.95\times 10^{7}$ & $6.86\times 10^{6}$ \\
6.66 & 1.21 & 1.77 & $5.65\times 10^{7}$ & $5.29\times 10^{6}$ \\
7.37 & 1.42 & 2.04 & $4.63\times 10^{7}$ & $3.96\times 10^{6}$ \\
8.15 & 1.65 & 2.34 & $3.82\times 10^{7}$ & $3.18\times 10^{6}$ \\
9.00 & 1.93 & 2.70 & $3.17\times 10^{7}$ & $2.48\times 10^{6}$ \\
9.95 & 2.27 & 3.11 & $2.65\times 10^{7}$ & $2.00\times 10^{6}$ \\
11.00 & 2.65 & 3.59 & $2.24\times 10^{7}$ & $1.65\times 10^{6}$ \\
12.17 & 3.11 & 4.14 & $1.91\times 10^{7}$ & $1.37\times 10^{6}$ \\
13.45 & 3.64 & 4.77 & $1.64\times 10^{7}$ & $1.16\times 10^{6}$ \\
14.87 & 4.26 & 5.49 & $1.42\times 10^{7}$ & $9.95\times 10^{5}$ \\
16.44 & 4.99 & 6.30 & $1.24\times 10^{7}$ & $8.66\times 10^{5}$ \\
18.17 & 5.80 & 7.17 & $1.10\times 10^{7}$ & $7.81\times 10^{5}$ \\
20.09 & 6.72 & -- & $9.74\times 10^{6}$ & -- \\
22.21 & 7.80 & -- & $8.72\times 10^{6}$ & -- \\
24.55 & 9.00 & -- & $7.86\times 10^{6}$ & -- \\
27.14 & 10.39 & -- & $7.13\times 10^{6}$ & -- \\
30.00 & 11.93 & -- & $6.51\times 10^{6}$ & -- \\
33.16 & 13.64 & -- & $5.98\times 10^{6}$ & -- \\
36.66 & 15.57 & -- & $5.53\times 10^{6}$ & -- \\
40.53 & 17.75 & -- & $5.13\times 10^{6}$ & -- \\
44.80 & 20.15 & -- & $4.78\times 10^{6}$ & -- \\
49.53 & 22.84 & -- & $4.48\times 10^{6}$ & -- \\
54.75 & 25.87 & -- & $4.22\times 10^{6}$ & -- \\
60.52 & 29.19 & -- & $3.98\times 10^{6}$ & -- \\
66.91 & 33.02 & -- & $3.77\times 10^{6}$ & -- \\
73.97 & 37.23 & -- & $3.58\times 10^{6}$ & -- \\
81.78 & 41.87 & -- & $3.42\times 10^{6}$ & -- \\
90.41 & 47.06 & -- & $3.27\times 10^{6}$ & -- \\
99.96 & 52.90 & -- & $3.13\times 10^{6}$ & -- \\
\hline
\end{tabular}
\end{table*}

\begin{table*}[ht!]
\caption{Same as Table \ref{tab:mesa_properties_014} for $Z=0.002$ models.}
\label{tab:mesa_properties_002}
\centering
\begin{tabular}{ccccc}
\hline\hline
$M_{\rm init} \ (M_\odot)$ & $M_{\rm He} \ (M_\odot)$ & $M_{\rm strip} \ (M_\odot)$ & $t_{\rm MS}$ (yr) & $t_{\rm strip}$ (yr) \\
\hline
2.00 & 0.25 & 0.39 & $8.38\times 10^{8}$ & -- \\
2.21 & 0.29 & 0.45 & $6.55\times 10^{8}$ & -- \\
2.44 & 0.33 & 0.51 & $5.13\times 10^{8}$ & -- \\
2.70 & 0.37 & 0.59 & $4.02\times 10^{8}$ & $8.99\times 10^{7}$ \\
2.99 & 0.42 & 0.67 & $3.16\times 10^{8}$ & $5.77\times 10^{7}$ \\
3.30 & 0.48 & 0.76 & $2.49\times 10^{8}$ & $3.95\times 10^{7}$ \\
3.65 & 0.55 & 0.87 & $1.97\times 10^{8}$ & $2.99\times 10^{7}$ \\
4.04 & 0.62 & 1.01 & $1.57\times 10^{8}$ & $1.92\times 10^{7}$ \\
4.46 & 0.72 & 1.15 & $1.26\times 10^{8}$ & $1.47\times 10^{7}$ \\
4.93 & 0.82 & 1.30 & $1.01\times 10^{8}$ & $1.20\times 10^{7}$ \\
5.45 & 0.94 & 1.50 & $8.18\times 10^{7}$ & $8.34\times 10^{6}$ \\
6.03 & 1.09 & 1.68 & $6.67\times 10^{7}$ & $6.05\times 10^{6}$ \\
6.66 & 1.26 & 1.93 & $5.47\times 10^{7}$ & $4.74\times 10^{6}$ \\
7.37 & 1.47 & 2.22 & $4.50\times 10^{7}$ & $3.64\times 10^{6}$ \\
8.15 & 1.71 & 2.54 & $3.73\times 10^{7}$ & $2.90\times 10^{6}$ \\
9.00 & 1.98 & 2.92 & $3.12\times 10^{7}$ & $2.34\times 10^{6}$ \\
9.95 & 2.32 & 3.37 & $2.63\times 10^{7}$ & $1.89\times 10^{6}$ \\
11.01 & 2.71 & 3.88 & $2.24\times 10^{7}$ & $1.58\times 10^{6}$ \\
12.17 & 3.18 & 4.47 & $1.91\times 10^{7}$ & $1.33\times 10^{6}$ \\
13.45 & 3.70 & 5.15 & $1.65\times 10^{7}$ & $1.13\times 10^{6}$ \\
14.87 & 4.33 & 5.91 & $1.43\times 10^{7}$ & $9.80\times 10^{5}$ \\
16.44 & 5.05 & 6.63 & $1.25\times 10^{7}$ & $8.65\times 10^{5}$ \\
18.17 & 5.85 & 7.26 & $1.11\times 10^{7}$ & $7.24\times 10^{5}$ \\
20.09 & 6.80 & -- & $9.85\times 10^{6}$ & -- \\
22.21 & 7.88 & -- & $8.82\times 10^{6}$ & -- \\
24.55 & 9.12 & -- & $7.95\times 10^{6}$ & -- \\
27.14 & 10.50 & -- & $7.22\times 10^{6}$ & -- \\
30.00 & 12.08 & -- & $6.59\times 10^{6}$ & -- \\
33.17 & 13.84 & -- & $6.05\times 10^{6}$ & -- \\
36.66 & 15.84 & -- & $5.58\times 10^{6}$ & -- \\
40.53 & 18.12 & -- & $5.18\times 10^{6}$ & -- \\
44.81 & 20.62 & -- & $4.83\times 10^{6}$ & -- \\
49.53 & 23.42 & -- & $4.52\times 10^{6}$ & -- \\
54.76 & 26.55 & -- & $4.25\times 10^{6}$ & -- \\
60.53 & 30.06 & -- & $4.01\times 10^{6}$ & -- \\
66.92 & 33.96 & -- & $3.80\times 10^{6}$ & -- \\
73.98 & 38.29 & -- & $3.61\times 10^{6}$ & -- \\
81.79 & 43.13 & -- & $3.44\times 10^{6}$ & -- \\
90.43 & 48.54 & -- & $3.28\times 10^{6}$ & -- \\
99.97 & 54.54 & -- & $3.15\times 10^{6}$ & -- \\
\hline
\end{tabular}
\end{table*}

\begin{table*}[ht!]
\caption{Same as Table \ref{tab:mesa_properties_014} for $Z=0.0002$ models.}
\label{tab:mesa_properties_0002}
\centering
\begin{tabular}{ccccc}
\hline\hline
$M_{\rm init} \ (M_\odot)$ & $M_{\rm He} \ (M_\odot)$ & $M_{\rm strip} \ (M_\odot)$ & $t_{\rm MS}$ (yr) & $t_{\rm strip}$ (yr) \\
\hline
2.00 & 0.26 & 0.42 & $7.05\times 10^{8}$ & -- \\
2.21 & 0.30 & 0.49 & $5.58\times 10^{8}$ & -- \\
2.44 & 0.34 & 0.57 & $4.43\times 10^{8}$ & $1.08\times 10^{8}$ \\
2.70 & 0.39 & 0.66 & $3.52\times 10^{8}$ & $6.72\times 10^{7}$ \\
2.99 & 0.44 & 0.74 & $2.80\times 10^{8}$ & $4.75\times 10^{7}$ \\
3.30 & 0.50 & 0.87 & $2.24\times 10^{8}$ & -- \\
3.65 & 0.57 & 1.00 & $1.79\times 10^{8}$ & $2.17\times 10^{7}$ \\
4.04 & 0.65 & 1.15 & $1.45\times 10^{8}$ & $1.56\times 10^{7}$ \\
4.46 & 0.74 & 1.30 & $1.17\times 10^{8}$ & $1.25\times 10^{7}$ \\
4.93 & 0.85 & 1.49 & $9.52\times 10^{7}$ & $8.93\times 10^{6}$ \\
5.45 & 0.98 & 1.70 & $7.76\times 10^{7}$ & $7.95\times 10^{6}$ \\
6.03 & 1.12 & 1.92 & $6.37\times 10^{7}$ & $5.59\times 10^{6}$ \\
6.66 & 1.29 & 2.20 & $5.26\times 10^{7}$ & $4.26\times 10^{6}$ \\
7.37 & 1.49 & 2.53 & $4.36\times 10^{7}$ & $3.45\times 10^{6}$ \\
8.15 & 1.74 & 2.90 & $3.64\times 10^{7}$ & $2.73\times 10^{6}$ \\
9.00 & 2.02 & 3.33 & $3.06\times 10^{7}$ & $2.22\times 10^{6}$ \\
9.95 & 2.35 & 3.81 & $2.59\times 10^{7}$ & $1.82\times 10^{6}$ \\
11.01 & 2.75 & 4.38 & $2.21\times 10^{7}$ & $1.53\times 10^{6}$ \\
12.17 & 3.20 & 5.04 & $1.89\times 10^{7}$ & $1.29\times 10^{6}$ \\
13.45 & 3.75 & 5.73 & $1.64\times 10^{7}$ & $1.12\times 10^{6}$ \\
14.87 & 4.36 & 6.36 & $1.43\times 10^{7}$ & -- \\
16.44 & 5.08 & 7.20 & $1.25\times 10^{7}$ & $8.51\times 10^{5}$ \\
18.17 & 5.90 & 7.92 & $1.10\times 10^{7}$ & $8.03\times 10^{5}$ \\
20.09 & 6.86 & -- & $9.83\times 10^{6}$ & -- \\
22.21 & 7.95 & -- & $8.81\times 10^{6}$ & -- \\
24.55 & 9.19 & -- & $7.95\times 10^{6}$ & -- \\
27.14 & 10.61 & -- & $7.21\times 10^{6}$ & -- \\
30.00 & 12.20 & -- & $6.58\times 10^{6}$ & -- \\
33.17 & 13.98 & -- & $6.04\times 10^{6}$ & -- \\
36.67 & 16.00 & -- & $5.57\times 10^{6}$ & -- \\
40.54 & 18.28 & -- & $5.17\times 10^{6}$ & -- \\
44.81 & 20.84 & -- & $4.82\times 10^{6}$ & -- \\
49.54 & 23.76 & -- & $4.51\times 10^{6}$ & -- \\
54.77 & 26.96 & -- & $4.24\times 10^{6}$ & -- \\
60.55 & 30.51 & -- & $4.00\times 10^{6}$ & -- \\
66.94 & 34.50 & -- & $3.79\times 10^{6}$ & -- \\
74.01 & 39.02 & -- & $3.60\times 10^{6}$ & -- \\
81.82 & 43.90 & -- & $3.43\times 10^{6}$ & -- \\
90.45 & 49.40 & -- & $3.28\times 10^{6}$ & -- \\
99.99 & 55.46 & -- & $3.14\times 10^{6}$ & -- \\
\hline
\end{tabular}
\end{table*}

\end{appendix}

\end{document}